\documentclass{article}

\usepackage[utf8]{inputenc}
\usepackage[fleqn]{amsmath}
\usepackage{amsfonts, amssymb}

\usepackage[pdftex,backref,colorlinks]{hyperref}
  \hypersetup{
     colorlinks,
     linkcolor={red!50!blue},
     citecolor={blue!50!green},
     urlcolor={blue!80!black}
 }
\usepackage[ntheorem]{empheq}
\usepackage[hyperref]{ntheorem}

\newtheorem{theorem}{Theorem}
\newtheorem{lemma}{Lemma}

\newtheorem{corollary}{Corollary}
\newtheorem{definition}{Definition}

\usepackage{extarrows}                        
\usepackage{units} 
\usepackage{bm} 
\usepackage{bbold}                    
\usepackage{tensor}                     
\usepackage{soul} 
\usepackage{cancel} 
\usepackage[retainorgcmds]{IEEEtrantools} 
\usepackage{graphicx}
\usepackage{fixltx2e} 
\usepackage{array} 
\newcolumntype{C}[1]{>{\centering\let\newline\\\arraybackslash\hspace{0pt}}m{#1}}

\usepackage[numbers,sort&compress]{natbib} 
\usepackage{color}
\usepackage{realboxes}
\usepackage{framed}
\usepackage{array} 
\usepackage{mathtools} 
\usepackage[version=3]{mhchem} 

\usepackage{enumerate} 
\usepackage{bookmark}                       
\usepackage{float}                    
\usepackage{rotating}
\usepackage{adjustbox}

\usepackage{ytableau} 
\usepackage{youngtab}

\ytableausetup{mathmode, centertableaux}
\usepackage{xargs}
\usepackage{letltxmacro}

\let\oldytableau\ytableau
\let\endoldytableau\endytableau
\renewenvironment{ytableau}{\begin{adjustbox}{scale=.78}\oldytableau}{\endoldytableau\end{adjustbox}}

\usepackage{xcolor}
\usepackage{tikz}
\usepackage{tkz-euclide}
\usetikzlibrary{matrix,fit,arrows,decorations}
\usepackage{pbox}
\newcommand{\diagram}[2][{}]{\pbox{\textwidth}{\includegraphics[#1]{{#2}}}}

\newcommand{\Tr}[1]{\mathrm{Tr}\left(#1\right)}

\newcommand{\FPic}[2][{}]{\hspace{-0.27mm}\pbox{\textwidth}{\includegraphics[#1]{{#2}}}\hspace{-0.27mm}}

\newcommand{\ybox}[1]{\begin{ytableau} #1 \end{ytableau}}

\newcommand{\SUN}{\mathsf{SU}(N)}
\newcommand{\MixedPow}[2]{V^{\otimes
    #1}\otimes\left(V^*\right)^{\otimes #2}}
\newcommand{\Pow}[1]{V^{\otimes #1}}
\newcommand{\Pows}[1]{{V^*}^{\otimes #1}}
\newcommand{\DPow}[1]{\left(V^*\right)^{\otimes #1}}
\newcommand{\Lin}[1]{\mathrm{Lin}\left( #1 \right)}
\newcommand{\API}[1]{\mathsf{API}\left( #1 \right)}

\def\smath#1{\text{\scalebox{.8}{$#1$}}}
\def\sfrac#1#2{\smath{\frac{#1}{#2}}}

\def\lmath#1{\text{\scalebox{1.2}{$#1$}}}
\def\lfrac#1#2{\lmath{\frac{#1}{#2}}}

\newcommand{\notimplies}{%
  \mathrel{{\ooalign{\hidewidth$\not\phantom{=}$\hidewidth\cr$\implies$}}}}

\makeatletter
\newcommand{\vast}{\bBigg@{3}}
\newcommand{\Vast}{\bBigg@{4}}
\makeatother

\tikzstyle basiclabel=[draw=none,fill=none,shape=rectangle,inner sep=2pt,scale=.8]
\tikzstyle leftlabel=[basiclabel,anchor=east]
\tikzstyle rightlabel=[basiclabel,anchor=west]

\newcommand{\qed}{\hfill\tikz{\draw[draw=black,line width=0.6pt] (0,0) rectangle (2.8mm,2.8mm);}\bigskip}

\renewenvironment{abstract}{\centering\begin{minipage}{.95\textwidth}
\sffamily{\bf Abstract:}}
{\end{minipage}\vskip 3em}
\makeatletter
\renewcommand\@maketitle{\hfill
\begin{minipage}{\textwidth}
\vskip 2em
\let\footnote\thanks 
{\LARGE \bf \@title \par }
\vskip 1.5em
{\large \@author \par}
\end{minipage}
\vskip 3em \par
}
\makeatother
\usepackage{authblk}

\usepackage[a4paper,left=2.0cm]{geometry}
\title{Compact Hermitian Young Projection Operators}
\author[1]{J. Alcock-Zeilinger}
\author[1]{H. Weigert}
\affil[1]{\small University of Cape Town; Dept. of Physics, Private Bag X3, Rondebosch 7701, South Africa}
\date{October 2016}

\begin{document}
\maketitle

\begin{abstract}
  \noindent In this paper, we describe a compact and practical
  algorithm to construct Hermitian Young projection operators for
  irreducible representations of the special unitary group $\SUN$, and
  discuss why ordinary non-Hermitian Young projection operators are
  unsuitable for physics applications. The proof of this construction
  algorithm uses the iterative method described by Keppeler and
  Sjödahl in~\cite{Keppeler:2013yla}. We further show that Hermitian
  Young projection operators share desirable properties with Young
  tableaux, namely a nested hierarchy when ``adding a particle''. We
  close by exhibiting the enormous advantage of the Hermitian Young
  projection operators constructed in this paper over those given by
  Keppeler and Sj\"odahl.
\end{abstract}

\setlength{\parskip}{2pt plus 2pt minus 1pt}
\tableofcontents
\setlength{\parskip}{7pt plus 2pt minus 1pt}

\section{Introduction \& outline}\label{sec:Introduction}

\subsection{Historical overview}\label{sec:IntroHistory}

More than a hundred years ago, the representation theory of compact,
semi-simple Lie groups, in particular also of $\SUN$, were a hot topic
of research. Most known to physicists is the work done by Clebsch and
Gordan, where the product representations of $\SUN$ can be classified
using the Clebsch-Gordan
coefficients~\cite{Tung:1985na,Fulton:2004,Peskin:1995ev}. This is the
textbook method for $N=2$ to find the irreducible representations of
spin of an $m$-particle configuration by giving an explicit change of
basis. While this method is perfectly adequate also for $N\neq 2$, it
requires one to choose the parameter $N$ at the start of the
calculation. Thus, this approach is of little use to us, as we mainly
strive to apply representation theory in a context of QCD, where it is
often essential for $N$ (representing $N_c$, the number of colors in
this case) to be a parameter to be varied at the end of the
calculation to get a better understanding of underlying
structures~\cite{Kovchegov:2008mk,Marquet:2010cf,Lappi:2016gqe}.

Shortly after the research by Clebsch and Gordan was conducted,
Eli{\'e} Cartan introduced another method of finding the irreducible
representations of Lie groups via finding certain subalgebras of the
associated Lie algebras~\cite{Cartan1894structure} since known as
Cartan subalgebras. This method is based on finding the highest
weights corresponding to the irreducible representations, and then
constructing all basis states within it. This process was used by
Gell-Mann in 1961~\cite{GellMann:1961ky,GellMann:1964xy} when he
introduced the \emph{eight-fold way} (here $N$ represents $N_f$ the
number of flavors) to order hadrons into flavor multiplets such as
the baryon octet and decuplet featuring prominently in any introductory
text on particle physics and as a motivation to study representation
theory in many a mathematical introduction to the
topic~\cite{Fulton:2004, KosmannSchwarzbach:2000,
  Griffiths:2008zz}. Usually, one also fixes the parameter $N$ from
the outset when using this method. While it is possible keep $N$ as a
parameter\footnote{This is an elusive piece of knowledge:
  Fulton~\cite[chapter 8.2, Lemma 4]{Fulton:1997}, for example,
  provides the basis for finding highest weight vectors directly from
  tableaux without fixing $N$.}  Cartan's method is of restricted use
in practical applications as it requires us to construct all $N^m$
associated basis elements to fully characterize the irreducible
representations required to span $\Pow{m}$: For an unspecified $N$,
this becomes a daunting task.

In 1928, approximately three decades after Cartan's work, Alfred Young
conceived a combinatorial method of classifying the idempotents on the
algebra of permutations~\cite{Young:1928}. This method was
subsequently used in the 1930's to establish a connection between
these idempotents and the irreducible representations of compact,
semi-simple Lie groups, now known as the \emph{Schur-Weyl
  duality},~\cite{Weyl:1946}. This duality is based on the
\emph{theory of invariants},~\cite{Cvitanovic:2008zz,Weyl:1946}, which
exploits the invariants (in particular the \emph{primitive
  invariants}) of a Lie group $G$ and constructs projection operators
corresponding to the irreducible representations of $G$. Since the
present paper will rely on the theory of invariants, a short overview
is in order: We will deal with a product representations of $\SUN$
constructed from its fundamental representation on a given vector
space $V$ with $\text{dim}(V) = N$, whose action will simply be denote
by $v\mapsto U v$ for all $U\in\SUN$ and $v\in V$. Choosing a basis
$\{e_{(i)} |i = 1, \ldots, \text{dim}(V)\}$ such that $v = v^i e_{(i)}$ this
becomes $v^i \mapsto \tensor{U}{^i_j} v^j$. This immediately induces a
product representation of $\SUN$ on $\Pow{m}$ if
one uses this basis of $V$ to induce a basis on $\Pow{m}$ so that a
general element $\bm v\in\Pow{m}$ takes the form
$\bm v = v^{i_1\ldots i_m} e_{(i_1)}\otimes\cdots\otimes e_{(i_m)}$:
\begin{align}
  \label{eq:SUN-Action}
  U\circ\bm v 
  = 
  U\circ v^{i_1\ldots i_m} e_{(i_1)}\otimes\cdots\otimes e_{(i_m)}
  :=
  \tensor{U}{^{i_1}_{j_1}} 
  \cdots 
  \tensor{U}{^{i_m}_{j_m}} 
  v^{j_1\ldots j_m} e_{(i_1)}\otimes\cdots\otimes e_{(i_m)}
\end{align} 
Since all the factors in $\Pow{m}$ are identical, the notion of
permuting the factors is a natural one and leads to a linear map on
$\Pow{m}$ according to
\begin{align}
  \label{eq:perm-facs-def}
  \rho\circ\bm v 
  = 
  \rho\circ v^{i_1\ldots i_m} e_{(i_1)}\otimes\cdots\otimes e_{(i_m)}
  :=
  v^{\rho(i_1)\ldots \rho(i_m)} e_{(i_1)}\otimes\cdots\otimes e_{(i_m)}
\end{align}
where $\rho$ is an element of $S_m$, the group of permutations of $m$
objects.\footnote{Permuting the basis vectors instead involves
  $\rho^{-1}$: $v^{\rho(i_1)\ldots \rho(i_m)} e_{(i_1)} \otimes \cdots
  \otimes e_{(i_m)} = v^{i_1\ldots i_m} e_{(\rho^{-1}(i_1))} \otimes
  \cdots \otimes e_{(\rho^{-1}(i_m))}$}{} From the
definitions~\eqref{eq:SUN-Action} and~\eqref{eq:perm-facs-def} one
immediately infers that the product representation commutes with all
permutations on any $\bm v\in\Pow{m}$:
\begin{equation}
\label{eq:InvariantsIntro1}
  U \circ \rho \circ \bm v  = \rho \circ U\circ  \bm v
  \ .
\end{equation}
In other words, any such permutation $\rho$ is an \emph{invariant} of
$\SUN$ (or in fact any Lie group $G$ acting on $V$):
\begin{align}
  \label{eq:per-invariant}
  U\circ\rho\circ U^{-1} = \rho
  \ .
\end{align}
It can further be shown that these permutations in fact span the space
of all linear invariants of $\SUN$ over
$\Pow{m}$~\cite{Cvitanovic:2008zz}. The permutations are thus referred
to as the \emph{primitive invariants} of $\SUN$ over $\Pow{m}$. The
full space of linear invariants is then given by
\begin{align}
  \label{eq:PI-def}
  \API{\SUN,\Pow{m}} 
  := 
  \Bigl\{ 
  \sum_{\sigma\in S_m} \alpha_\sigma \sigma \Big| 
  \alpha_\sigma\in \mathbb R, \sigma\in S_m \Bigr\}
  \subset
  \Lin{\Pow{m}}
  \ .
\end{align}
Note we are exclusively focusing on $\API{\SUN,\Pow{m}}$ and make no
efforts to directly discuss the invariants on
$\Pow{m}\otimes\Pows{m'}$. For $\SUN$ these are implicitly included
due to the presence of $\epsilon^{i_1 \ldots i_N}$ as a second
invariant besides $\delta^i_{\ j}$ -- the construction of explicit
algorithms tailored to expose this structure are beyond the scope of
this paper. For a more comprehensive introduction to invariant theory,
readers are referred to~\cite{Weyl:1946, Bourbaki7-9:2000, Cvitanovic:2008zz,
  Tung:1985na}.

If we denote by $\mathcal Y_m$ the complete set of irreducible
representations of $\SUN$ in $\Pow{m}$, (according to Young this is in
fact the set of Young tableaux with $m$-boxes), then a meaningful set
of projection operators $\{ L_\Theta \in\API{\SUN,\Pow{m}} |
\Theta\in\mathcal Y_m \}$ onto the invariant subspaces must satisfy
the following three properties
\begin{subequations}
  \label{eq:LY-class}
\begin{enumerate}
\item The projetors must be 
  idempotent, that is they satisfy 
  \begin{align}
    \label{eq:LY-idem}
    L_\Theta\cdot L_\Theta = L_\Theta
    \ .
  \end{align}
\item The operators are mutually orthogonal in the sense that
  \begin{align}
    \label{eq:LY-orth}
    L_\Theta\cdot L_\Phi = 0 \hspace{1cm} \text{ if $\Theta\neq\Phi$.}
  \end{align}

\item The complete set of projection
  operators for $\SUN$ over $\Pow{m}$ sum up to the identity element
  of $\Pow{m}$,
  \begin{align}
    \label{eq:LY-decom-unity}
    \sum\limits_{\Theta\in\mathcal Y_m} L_\Theta = \mathrm{id}_{\Pow{m}}
    \ .
  \end{align}
\end{enumerate}
\end{subequations}
Projection operators in $\API{\SUN,\Pow{m}}$, derived from Young
tableaux that satisfy all three conditions without restrictions,
together with equation~\eqref{eq:InvariantsIntro1} then classify all
irreducible representations of
$\SUN$,~\cite{Littlewood:1950,Tung:1985na,Cvitanovic:2008zz,Fulton:2004,Keppeler:2013yla}.

Young projectors $Y_\Theta$ are suitable for this purpose only for
$m=1,2,3,4$: They fail to satisfy conditions~\eqref{eq:LY-orth}
and~\eqref{eq:LY-decom-unity} from $m = 5$ onwards so that
additional work is needed to ensure that the theory adresses
\emph{all} irreducible representations contained in $\Pow{m}$.
In~\cite[sec.~5.4]{Littlewood:1950}, Littlewood describes how to
``correct'' Young projectors $Y_\Theta$ for $m \ge 5$ to restore
conditions~\eqref{eq:LY-orth} and~\eqref{eq:LY-decom-unity}. We
call the resulting operators Littlewood-Young (LY) projectors and denote
them $L_\Theta$ to distinguish them form the original Young projectors
$Y_\Theta$. A short account of their construction using our notation
is given in appendix~\ref{sec:littl-young-proj} for completeness.

For classification purposes one does \emph{not} require the operators
$L_\Theta$ to be Hermitian and from $m=3$ a growing fraction of the
LY projectors lack that often useful feature.

On the positive side, the LY projection operators are
compact and can be constructed keeping $N$ as a parameter, both
desirable properties for the practitioner.

With Young's (and Littlewood's) contributions, the representation
theory of compact, semi-simple Lie groups was considered a fully
understood and complete theory from approximately 1950 onward, even
though many misconception, especially about the full extent of the
theory remained, in particular among casual practitioners.

In the 1970's, Penrose devised a graphical method of dealing with
primitive invariants of Lie groups including Young projection
operators~\cite{Penrose1971Mom,Penrose1971Com}, which was
subsequently applied in a collaboration with
MacCallum~\cite{Penrose:1972ia}. This graphical method, now dubbed
the \emph{birdtrack formalism}, was modernized and further developed
by Cvitanovi{\'c}~\cite{Cvitanovic:2008zz} in recent years. The
immense benefit of the birdtrack formalism is that it makes the
actions of the operators visually accessible and thus more intuitive.
For illustration, we give as an example the permutations of $S_3$
written both in their cycle notation (see~\cite{Tung:1985na} for a
textbook introduction) as well as birdtracks:
\begin{equation}
\label{eq:S3-Birdtracks}
  \underbrace{\FPic{3ArrLeft}\FPic{3IdSN}\FPic{3ArrRight}}_{\mathrm{id}}\; , \quad \underbrace{\FPic{3ArrLeft}\FPic{3s12SN}\FPic{3ArrRight}}_{(12)}\; , \quad \underbrace{\FPic{3ArrLeft}\FPic{3s13SN}\FPic{3ArrRight}}_{(13)}\; , \quad \underbrace{\FPic{3ArrLeft}\FPic{3s23SN}\FPic{3ArrRight}}_{(23)}\; , \quad \underbrace{\FPic{3ArrLeft}\FPic{3s123SN}\FPic{3ArrRight}}_{(123)}\; , \quad \underbrace{\FPic{3ArrLeft}\FPic{3s132SN}\FPic{3ArrRight}}_{(132)}\; .
\end{equation}
The action of each of the above permutations on a tensor product
$v_1\otimes v_2\otimes v_3$ is clear, for example
\begin{equation}
  (123) \left(v_1\otimes v_2\otimes v_3\right) = v_3\otimes v_1\otimes v_2.
\end{equation}
In the birdtrack formalism, this equation is written as
\begin{equation}
  \FPic{3ArrLeft}\FPic{3s123SN}\FPic{3ArrRight}\FPic{3v123Labels} \; = \; \FPic{3v312Labels} \; ,
\end{equation}
where each term in the product $v_1\otimes v_2\otimes v_3$ (written as
a tower $\FPic{3v123Labels}$) can be thought of as being moved along
the lines of
$\FPic{3ArrLeft}\FPic{3s123SN}\FPic{3ArrRight}$ . Birdtracks are thus
naturally read from right to left as is also indicated by the arrows
on the legs.\footnote{The direction of arrows thus encodes whether the
  leg is acting on $V$ or its dual $V^*$, \emph{c.f.} section~\ref{sec:HermitianLinearMapsSection}.}

The representation theory of $\SUN$ found a short-lived revival in
2014, when Keppeler and Sjödahl presented a construction algorithm for
Hermitian projection operators (based on the idempotents already found
by Young)~\cite{Keppeler:2013yla}. This paper arose out of a need for
\emph{Hermitian} operators in a physics context. In their paper, the
birdtrack formalism was used to devise a recursive construction
algorithm for Hermitian projection operators. This algorithm does
produce Hermitian operators satisfying the
requirements~\eqref{eq:LY-class} (idempotency, operator
orthogonality, and decomposition of unity) for a classification tool
and allows to keep $N$ as a parameter just as the Littlewood-Young
operators devised much earlier.
 
% Keppeler and Sjödahl's operators start to differ from the
% Young-Littlewood projection operators from $m = 3$, where the latter
% cease to be hermitian in general. As we will point out below (see
% section~\ref{sec:IntroHermitianOps}), hermiticity is
% directly tied with other structural information one inherently
% associates with product representations in the sense of a refinement
% of~\eqref{eq:LY-decom-unity}.

We will argue below that Hermiticity provides more than just cosmetic
advantages: We show that Hermitean Young projection operators mimic
the tableau hierarchy of Young tableaux -- a fact that has been
overlooked by KS. We directly trace this back to Hermiticity -- the
Littlewood-Young projection operators like any non-Hermitian
corrected version of Young projectors must fail in this regard.

To the practitioner this comes at a high price: the expressions
created by Keppeler and Sjödahl's algorithm soon become extremely long
and thus computationally expensive and impractical.  In this paper, we
give a considerably more efficient and thus more \emph{practical}
construction algorithm for Hermitian Young projection operators
yielding compact expressions.

The remainder of this present section~\ref{sec:Introduction} gives a
detailed outline of this paper and lists all goals that will be
achieved along the way.

\subsection{Where non-Hermitian projection
  operators fail to deliver}\label{sec:IntroHermitianOps}

Among practitioners, many misconceptions still exist with regards to
Young projection operators and their corrected forms. The probably
most generic one stems from the presentation of Young tableaux and
their corresponding projection operators in the literature: It is usually explained
in parallel that
\begin{enumerate}
\item Young tableaux follow a
progressive hierarchy, in the sense that tableaux consisting of $n$
boxes can be obtained from Young tableaux of $(n-1)$ boxes merely by
adding the box \ybox{n} in the appropriate place. For example, the
tableaux $\scalebox{0.75}{\begin{ytableau} 1 & 2 \\ 3\end{ytableau}}$
and $\scalebox{0.75}{\begin{ytableau} 1 & 2 & 3\end{ytableau}}$ can
both be obtained from the tableau $\scalebox{0.75}{\begin{ytableau} 1
    & 2 \end{ytableau}}$ ,
\begin{equation}
\label{eq:Intro0}
\diagram[scale=.9]{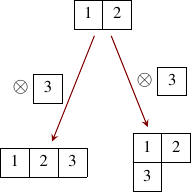}
\; 
\hspace{1.2cm} \text{and also} \hspace{1.2cm}
\diagram[scale=.9]{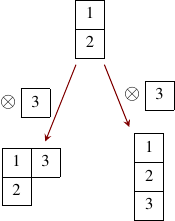}
\; .
\end{equation}
Since this is a key concept, we will need some notation and
nomenclature to refer to it. In general, for a \emph{particular} Young tableau $\Theta$
with $(n-1)$ boxes, we will denote the set of all Young
tableaux that can be obtained from $\Theta$ by adding the box \ybox{n}
by
\begin{equation}
  \label{eq:ChildSetDef}
  \Bigl\lbrace \Theta \otimes \ybox{n} \Bigr\rbrace;
\end{equation}
this set will also be referred to as the \emph{child-set} of $\Theta$.

\item This is complemented by the fact that the \emph{corrected Young
    projection operators (e.g. the LY-operators) span the full space}, that is
\begin{equation}
  \label{eq:YSpanFullSpace}
  \sum_{\Theta \in \mathcal{Y}_n} L_{\Theta} = \mathrm{id}_n %\hspace{1cm} n=1,2,3,4,
\end{equation}
where $\mathcal{Y}_n$ is understood to be the set of \emph{all} Young
tableaux consisting of $n$ boxes (for a fixed $n$), and
$\mathrm{id}_n$ is the identity operator on the space $\Pow{n}$. (As
mentioned earlier this also holds for the standard Young projection
operators for $m\leq4$.) Equation~\eqref{eq:YSpanFullSpace} is also
known as the \emph{completeness relation} of Littlewood-Young projection
operators. In particular for $n=2,3$ (where the $L_\Theta$ reduce to
$Y_\Theta$, \emph{c.f.} appendix~\ref{sec:littl-young-proj}),
\ytableausetup{mathmode, boxsize=0.7em}%
\begin{equation}
  \label{eq:Y2Span2}
  Y_{\begin{ytableau} \scriptstyle 1 & \scriptstyle 2 \end{ytableau}}
  + Y_{\begin{ytableau} \scriptstyle 1 \\ \scriptstyle
      2 \end{ytableau}} = \mathrm{id}_2
\end{equation}
and
\begin{equation}
  \label{eq:Y3Span3}
  Y_{\begin{ytableau} \scriptstyle 1 & \scriptstyle 2 & \scriptstyle 3 \end{ytableau}}
  + Y_{\begin{ytableau} \scriptstyle 1 & \scriptstyle 2 \\
      \scriptstyle 3 \end{ytableau}} 
  + Y_{\begin{ytableau} \scriptstyle 1 & \scriptstyle 3 \\
      \scriptstyle 2 \end{ytableau}} 
  + Y_{\begin{ytableau} \scriptstyle 1 \\ \scriptstyle
      2 \\ \scriptstyle 3 \end{ytableau}} = \mathrm{id}_3.
\end{equation}
The completeness relations offer decompositions of unity in both cases.

\end{enumerate}

The hierarchy relation~\eqref{eq:Intro0} of Young tableaux and the
completeness relation~\eqref{eq:YSpanFullSpace} of LY projection
operators then might lead the unwary reader to (incorrectly) infer
that the tableau hierarchy~\eqref{eq:Intro0} automatically implies
that this decomposition of unity is in fact \emph{nested}, i.e. that
the child tableaux correspond to projection operators that furnish
decompositions of their parent projectors so that the identities
\begin{equation}\label{eq:Intro1}
  Y_{\begin{ytableau}
    \scriptstyle 1 & \scriptstyle 2 & \scriptstyle 3 \end{ytableau}} + 
Y_{\begin{ytableau} \scriptstyle 1 & \scriptstyle 2 \\
    \scriptstyle 3 \end{ytableau}} 
 \overset{?}{=} Y_{\begin{ytableau} \scriptstyle 1 & \scriptstyle
    2 \end{ytableau}}
\hspace{1cm}\text{ and }\hspace{1cm}
Y_{\begin{ytableau} \scriptstyle 1 & \scriptstyle 3 \\
      \scriptstyle 2 \end{ytableau}} 
  + Y_{\begin{ytableau} \scriptstyle 1 \\ \scriptstyle
      2 \\ \scriptstyle 3 \end{ytableau}} \overset{?}= Y_{\begin{ytableau} \scriptstyle 1 \\ \scriptstyle
      2 \end{ytableau}}
\end{equation}
would hold. Both of these ``equations'' can easily be shown to be
false by direct calculation. The authors have not found any literature
that clearly states that relation~\eqref{eq:Intro0} holds for Young
tableaux only, and does not have a counterpart in terms of Littlewood-Young
projection operators.

In physics applications,~\eqref{eq:YSpanFullSpace} is often not
sufficient, and we require a counterpart of the structure
in~\eqref{eq:Intro0} for a suitable set of projection operators, thus
repairing the failure of ``equation''~\eqref{eq:Intro1}. The desired
analogue exists, it is given by the Hermitian Young projection
operators introduced by Keppeler and Sjödahl
(KS)~\cite{Keppeler:2013yla}, although this practically crucial
observation was \emph{not} mentioned by KS in their paper. In the
present paper, we will explicitly demonstrate that the tableau
hierarchy \eqref{eq:Intro0} can be transferred to the KS-operators in
the desired manner: Using $P_{\Theta}$ to denote the Hermitian Young
projection operator corresponding to the tableau $\Theta$, it turns
out that the decompositions in our example are indeed nested, so that
\begin{equation}\label{eq:Intro2}
  P_{\begin{ytableau}
    \scriptstyle 1 & \scriptstyle 2 & \scriptstyle 3 \end{ytableau}} + 
  P_{\begin{ytableau} \scriptstyle 1 & \scriptstyle 2 \\
    \scriptstyle 3 \end{ytableau}} 
  = 
  P_{\begin{ytableau} \scriptstyle 1 & \scriptstyle
    2 \end{ytableau}}
\hspace{1cm}\text{ and }\hspace{1cm}
  P_{\begin{ytableau} \scriptstyle 1 & \scriptstyle 3 \\
      \scriptstyle 2 \end{ytableau}} 
  + P_{\begin{ytableau} \scriptstyle 1 \\ \scriptstyle
      2 \\ \scriptstyle 3 \end{ytableau}} 
  = 
  P_{\begin{ytableau} \scriptstyle 1 \\ \scriptstyle
      2 \end{ytableau}}
\end{equation}
hold, and that this generalizes to all Hermitian projectors
corresponding to Young tableaux. Thus, the first goal of this paper will be to show that
this pattern holds in general:

\newtheorem{OpsSpanGoal}{Goal}
\begin{OpsSpanGoal}\label{thm:OpsSpanGoal}
  We are interested in a nested decomposition of projection operators
  in analogy to the the hierarchy relation of Young tableaux (exemplified
  in eq.~\eqref{eq:Intro0}) to operators,
  thus generalizing eq.~\eqref{eq:Intro2} to
  \begin{equation}
    \label{eq:Outline1}
    \sum_{\Phi \in \lbrace\Theta\otimes\ybox{\scriptstyle n}\rbrace}
    P_{\Phi} = P_{\Theta}
\ .
  \end{equation}
  \begin{enumerate}
  \item\label{itm:Goal1Part1} We will look at two particular examples
    which illustrate that the summation property~\eqref{eq:Outline1}
    does not hold for Young projection operators,
    section~\ref{NonHermitianSection}. In particular, we will find
    that assuming~\eqref{eq:Outline1} holds for Young projectors over
    $\Pow{n}$ and all their ancestors (up to some value of $n$) forces
    one to falsely conclude that these projectors are Hermitian. This
    serves as a motivation that the summation property should hold for
    \emph{Hermitian} Young projection operators.
    \ytableausetup{mathmode, boxsize=1.1em}%
\item\label{itm:Goal1Part2} In section~\ref{sec:HermitianSubspaceSpan}
  we will find our intuition restored when we prove that
  eq.~\eqref{eq:Outline1} indeed holds for all Hermitian Young
  projectors. This will be accomplished by using
    a shortened version of the KS-operators; a construction principle
    for these shortened operators is given in
    section~\ref{sec:ShortKSOperators}.
  % \end{enumerate}
  % Thus, having shown both implications ((eq.~\eqref{eq:Outline1}
  % holds)$\Longleftrightarrow$($P$ Hermitian)), we conclude that the
  % nested tableau hierarchy of Young tableaux transfers to their
  % corresponding projection operators if and only if they are
  % Hermitian.  \ytableausetup{mathmode, boxsize=0.7em}%
  % \begin{enumerate}[resume]
\item\label{itm:Goal1Part3} This then automatically extends over
  \emph{any number of generations} of tableaux, \emph{c.f.}
  eq.~\eqref{eq:SpanSubspaces2},
\begin{equation}
  \label{eq:Outline2}
\sum_{\Phi \in \lbrace\Theta\otimes\ybox{\scriptstyle
    m}\otimes\cdots\otimes\ybox{\scriptstyle n}\rbrace}
P_{\Phi}
=
P_{\Theta}
\qquad
\text{for $\Phi\in\mathcal{Y}_n$ and $\Theta\in\mathcal{Y}_{m-1}$, $m<n$}
\ ,
\end{equation}
section~\ref{sec:HermitianSubspaceSpan}.
  \end{enumerate}
\end{OpsSpanGoal}

\subsection{Shorter is better}

Having motivated the necessity of Hermitian Young projection
operators, we will now shift our focus to their application. In
particular, the authors of this paper are foremost interested in
applications in a QCD context such as is laid out
in~\cite{AlcockZeilinger2016Singlets}. With this objective in mind,
the Hermitian Young projection operators conceived by
KS~\cite{Keppeler:2013yla} soon lose all practical usefulness as the
number of factors in $\Pow{m}$ grows: the expressions become too
long and thus computationally expensive; a quality, that is explained
in section \ref{sec:MOLDAdvantage}.

An array of practical
tools~\cite{Alcock-Zeilinger:2016bss} particularly
suited for the \emph{birdtrack formalism}~\cite{Cvitanovic:2008zz}, in
which the Hermitian Young projection operators by KS were constructed,
allows to devise a new construction principle for Hermitian Young
projection operators, which we could not resist to dub
\emph{MOLD construction}. As elements in the algebra of invariants,
the MOLD-operators are identical to the KS-operators, however their
expressions in terms of symmetrizers and antisymmetrizers as well as
the number of steps used in the construction is shorter, often
dramatically so. We gain access to all the desired properties of the
KS-operators at a much lower computational cost: their
\emph{idempotency}, their \emph{mutual orthogonality}, their
\emph{completeness relation}\footnote{All of these properties of the
  KS-operators are described in Theorem~\ref{thm:KSProjectors} and
  in~\cite{Keppeler:2013yla}.}, and also the hierarchy
relation~\eqref{eq:Outline1}. A clear comparison between the MOLD and
the KS constructions and the resulting expressions for the Hermitian
Young projection operators can be found in section
\ref{sec:MOLDAdvantage}. This constitutes the second goal of this paper:

\newtheorem{MOLDOpsGoal}[OpsSpanGoal]{Goal}
\begin{MOLDOpsGoal}\label{thm:MOLDOpsGoal}
  We will provide a construction principle for Hermitian Young
  projection operators that produces \emph{compact}, and thus
  practically useful expression for these operators, section
  \ref{sec:CompactHermitianOps}. An explicit comparison of projection
  operators obtained from the MOLD- and the KS-algorithms is given
  in~\ref{sec:MOLDAdvantage}.
\end{MOLDOpsGoal}

\section{Tableaux, projectors, birdtracks, and conventions}\label{sec:Background}

Before we set out to achieve Goals~\ref{thm:OpsSpanGoal}
and~\ref{thm:MOLDOpsGoal}, we will provide a short sketch of
birdtracks and the way in which they relate to Young tableaux in
section~\ref{sec:BirdtracksSection}, mainly to prepare for
section~\ref{sec:Notation} where we establish the notation used in
this paper. For a more comprehensive introduction to the birdtrack
formalism refer to~\cite{Cvitanovic:2008zz}.

\ytableausetup
{mathmode, boxsize=normal}

\subsection{Birdtracks \& projection operators}\label{sec:BirdtracksSection}

Our aim in this section is to establish a link between Young
tableaux~\cite{Tung:1985na} and birdtracks~\cite{Penrose1971Mom,
  Penrose1971Com, Penrose:1972ia, Cvitanovic:2008zz}, as it is our
ultimate goal is to use the tools presented in this paper in a QCD
context where $\SUN$ with $N=N_c=3$ is the gauge group of the
theory~\cite{AlcockZeilinger2016Singlets} in a manner that allows us
to keep $N$ as a parameter in order to have direct access to
additional structure, not least the large $N_c$ limit.

As mentioned earlier, one way to generate the projection
operators corresponding to the irreducible representations of $\SUN$
without being forced to choose a numerical value for $N$ at the outset
is via the method of Young projection operators, which can be
constructed from Young tableaux, see for example~\cite{Tung:1985na,
  Fulton:2004,Fulton:1997,Sagan:2000} and other standard textbooks.

We therefore begin with a short memory-refresher on Young tableaux,
our main source for this will be~\cite{Tung:1985na}. A \emph{Young
  tableau} is defined to be an arrangement of $m$ boxes which are
left-aligned and top-aligned, and each box is filled with a unique
integer between $1$ and $m$ such that the numbers increase from left
to right in each row and from top to bottom in each column\footnote{In
  some references, the presently described tableau may also be
  referred to as a \emph{standard} Young tableau~\cite{Fulton:1997,Sagan:2000}.}. For example, among
the two conglomerations of boxes
\begin{equation}
  \Theta =
  \begin{ytableau}
    1 & 3 & 6 \\
    2 & 5 & 7 \\
    4
  \end{ytableau} \quad \text{and} \quad
\tilde{\Theta} =
\begin{ytableau}
  3 & 4 & 1 & \none \\
  2 & 6 & 7 & 5
\end{ytableau} 
\ ,
\end{equation}
$\Theta$ is a Young tableau but $\tilde{\Theta}$ is not since
$\tilde{\Theta}$ is neither top aligned nor are the numbers
increasing within each column and row. The study of Young tableaux is
the topic of several books, e.g.~\cite{Fulton:1997}, and
is thus too vast a topic to fully explore here.

Throughout this paper, $\mathcal{Y}_n$ will denote the set
of all Young tableaux consisting of $n$ boxes. For example,
\begin{equation}
  \mathcal{Y}_3 := \vast\lbrace
  \begin{ytableau}
    1 & 2 & 3
  \end{ytableau}, \quad
  \begin{ytableau}
    1 & 2 \\
    3
  \end{ytableau}, \quad
  \begin{ytableau}
    1 & 3 \\
    2
  \end{ytableau}, \quad
  \begin{ytableau}
    1 \\
    2 \\
    3
  \end{ytableau}
\vast\rbrace = \Biggl\lbrace
  \begin{ytableau}
    1 & 2
  \end{ytableau} \otimes \ybox{3} \Biggr\rbrace \cup \Biggl\lbrace
  \begin{ytableau}
    1 \\
    2
  \end{ytableau} \otimes \ybox{3} \Biggr\rbrace.
\end{equation}
We will denote a
particular Young tableau by an upper case Greek letter, usually
$\Theta$ or $\Phi$.

To establish the connection with birdtrack notation, let us consider a symmetrizer over elements $1$ and $2$,
$\bm{S}_{12}$, corresponding to a Young tableau
\begin{equation}
  \begin{ytableau}
    1 & 2
  \end{ytableau}
\ ,
\end{equation}
as symmetrizers always correspond to rows of Young tableaux~\cite{Tung:1985na}. We know that this symmetrizer $\bm{S}_{12}$ is given by
$\tfrac{1}{2}\left(\mathrm{id}+(12)\right)$, where $\mathrm{id}$ is
the identity and $(12)$ denotes
the transposition that swaps elements $1$ and $2$. Graphically, we would
denote this linear combination as~\cite{Cvitanovic:2008zz}
\begin{equation}
  \bm{S}_{12} 
  = \; \sfrac{1}{2} \left(    \;\FPic{2ArrLeft}\FPic{2IdSN}\FPic{2ArrRight} 
    + 
    \FPic{2ArrLeft}\FPic{2s12SN}\FPic{2ArrRight}
    \;\right).
\end{equation}
This operator is read from right to left\footnote{This is no longer
  strictly true for birdtracks representing primitive invariants of
  $\SUN$ over  $\MixedPow{m}{n}$, which includes dual
  vector spaces. A more informative discussion on this is out of the
  scope of this paper; readers are referred to~\cite{Cvitanovic:2008zz}.}, as it is viewed to act as a linear map
from the space $V\otimes V$ into itself. In this paper, permutations
and linear combinations thereof will always be interpreted as elements
of $\Lin{\Pow{n}}$, where $\Lin{\Pow{n}}$ denotes the space of linear
maps over $\Pow{n}$. In particular, we will denote the sub-space of
$\Lin{\Pow{n}}$ that is spanned by the primitive invariants of $\SUN$
by $\API{\SUN,\Pow{n}}$.

Following~\cite{Cvitanovic:2008zz}, we will denote a symmetrizer over
an index-set $\mathcal{N}$, $\bm{S}_{\mathcal{N}}$, by an empty
(white) box over the index lines in $\mathcal{N}$. Thus, the
symmetrizer $\bm{S}_{12}$ is denoted by
\FPic{2ArrLeft}\FPic{2Sym12SN}\FPic{2ArrRight}. Similarly, an
antisymmetrizer over an index-set $\mathcal{M}$,
$\bm{A}_{\mathcal{M}}$, is denoted by a filled (black) box over the
appropriate index lines. For example,
\begin{equation}
  \bm{A}_{12}=\FPic{2ArrLeft}\FPic{2ASym12SN}\FPic{2ArrRight}
  \quad \text{corresponds to
    the Young tableau} \quad
  \begin{ytableau}
    1 \\
    2
  \end{ytableau} \; ,
\end{equation}
since antisymmetrizers correspond to columns of Young
tableaux~\cite{Tung:1985na}. For any Young tableau $\Theta$, one can
form an idempotent, the so-called Young projection operator
corresponding to
$\Theta$~\cite{Young:1928,Cvitanovic:2008zz,Fulton:1997,Sagan:2000,Tung:1985na}:
Let $\bm{S}_{\mathcal{R}_i}$ denote the symmetrizer corresponding to
the $i^{th}$ row of the tableau $\Theta$, and let
$\mathbf{S}_{\Theta}$ denotes the set (or product, it does not matter
since the symmetrizers $\bm{S}_{\mathcal{R}_i}$ are disjoint by the
definition of a Young tableau) of all symmetrizers
$\bm{S}_{\mathcal{R}_i}$,
\begin{equation}
  \mathbf{S}_{\Theta} = \bm{S}_{\mathcal{R}_1} \cdots \bm{S}_{\mathcal{R}_k} \ .
\end{equation}
Similarly, let
\begin{equation}
    \mathbf{A}_{\Theta} = \bm{A}_{\mathcal{C}_1} \cdots \bm{A}_{\mathcal{C}_l}
\end{equation}
where $\bm{A}_{\mathcal{C}_j}$ corresponds to the $j^{th}$ column of
$\Theta$. Then, the object 
\begin{equation}
  \label{eq:Young-Def}
Y_{\Theta} := \alpha_{\Theta} \cdot \mathbf{S}_{\Theta} \mathbf{A}_{\Theta}
\end{equation}
is an idempotent, where $\alpha_{\Theta}$ is a combinatorial factor
involving the hook length of the tableau
$\Theta$~\cite{Fulton:1997,Sagan:2000}. $Y_{\Theta}$ is called the
\emph{Young projection operator} corresponding to $\Theta$. Besides
being idempotent\footnote{This property is surprisingly hard
  to prove without the simplification rules paraphrased in section~\ref{sec:CancellationRules}~\cite{Alcock-Zeilinger:2016bss}.}
\begin{subequations}
\label{eq:YoungProperties}
    \begin{equation}
    \label{eq:YoungIdempotency}
Y_{\Theta} \cdot  Y_{\Theta} = Y_{\Theta}
\ ,
  \end{equation}
  Young projection operators on $\Pow{n}$ are also mutually orthogonal if $n < 5$: If $\Theta$
  and $\Phi$ are two Young tableaux consisting of the same number of
  boxes, then
  \begin{equation}
    \label{eq:YoungOrthogonality}
    Y_{\Theta} \cdot
  Y_{\Phi} = \delta_{\Theta \Phi} Y_{\Theta} \hspace{1cm} n=1,2,3,4
\ .
  \end{equation}
and for general $n$ provided the shapes of the associated tableaux are different.

Furthermore, again for $m < 5$, Young projection operators satisfy a
completeness relation, that is, the Young projection operators
corresponding to the tableaux in $\mathcal{Y}_n$ sum up to the
identity operator on the space $\Pow{n}$,
\begin{equation}
    \label{eq:YoungCompleteness}
\sum_{\Theta\in\mathcal{Y}_n} Y_{\Theta} = \mathbb{1}_n \hspace{1cm} n=1,2,3,4
\ .
\end{equation}
\end{subequations}
Both properties~\eqref{eq:YoungOrthogonality}
and~\eqref{eq:YoungCompleteness} can be restored for $m \ge 5$ by
suitable subtractions without
compromising~\eqref{eq:YoungIdempotency} in the process (see~\cite[sec.~5.4]{Littlewood:1950} or
in~\cite[sec.~II.3.6]{Schensted:1976}), \emph{c.f.}
eqns.~\eqref{eq:LY-class} and
appendix~\ref{sec:littl-young-proj}.  These three properties allow the
(corrected) Young projection operators associated to the Young
tableaux in $\mathcal{Y}_n$ to fully classify the irreducible
representations of $\SUN$ over $\Pow{n}$~\cite{Weyl:1946,
  Bourbaki7-9:2000, Cvitanovic:2008zz,Tung:1985na}. The Hermitian
replacements for these operators given
in~\cite{Keppeler:2013yla,Sjodahl:2013hra,Keppeler:2012ih} share all
three properties without restrictions on $n$ and are formulated in
terms of $Y_\Theta$ entirely, relying on the unrestricted nature
of~\eqref{eq:YoungIdempotency}. This remains to be the case for the
improved algorithms presented here.

% Littlewood devised a method that restores both properties
% ~\eqref{eq:YoungOrthogonality} and~\eqref{eq:YoungCompleteness}
% $m\geq5$ by suitable suitable subtractions without
% compromising~\eqref{eq:YoungIdempotency} in the process
% (see~\cite[sec.~5.4]{Littlewood:1950} or
% in~\cite[sec.~II.3.6]{Schensted:1976})
% \begin{subequations}
% \label{eq:LYProperties}
% \begin{align}
% L_{\Theta} \cdot  L_{\Theta} & = L_{\Theta}
% \label{eq:LYIdempotency}
% \\
% L_{\Theta} \cdot L_{\Phi} & = \delta_{\Theta \Phi} L_{\Theta}
% \label{eq:LYOrthogonality} 
% \\
% \sum_{\Theta\in\mathcal{Y}_n} L_{\Theta} & = \mathbb{1}_n
% \label{eq:LYCompleteness}
% \ ;
% \end{align}
% \end{subequations}
% we summarize Littlewood's method of constructing the operators
% $L_{\Theta}$ from the Young projectors section~\ref{sec:littl-young-proj}. These three properties allow the
% Littlewood-Young projection operators $L_{\Theta}$ associated to the
% Young tableaux in $\Theta\in\mathcal{Y}_n$ to fully classify the
% irreducible representations of $\SUN$ over $\Pow{n}$~\cite{Weyl:1946,
%   Bourbaki7-9:2000, Cvitanovic:2008zz,Tung:1985na}. The Hermitian
% replacements for these operators given
% in~\cite{Keppeler:2013yla,Sjodahl:2013hra,Keppeler:2012ih} share all
% three properties without restrictions on $n$ and are formulated in
% terms of $Y_\Theta$ entirely, relying on the unrestricted nature
% of~\eqref{eq:YoungIdempotency}. This remains to be the case for the
% improved algorithms presented here.

It should be noted that, since all symmetrizers in
$\mathbf{S}_{\Theta}$ (resp. antisymmetrizers in
$\mathbf{A}_{\Theta}$) are disjoint, each index line enters \emph{at
  most} one symmetrizer (resp. antisymmetrizer) in birdtrack
notation. Thus, one may draw all symmetrizers (resp. antisymmetrizers)
underneath each other.

As an example, we  construct the birdtrack Young projection
operator corresponding to the following Young tableau,
\begin{equation}
  \label{eq:SymsEx1}
  \Theta =
  \begin{ytableau}
    1 & 3 & 4 \\
    2 & 5
  \end{ytableau} \; .
\end{equation}
$Y_{\Theta}$ is given by
\begin{equation}
  Y_{\Theta}=2\cdot\bm{S}_{134}\bm{S}_{25}\bm{A}_{12}\bm{A}_{35}
\ ,
\end{equation}
where the constant $2$ is the combinatorial factor that ensures the
idempotency of $Y_{\Theta}$. In birdtrack notation, the Young projection operator
$Y_{\Theta}$ becomes
\begin{equation}
  Y_{\Theta} = 2 \cdot
    \FPic{5ArrLeft}    \FPic{5s234N}\FPic{5Sym123Sym45N}    \FPic{5s2453N}\FPic{5ASym12ASym34N}    \FPic{5s45N}\FPic{5ArrRight}
  \; ,
\end{equation}
where we were able to draw the two symmetrizers in
$\mathbf{S}_{\Theta}$ and the antisymmetrizers in
$\mathbf{A}_{\Theta}$ underneath each other, as claimed.

\subsection{Notation \& conventions}\label{sec:Notation}

In the literature, there is a great multitude of (sometimes
conflicting) conventions and notations regarding birdtracks, Young
symmetrizers and other quantities used in this paper. We will devote
this section to laying down the conventions that will be used
here.

\subsubsection{Structural relationships between Young tableaux of
  different sizes}\label{sec:YoungTableauxStructure}

Throughout this paper, $Y_{\Theta}$ shall denote the normalized Young
projection operator corresponding to a Young tableau $\Theta$, and
$P_{\Theta}$ will refer to the normalized Hermitian Young projection
operator corresponding to $\Theta$. Furthermore, for any operator $O$
consisting of symmetrizers and anti-symmetrizers, the symbol $\bar{O}$
will refer to a product of symmetrizers and antisymmetrizers without
any \emph{additional} scalar factors. For example,
\begin{equation}
  Y_{\Theta} := \underbrace{\frac{4}{3}}_{=: \alpha_{\Theta}} \;
  \underbrace{\FPic{3ArrLeft}\FPic{3Sym13ASym12}\FPic{3ArrRight}}_{    =: \bar{Y}_{\Theta}}.
\end{equation}
Thus, for a birdtrack operator $O$, comprised solely of symmetrizers
and antisymmetrizers, $\bar{O}$ denotes the \emph{graphical}
part alone,
\begin{equation}
  \label{eq:BirdtrackPart}
  O := \omega \bar{O},
\end{equation}
where $\omega$ is some scalar. The benefit of this notation is that
the barred operator stays unchanged under multiplication with a
non-zero scalar $\lambda$,
\begin{equation}
\label{eq:Bar-Notation}
  \lambda \cdot O \neq O \qquad \text{but} \qquad \lambda \cdot
  \bar{O} = \bar{O}
\ .
\end{equation}
It should be noted that $\bar{Y}_{\Theta}$ and $\bar{P}_{\Theta}$ are
only quasi-idempotent, while $Y_{\Theta}$ and $P_{\Theta}$ are
idempotent. We will denote the normalization constants of $Y_{\Theta}$
and $P_{\Theta}$ by $\alpha_{\Theta}$ and $\beta_{\Theta}$
respectively, such that
\begin{equation}
  \label{eq:NormalizationConstant}
Y_{\Theta} := \alpha_{\Theta} \bar{Y}_{\Theta} \qquad \text{and}
\qquad P_{\Theta} := \beta_{\Theta} \bar{P}_{\Theta}
\ ;
\end{equation}
where $\alpha_{\Theta}$ can be obtained from the hook length
formula~\cite{Sagan:2000,Fulton:1997,Cvitanovic:2008zz} and the
normalization constant $\beta_{\Theta}$ is given together with the
appropriate construction principle for $P_{\Theta}$ (we encounter
three different versions, one each for the original KS construction in
Theorem~\ref{thm:KSProjectors}, the simplified KS construction in
Corollary~\ref{thm:ShortKSProjectors}, and the MOLD version in
Theorem~\ref{thm:MOLDConstruction}).

It is a well-known fact that Young tableaux in $\mathcal{Y}_n$ can be
built from Young tableaux in $\mathcal{Y}_{n-1}$ by adding the box
\ybox{n} at an appropriate place as to not destroy the properties of
Young tableaux; such places are referred to as \emph{outer
  corners}~\cite{Sagan:2000}. In this way, the Young tableau
$\Theta = \scalebox{0.75}{\begin{ytableau} 1 & 2 & 3
    \\4 \end{ytableau}}$
generates the subset $\lbrace\Theta\otimes\ybox{5}\rbrace$ of
$\mathcal{Y}_5$,
\begin{equation}
  \label{eq:YT4toYT5}
\begin{tikzpicture}[baseline=(current bounding box.west),every node/.style={inner sep=0pt, outer sep=3pt}]
\node(T4)at(0,2.5){$ \Theta = \;
  \begin{ytableau}
    1 & 2 & 3 \\
    4
  \end{ytableau}
$};
\node[anchor=west] (Y4) at (4,2.5) {$\in\mathcal{Y}_4$};
\node(T51)at(-2.5,0){$
  \begin{ytableau}
    1 & 2 & 3 \\
    4 \\
    *(magenta!40) 5
  \end{ytableau}
$};
\node(T52)at(0,0){$
  \begin{ytableau}
    1 & 2 & 3 \\
    4 & *(magenta!40) 5
  \end{ytableau}
$};
\node(T53)at(2.5,0){$
  \begin{ytableau}
    1 & 2 & 3 & *(magenta!40) 5 \\
    4 
  \end{ytableau}
$};
\node[anchor=west] (Y5) at (4,0) {$\in \left\lbrace\Theta\otimes
  \begin{ytableau}
    5
  \end{ytableau}\right\rbrace \subset
\mathcal{Y}_5$};
\draw[-{stealth},color=red!50!black] (T4) to (T51);
\draw[-{stealth},color=red!50!black] (T4) to (T52);
\draw[-{stealth},color=red!50!black] (T4) to (T53);
\end{tikzpicture}\end{equation}
This operation is not a map in the mathematical sense as it does not
yield a unique result. The reverse operation, taking away the box with
the highest entry, is a map; let us denote this map by
$\pi$. $\pi$ can then repeatedly be applied to the resulting
tableau,
\begin{equation}
  \begin{ytableau}
    1 & 3 & *(magenta!45) 6 \\
    2 & *(magenta!30) 5 \\
    *(magenta!15) 4
  \end{ytableau} \quad \xlongrightarrow[]{\pi} \quad
  \begin{ytableau}
    1 & 3 \\
    2 & *(magenta!30) 5 \\
    *(magenta!15) 4
  \end{ytableau} \quad \xlongrightarrow[]{\pi} \quad
  \begin{ytableau}
    1 & 3 \\
    2 \\
    *(magenta!15) 4
  \end{ytableau} \quad \xlongrightarrow[]{\pi} \quad
  \begin{ytableau}
    1 & 3 \\
    2 
  \end{ytableau}
  \ .
\end{equation}

\begin{definition}[parent map and ancestor tableaux]
\label{ParentMap}
Let $\Theta \in \mathcal{Y}_n$ be a Young tableau. We define its
\emph{parent tableau} $\Theta_{(1)} \in \mathcal{Y}_{n-1}$ to be the
tableau obtained from $\Theta$ by removing the box \ybox{n} of
$\Theta$.\footnote{We note that the tableau $\Theta_{(1)}$ is always a
  Young tableau if $\Theta$ was a Young tableau, since removing the
  box with the highest entry cannot possibly destroy the properties of
  $\Theta$ (and thus $\Theta_{(1)}$) that make it a Young tableau.}
Furthermore, we will define a \emph{parent map} $\pi$ from
$\mathcal{Y}_n$ to $\mathcal{Y}_{n-1}$, for a particular $n$,
\begin{subequations}
\begin{equation}
   \label{eq:ParentMap1}
  \pi: \mathcal{Y}_n \rightarrow \mathcal{Y}_{n-1},
\end{equation}
which acts on $\Theta$ by removing the box \ybox{n} from
$\Theta$,
\begin{equation}
   \label{eq:ParentMap2}
   \pi: \Theta \mapsto \Theta_{(1)}.
\end{equation}
\end{subequations}
In general, we define the successive action of the parent map $\pi$ by
\begin{subequations}
\begin{equation}
  \label{eq:ParentMapSucc1}
  \mathcal{Y}_n \; \xlongrightarrow[]{\pi} \; \mathcal{Y}_{n-1} \;
  \xlongrightarrow[]{\pi} \; \mathcal{Y}_{n-2} \;
  \xlongrightarrow[]{\pi} \; \ldots \; \xlongrightarrow[]{\pi} \; \mathcal{Y}_{n-m},
\end{equation}
and denote it by $\pi^m$,
\begin{equation}
  \label{eq:ParentMapSucc2}
  \pi^m: \mathcal{Y}_n \rightarrow \mathcal{Y}_{n-m}, \qquad  \pi^m:= \mathcal{Y}_n \; \xlongrightarrow[]{\pi} \; \mathcal{Y}_{n-1} \;
  \xlongrightarrow[]{\pi} \; \mathcal{Y}_{n-2} \;
  \xlongrightarrow[]{\pi} \; \ldots \; \xlongrightarrow[]{\pi} \; \mathcal{Y}_{n-m}
\end{equation}
We will further denote
the unique tableau obtained from $\Theta$ by applying the map $\pi$ $m$
times, $\pi^m(\Theta)$, by $\Theta_{(m)}$, and refer to it as the
\emph{ancestor tableau} of $\Theta$ $m$ generations back,
\begin{equation}
  \label{eq:ParentMapSucc3}
  \pi^m: \Theta \mapsto \Theta_{(m)}
  \ .
\end{equation}
\end{subequations}
\end{definition}
\ytableausetup
{mathmode, boxsize=0.7em}

\subsubsection{Embeddings and images of linear operators}\label{sec:OperatorsNotation}

Any operator $O \in \mathrm{Lin}\left(V^{\otimes n}\right)$
can be embedded into $\mathrm{Lin}\left(V^{\otimes m}\right)$ for $m>n$ in several ways, simply by letting the embedding act as the
identity on $(m-n)$ of the factors; how to select these factors is a
matter of what one plans to achieve.  The most useful convention for
our purposes is to let $O$ act on the first $n$ factors and
operate with the identity on the last $(m-n)$ factors. We will
call this the \emph{canonical embedding}.  On the level of birdtracks,
this amounts to letting the index lines of $O$ coincide with
the top $n$ index lines of $\mathrm{Lin}\left(V^{\otimes m}\right)$,
and the bottom $(m-n)$ lines of the embedded operator constitute the
identity birdtrack of size $(m-n)$. For example, the operator
$\Bar Y_{\begin{ytableau} \scriptstyle 1 & \scriptstyle 2 \\ \scriptstyle
    3 \end{ytableau}}$ is canonically embedded into
$\mathrm{Lin}\left(V^{\otimes 5}\right)$ as
\begin{equation}
  \label{eq:CanonicalEmbedding}
  \FPic{3ArrLeft}\FPic{3Sym12ASym13}\FPic{3ArrRight} 
  \; \hookrightarrow \; 
  \FPic{5ArrLeft}\FPic{5IdN}\FPic{5Sym12N}\FPic{5s23N}\FPic{5ASym12N}\FPic{5s23N}\FPic{5ArrRight}
\ .
\end{equation}
Furthermore, we will use the same
symbol $O$ for the operator as well as for its embedded
counterpart. Thus, $\Bar Y_{\begin{ytableau} \scriptstyle 1 & \scriptstyle
    2 \\ \scriptstyle 3 \end{ytableau}}$ shall denote both the
operator on the left as well as on the right hand side of the
embedding~\eqref{eq:CanonicalEmbedding}.  \ytableausetup {mathmode, boxsize=normal}

Lastly, if a \emph{Hermitian} projection operator $A$ projects onto a
subspace completely contained in the image of a projection
operator $B$, then we denote this as $A\subset B $, transferring the
familiar notation of sets to the associated projection operators. In
particular, $A\subset B$ if and only if
\begin{equation}
  \label{eq:OperatorInclusion1}
  A \cdot B = B \cdot A = A
\end{equation}
for the following reason: If the subspaces obtained by the consecutive
application of the operators $A$ and $B$ in any order is the same as that
obtained by merely applying $A$, then not only need the subspaces onto
which $A$ and $B$ project overlap (as otherwise $A\cdot B=B\cdot A=0$),
but the subspace corresponding to $A$ must be completely contained in
the subspace of $B$ - otherwise the last equality
of~\eqref{eq:OperatorInclusion1} would not hold. Hermiticity is crucial
for these statements -- they thus do not apply to most Young
projection operators on $\Pow{m}$ if $m \ge 3$. A familiar example for
this situation is the relation between symmetrizers of different
length: a symmetrizer $\bm{S}_{\mathcal{N}}$ can be absorbed into a
symmetrizer $\bm{S}_{\mathcal{N'}}$, as long as the index set
$\mathcal{N}$ is a subset of $\mathcal{N'}$, and the same statement
holds for antisymmetrizer,~\cite{Cvitanovic:2008zz}. For example,
\begin{equation}
  \FPic{3ArrLeft}\FPic{3Sym12SN}\FPic{3Sym123SN}\FPic{3ArrRight} 
  \; = \;
  \FPic{3ArrLeft}\FPic{3Sym123SN}\FPic{3ArrRight} 
  \; = \; 
  \FPic{3ArrLeft}\FPic{3Sym123SN}\FPic{3Sym12SN}\FPic{3ArrRight}
\ .
\end{equation}
Thus, by the above notation,
$\bm{S}_{\mathcal{N'}}\subset\bm{S}_{\mathcal{N}}$, if
$\mathcal{N}\subset\mathcal{N'}$. Or, as in our example,
\begin{equation}
  \FPic{3ArrLeft}\FPic{3Sym123SN}\FPic{3ArrRight} 
  \; \subset \; 
  \FPic{3ArrLeft}\FPic{3Sym12SN}\FPic{3ArrRight}
\ .
\end{equation}
In particular, it follows immediately from the definition of the
ancestor tableau (Definition~\ref{ParentMap}, eq.~\eqref{eq:ParentMapSucc3}) that
\begin{equation}
  \label{eq:Inclusion-Ancestor-Ops}
\mathbf{S}_{\Theta_{(k)}}\mathbf{S}_{\Theta} 
= \mathbf{S}_{\Theta}
= \mathbf{S}_{\Theta}\mathbf{S}_{\Theta_{(k)}}
\qquad \text{and} \qquad
\mathbf{A}_{\Theta_{(k)}}\mathbf{A}_{\Theta} 
= \mathbf{A}_{\Theta}
= \mathbf{A}_{\Theta}\mathbf{A}_{\Theta_{(k)}}
\end{equation}
for every ancestor tableau $\Theta_{(k)}$ of $\Theta$.

\subsection{Cancellation rules}\label{sec:CancellationRules}

We suspect that one of the reasons why the birdtrack formalism has not
yet gained as much popularity as it ought is because there exist
virtually no practical rules which allow easy manipulation of
birdtrack operators in the
literature. In~\cite{Alcock-Zeilinger:2016bss}, we establish various
rules designed to easily manipulate birdtrack operators comprised of
symmetrizers and anti-symmetrizers. Since all operators considered in
this paper are of this form, the simplification rules
of~\cite{Alcock-Zeilinger:2016bss} are immediately applicable: None of
the proofs of the construction algorithms in this paper would have
been possible without these rules. Thus, we choose to summarize the
most important results of~\cite{Alcock-Zeilinger:2016bss} here. For
the proofs of these rules, readers are referred
to~\cite{Alcock-Zeilinger:2016bss}.

The simplification rules of~\cite{Alcock-Zeilinger:2016bss} fall into
two classes:
\begin{enumerate}
\item \emph{Cancellation rules}
  (Theorems~\ref{thm:CancelWedgedParentOp} and
  \ref{thm:CancelMultipleSets}, section~\ref{sec:CancellationRules}):
  these rules are to cancel large chunks of birdtrack operators, thus
  making them shorter (often significantly so) and more practical to
  use. The cancellation rules are used in several places in this paper, in
  particular in the proof of the shortened KS-operators
  (Corollary~\ref{thm:ShortKSProjectors}) and the proof of the
  construction of MOLD-operators
  (Theorem~\ref{thm:MOLDConstruction}). Since the cancellation rules
  are used multiple times throughout this paper, we recapitulate these
  rules in this present section.
\item \emph{Propagation rules} (Theorem~\ref{thm:PropagateSyms},
  section~\ref{sec:PropagationRules}): these rules allow one to
  commute (sets of) symmetrizers through (sets of) antisymmetrizers
  and vice versa. The propagation rules come in handy when trying to expose the
  implicit Hermiticity of a birdtrack operator. In this paper, we use
  these rules in the proof of Theorem~\ref{thm:LexicalConstruction},
  which is why we defer the re-statement of the propagation rules to
  appendix~\ref{sec:ProofsLexical} where also the proof of
  Theorem~\ref{thm:LexicalConstruction} can be found.
\end{enumerate}

\begin{theorem}[cancellation of wedged ancestor-operators]
  \label{thm:CancelWedgedParentOp} 
  Consider two Young tableaux $\Theta$ and $\Phi$ such that they have
  a common ancestor tableau $\Gamma$. Let $Y_{\Theta}$, $Y_{\Phi}$ and
  $Y_{\Gamma}$ be their respective Young projection operators, all
  embedded in an algebra that is able to contain all
  three. Then
    \begin{equation}
      \label{eq:CancelWedgedParentOp}
Y_{\Theta} Y_{\Gamma} Y_{\Phi} = Y_{\Theta} Y_{\Phi}.
    \end{equation}
  \end{theorem}

\noindent For example, consider the Young tableaux
\begin{equation}
  \Theta =
\begin{ytableau}
*(blue!25) 1 & *(blue!25) 2 & 5 \\
*(blue!25) 3 & 4
\end{ytableau}
\qquad \text{and} \qquad \Phi =
\begin{ytableau}
  *(blue!25) 1 & *(blue!25) 2 & 4 \\
  *(blue!25) 3
\end{ytableau},
\end{equation}
which have the common ancestor
\begin{equation}
  \Gamma =
\begin{ytableau}
*(blue!25) 1 & *(blue!25) 2 \\
*(blue!25) 3
\end{ytableau}.
\end{equation}
Then, the product $Y_{\Theta}Y_{\Gamma}Y_{\Phi}$ is
given by
\begin{equation}
  Y_{\Theta}Y_{\Gamma}Y_{\Phi} = 4 \cdot
  \scalebox{0.75}{$\underbrace{\FPic{5s354SN}\FPic{5Sym123Sym45N}\FPic{5s2453N}\FPic{5ASym12ASym34N}\FPic{5s23N}}_{\mbox{\normalsize
        $\bar{Y}_{\Theta}$}}\underbrace{\FPic{5Sym12N}\FPic{5s23N}\FPic{5ASym12N}\FPic{5s23N}}_{\mbox{\normalsize
        $\bar{Y}_{\Gamma}$}} \underbrace{\FPic{5s34N}\FPic{5Sym123N}\FPic{5s243N}\FPic{5ASym12N}\FPic{5s23N}}_{\mbox{\normalsize $\bar{Y}_{\Phi}$}}$},
\end{equation}
where $\alpha_{\Theta}\alpha_{\Gamma}\alpha_{\Phi}=4$.
According to Theorem~\ref{thm:CancelWedgedParentOp}, we are
allowed to cancel the operator $Y_{\Gamma}$ hence reducing the above product to
\begin{equation}
  Y_{\Theta}Y_{\Gamma}Y_{\Phi} = \; 4 \cdot
  \scalebox{0.75}{$\underbrace{\FPic{5s354SN}\FPic{5Sym123Sym45N}\FPic{5s2453N}\FPic{5ASym12ASym34N}\FPic{5s23N}}_{\mbox{\normalsize $\bar{Y}_{\Theta}$}}\cancel{\FPic{5Sym12N}\FPic{5s23N}\FPic{5ASym12N}\FPic{5s23N}}\underbrace{\FPic{5s34N}\FPic{5Sym123N}\FPic{5s243N}\FPic{5ASym12N}\FPic{5s23N}}_{\mbox{\normalsize $\bar{Y}_{\Phi}$}}$}
  \; = \; 3 \cdot \scalebox{0.75}{$\underbrace{\FPic{5s354SN}\FPic{5Sym123Sym45N}\FPic{5s2453N}\FPic{5ASym12ASym34N}\FPic{5s23N}}_{\mbox{\normalsize       $\bar{Y}_{\Theta}$}}\underbrace{\FPic{5s34N}\FPic{5Sym123N}\FPic{5s243N}\FPic{5ASym12N}\FPic{5s23N}}_{\mbox{\normalsize $\bar{Y}_{\Phi}$}}$},
\end{equation}
where $\alpha_{\Theta}\alpha_{\Phi}=3$.

A more general cancellation-Theorem is:

\begin{theorem}[cancellation of parts of the operator]\label{thm:CancelMultipleSets}
  Let $\Theta\in\mathcal{Y}_n$ be a Young tableau and
  $M$ be an element of $\API{\SUN,\Pow{n}}$. Then,
  there exists a (possibly vanishing) constant $\lambda$ such that
  \begin{equation}
    \label{eq:Cancel-General-O}
O := \mathbf{S}_{\Theta} \; M \; \mathbf{A}_{\Theta} =
\lambda \cdot Y_{\Theta}
\ .
  \end{equation}
If furthermore the operator $O$ is non-zero, then $\lambda\neq 0$. One
instance in which $O$ is guaranteed to be non-zero is if $M$ is of the
form
\begin{equation}
  \label{eq:CancelWedgedParent1}
  M = \; \mathbf{A}_{\Phi_1} \;
  \mathbf{S}_{\Phi_2} \; \mathbf{A}_{\Phi_3} \;
  \mathbf{S}_{\Phi_4} \; \cdots \; \mathbf{A}_{\Phi_{k-1}} \;
  \mathbf{S}_{\Phi_k}
\ ,
\end{equation}
where $\mathbf{A}_{\Phi_i}\supset\mathbf{A}_{\Theta}$ for every
$i\in\lbrace 1,3,\ldots k-1\rbrace$ and
$\mathbf{S}_{\Phi_j}\supset\mathbf{S}_{\Theta}$ and for every
$j\in\lbrace 2,4,\ldots k\rbrace$.
\end{theorem}

As an example, consider the operator
\begin{equation}
  O := \; \FPic{5s354N}\FPic{5Sym123Sym45N}\FPic{5s243N}\FPic{5ASym12N}\FPic{5s23s45N}\FPic{5Sym12Sym34N}\FPic{5s23N}\FPic{5ASym12ASym34N}\FPic{5s23N}
\;  = \; \lbrace\bm{S}_{125},\bm{S}_{34}\rbrace \cdot
  \lbrace\bm{A}_{13}\rbrace \cdot
  \lbrace\bm{S}_{12},\bm{S}_{34}\rbrace \cdot
  \lbrace\bm{A}_{13},\bm{A}_{24}\rbrace
\ .
\end{equation}
This operator meets all conditions of the above Theorem
\ref{thm:CancelMultipleSets}: The sets
$\lbrace\bm{S}_{125},\bm{S}_{34}\rbrace$ and
$\lbrace\bm{A}_{13},\bm{A}_{24}\rbrace$ together constitute the birdtrack of
a Young projection operator $\bar{Y}_{\Theta}$ corresponding to the
tableau
\begin{equation}
  \Theta :=
  \begin{ytableau}
    1 & 2 & 5 \\
    3 & 4
  \end{ytableau}
\ .
\end{equation}
The set $\lbrace\bm{A}_{13}\rbrace$ corresponds to the ancestor
tableau $\Theta_{(2)}$, and the set
$\lbrace\bm{S}_{12},\bm{S}_{34}\rbrace$ corresponds to the ancestor
tableau $\Theta_{(1)}$ and can thus be absorbed into
$\mathbf{A}_{\Theta}$ and $\mathbf{S}_{\Theta}$ respectively,
\emph{c.f.} eq.~\eqref{eq:Inclusion-Ancestor-Ops}. Hence $O$ can be
written as
\begin{equation}
  O = \; \mathbf{S}_{\Theta} \; \mathbf{A}_{\Theta_{(2)}} \;
  \mathbf{S}_{\Theta_{(1)}} \; \mathbf{A}_{\Theta}
  \ .
\end{equation}
Then, according to the above Cancellation-Theorem~\ref{thm:CancelMultipleSets}, we may cancel the wedged ancestor sets
$\mathbf{A}_{\Theta_{(2)}}$ and $\mathbf{S}_{\Theta_{(1)}}$ at the
cost of a non-zero constant $\tilde{\lambda}$. In particular, we find that
\begin{equation}
 O = \tilde{\lambda} \cdot 
 \underbrace{
\FPic{5s354SN}\FPic{5Sym123Sym45N}\FPic{5s2453N}\FPic{5ASym12ASym34N}\FPic{5s23N}}_{\bar{Y}_{\Theta}}
\ ,
\end{equation}
which is proportional to $Y_{\Theta}$.

\section{Hermitian Young projection
 operators}\label{HermitianYoungProjectorsSection}

Throughout this paper, we will be working with linear maps over linear
spaces, in particular with maps in
$\API{\SUN,\Pow{m}}\subset\Lin{\Pow{m}}$. All the familiar tools from
linear algebra (as can be found in \cite{Lang:1987} and other standard
textbooks) apply but will likely \emph{look} unfamiliar when employed
in the language of birdtracks. We thus devote this section to
translate the most important tools for this paper into the
birdtrack formalism.

\subsection{Hermitian conjugation of linear maps in birdtrack
  notation}\label{sec:HermitianLinearMapsSection}

We begin by recalling the definition of Hermitian conjugation for
linear maps. Let $U$ and $W$ be linear spaces, and let
$\langle\cdot,\cdot\rangle_U:U\rightarrow\mathbb{F}$, where
$\mathbb{F}$ is a field usually taken to be $\mathbb{C}$ or
$\mathbb{R}$, denote the scalar product defined on $U$, and similarly
for $W$. Furthermore, let $P:U\rightarrow W$ be an operator. The
scalar products then furnish a definition of the Hermitian conjugate
of $P$ (denoted by $P^{\dagger}:W\rightarrow U$) in the standard way:
\begin{equation}\label{eq:HermitianConjugateDef}
  \langle w, P u \rangle_W \stackrel{!}{=} \langle P^{\dagger} w, u \rangle_U
\end{equation}
for any $u\in U$ and $w\in W$~\cite{Lang:1987}. In our case,
$u$ and $w$ will be elements of
$V^{\otimes m}$, both $u$ and
$w$ appear  as tensors with $m$ upper indices,\footnote{At this
  point we recall that the basis vectors $\mathbf{e}$ of $\Pow{m}$ are
denoted with lower indices; therefore $u$ and $w$ act on the
$\mathbf{e}$'s as linear maps.}
\begin{equation}
 u^{j_1 \ldots j_m} \qquad \text{and} \qquad  w^{i_1 \ldots i_m}.
\end{equation}
Eq.~\eqref{eq:HermitianConjugateDef} is equivalent to
requiring the following diagram to commute, 
\begin{equation}
\label{eq:Commute-Diagram-Herm}
  \FPic{CD-mHermiteanConjugate} \ .
\end{equation}
The scalar product between these maps
is then defined in the usual way, $\langle
u,w\rangle=u^{\dagger}w\in\mathbb{C}$. The map $u^{\dagger}$ is an element of
the dual space $\DPow{m}$ and thus needs to be equipped with
\emph{lower} indices,\footnote{Since basis vectors $\mathbf{\omega}$ of
  $\DPow{m}$ have upper indices.}
\begin{equation}
  \label{eq:uHC}
  u^{\dagger} = \left(u^{j_1 \ldots j_m}\right)^* =: u_{j_1 \ldots j_m}.
\end{equation}
Complex conjugation $^*$ is necessary, since the vector space $V$ may
be complex. Hence, we have that
\begin{equation}
  \langle u, w \rangle = u^{\dagger} w = u_{i_1 \ldots i_m} w^{i_1 \ldots i_m} \in \mathbb{C}.
\end{equation}
In the above, all indices of $u^{\dagger}$ and $w$ were contracted so
that the outcome of the scalar product lies in a field, in our case
$\mathbb{C}$. Graphically, let us represent a tensor with $j$ lower
indices and $i$ upper indices by a box which has $j$ legs
exiting on the right and $i$ legs exiting on the left,
\begin{equation}
  \tensor{T}{^{a_1 a_2 \ldots a_i}_{b_1 b_2 \ldots b_j}} \rightarrow
  \scalebox{0.75}{\FPic{TensorTlabel}}
\ .
\end{equation}
Therefore, the tensors $u_{i_1 \ldots i_m}$ and $w^{i_1
  \ldots i_m}$ will have $m$ legs exiting on the right and left
respectively,
\begin{equation}
 u_{i_1 \ldots i_m} \rightarrow \;
 \scalebox{0.75}{\raisebox{-0.4\height}{\includegraphics{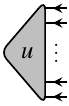}\includegraphics{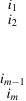}}} \qquad \text{and}
 \qquad  w^{i_1 \ldots i_m} \rightarrow
 \scalebox{0.75}{\raisebox{-0.4\height}{\includegraphics{TensoriLabels}\includegraphics{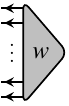}}}
\ ;
\end{equation}
from now on, we will suppress the index labels of the birdtracks
corresponding to the tensors in question. The scalar product $\langle
u,w\rangle$ is diagrammatically represented as 
\begin{equation}
\label{eq:uw-scalar-product-Birdtrack}
  \langle u, w \rangle 
  \; = \; 
  \scalebox{0.75}{\FPic{TensoruDag}\FPic{Tensorw}}
\ ,
\end{equation}
where the contraction of indices is indicated via the connection of
corresponding index lines in the
birdtrack. In~\eqref{eq:uw-scalar-product-Birdtrack}, we see that the
birdtrack corresponding to $\langle u,w\rangle$ does not have any
index lines exiting on either the right or the left, indicating that
it is indeed a scalar. We will now consider a scalar product
$\langle u,Pw\rangle$, where $P:\Lin{\Pow{m}}\rightarrow\Lin{\Pow{m}}$
is an operator, to find its Hermitian dual $P^{\dagger}$. $P$ must
have $m$ lower and $m$ upper indices,
\begin{equation}
  \tensor{P}{^{i_1 \ldots i_m}_{j_1 \ldots j_m}} \rightarrow
  \scalebox{0.75}{\raisebox{-0.4\height}{\includegraphics{TensoriLabels}\includegraphics{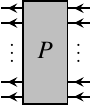}\includegraphics{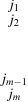}}}
\ ,
\end{equation}
where the $j$-indices act on an element of $\Lin{\Pow{m}}$ via index contraction.
The scalar product $\langle u,Pw\rangle$ will then be given by
\begin{equation}
  \label{eq:uPw}
 \langle u,Pw\rangle = \left(u^{i_1 \ldots i_m}\right)^*
 \tensor{P}{^{i_1 \ldots i_m}_{j_1 \ldots j_m}} w^{j_1 \ldots j_m}
\rightarrow \; \scalebox{0.75}{\FPic{TensoruDag}\FPic{TensoriLabels}\FPic{TensorP}\FPic{TensorjLabels}\FPic{Tensorw}}
  \; = \;
  \scalebox{0.75}{\FPic{TensoruDag}\FPic{TensorP}\FPic{Tensorw}}
\ .
\end{equation}
The adjoint $P^{\dagger}$ of $P$ is defined to be the object such
that relation~\eqref{eq:HermitianConjugateDef} holds. Thus,
$P^{\dagger}$ acts on the dual space,
$P^{\dagger}:\Lin{\DPow{m}}\rightarrow\Lin{\DPow{m}}$, which again
means that it has $m$ upper and $m$ lower indices, but now the
$j$-indices act on the element of $\Lin{\DPow{m}}$,
\begin{equation}
  \label{eq:raising-lowering-index}
  P^{\dagger} = \left(\tensor{P}{^{i_1 \ldots i_m}_{j_1 \ldots
        j_m}}\right)^{*} = \tensor{P}{_{i_1 \ldots i_m}^{j_1 \ldots j_m}}.
\end{equation}
It should be noted that once again, the raising and lowering of
indices induces a complex conjugation of the tensor components, as we
have already seen for $u$ in~\eqref{eq:uHC}.\footnote{The projection
  operators considered in this paper are real and thus remain
  unaffected by complex conjugation. This is no longer true for group
  elements or representations.} The same caveat applies to the
associated birdtrack diagrams in which we have to mirror the operator
about its vertical axis and reverse the direction of the
arrows:
\begin{equation}
  \label{eq:ScalarProductuPw1}
  \left(\scalebox{0.75}{\raisebox{-0.4\height}{\includegraphics{TensoriLabels}\includegraphics{TensorP}\includegraphics{TensorjLabels}}}\right)^{\dagger}
  \; = \;
  \scalebox{0.75}{\raisebox{-0.4\height}{\includegraphics{TensorjLabels}\includegraphics{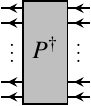}\includegraphics{TensoriLabels}}}.
\end{equation}
This results in all contracted index lines lining up correctly.
For example,
\ytableausetup
{mathmode, boxsize=0.7em}
\begin{IEEEeqnarray}{rClCrCl}
  \label{eq:Y13-2}
  Y_{\begin{ytableau} \scriptstyle 1 & \scriptstyle 2 \\
      \scriptstyle 3 \end{ytableau}} 
& = &
  \frac{4}{3} \cdot \FPic{3ArrLeft}\FPic{3Sym12ASym13}\FPic{3ArrRight} 
& \quad \Rightarrow \quad & 
  Y^{\dagger}_{\begin{ytableau} \scriptstyle
      1 & \scriptstyle 2 \\ \scriptstyle 3 \end{ytableau}} 
& = & \frac{4}{3} \cdot \FPic{3ArrLeft}\FPic{3ASym13Sym12}\FPic{3ArrRight}  \\
& = & \frac{1}{3} \left(
  \FPic{3ArrLeft}\FPic{3IdSN}\FPic{3ArrRight} +
  \FPic{3ArrLeft}\FPic{3s12SN}\FPic{3ArrRight} -
  \FPic{3ArrLeft}\FPic{3s13SN}\FPic{3ArrRight} -
  \FPic{3ArrLeft}\FPic{3s132SN}\FPic{3ArrRight} \right)
& &
& = & \frac{1}{3} \left(
  \FPic{3ArrLeft}\FPic{3IdSN}\FPic{3ArrRight} +
  \FPic{3ArrLeft}\FPic{3s12SN}\FPic{3ArrRight} -
  \FPic{3ArrLeft}\FPic{3s13SN}\FPic{3ArrRight} -
  \FPic{3ArrLeft}\FPic{3s123SN}\FPic{3ArrRight} \right) \ .
\nonumber
\end{IEEEeqnarray}
Therefore,
\begin{equation}
 \langle P^{\dagger}u,w\rangle = \left(u^{j_1 \ldots j_m}\right)^*
 \left(\tensor{P}{^{i_1 \ldots i_m}_{j_1 \ldots j_m}}\right)^* w^{i_1 \ldots i_m}
\rightarrow \; 
\scalebox{0.75}{\raisebox{-0.4\height}{%
\includegraphics{TensoruDag}%
\includegraphics{TensorjLabels}%
\includegraphics{TensorPDag}%
\includegraphics{TensoriLabels}%
\includegraphics{Tensorw}%
}}
  \; = \; 
  \scalebox{0.75}{\FPic{TensoruDag}\FPic{TensorPDag}\FPic{Tensorw}}
\ ;
\end{equation}
in direct correspondence with equation~\eqref{eq:uPw}. For birdtrack operators, the
Hermitean conjugate can thus be graphically formed by reflecting the birdtrack about
its vertical axis and reversing the arrows, for example
\begin{align}
  \label{eq:conj-example}
  \FPic{3ArrLeft}\FPic{3s123SN}\FPic{3ArrRight}
  \xrightarrow{\text{reflect}} 
  \reflectbox{\FPic{3ArrLeft}\FPic{3s123SN}\FPic{3ArrRight} }
  \xrightarrow{\text{rev. arr.}}
  \FPic{3ArrLeft}\FPic{3s132SN}\FPic{3ArrRight}
  \hspace{1cm}\text{i.e} \hspace{1cm}
  \left(
\FPic{3ArrLeft}\FPic{3s123SN}\FPic{3ArrRight}
\right)^\dagger 
= \FPic{3ArrLeft}\FPic{3s132SN}\FPic{3ArrRight}
\ .
\end{align}

The mirroring of birdtracks under Hermitian conjugation immediately
implies the unitarity of the primitive invariants (and thus that we
are dealing with a unitary representation of $S_m$ on $\Pow{m}$): the
inverse permutation of any primitive invariant $\rho\in S_m$ is
obtained by traversing the lines of the birdtrack corresponding to
$\rho$ in the opposite direction~\cite{Cvitanovic:2008zz}, for example
\begin{equation}
  \FPic{3ArrLeft}\FPic{3s123SN}\FPic{3ArrRight} \cdot \FPic{3ArrLeft}\FPic{3s132SN}\FPic{3ArrRight} 
  \; = \; 
  \FPic{3ArrLeft}\FPic{3IdSN}\FPic{3ArrRight} 
  \qquad \Longrightarrow \qquad 
  \FPic{3ArrLeft}\FPic{3s123SN}\FPic{3ArrRight} 
  \; = \; 
  \left(\FPic{3ArrLeft}\FPic{3s132SN}\FPic{3ArrRight}\right)^{-1}.
\end{equation}
However, since ``traversing the lines in the opposite direction''
clearly corresponds to flipping the birdtrack about its vertical axis
and reversing the direction of the arrows, we have that
\begin{equation}
  \label{eq:rho-unitary}
  \rho^{-1} = \rho^\dagger \hspace{1cm} \forall \rho\in S_m
  \ ;
\end{equation}
the primitive invariants are unitary.

These obvious Hermiticity properties of the primitive invariants make
it easy to judge Hermiticity of an operator once it is expanded in
this basis set. This is no longer the case in other representations:
While any mirror symmetric birdtrack represents a Hermitian operator,
the converse is not true in all representations. Despite a lack of
apparent mirror symmetry, the product birdtrack
\begin{equation}
\label{eqHerm-Op-Ex1}
  \FPic{4ArrLeft}\FPic{4Sym12Sym34N}    \FPic{4s23N}    \FPic{4ASym12ASym34N}    \FPic{4s23N}    \FPic{4Sym12N}\FPic{4ArrRight}
\end{equation}
is Hermitian, as can be shown by either using the simplification rules
of Theorem~\ref{thm:PropagateSyms} (app.~\ref{sec:PropagationRules})
which allow us to recast~\eqref{eqHerm-Op-Ex1} in an
explicitly mirror symmetric form, or by expanding it fully in terms of
primitive invariants.

In this paper, we will always consider birdtrack operators with lines
directed from right to left (as is indicated by the arrows on the
legs). To reduce clutter, we will from now on suppress the arrows and
(for example) simply write
\begin{equation}
  \FPic{3s123SN} \quad \text{when we mean} \quad
  \FPic{3ArrLeft}\FPic{3s123SN}\FPic{3ArrRight} \; .
\end{equation}

\subsection{Why equation~\texorpdfstring{\eqref{eq:Intro1}}{eqIntro1}
  and its generalization cannot hold for Young projection operators}\label{NonHermitianSection}

In this section, we will have a look at the summation property 
\begin{equation}
  \label{eq:summation-ancestor-notYoung}
  \sum_{\Phi\in\lbrace\Theta\otimes\ybox{\scriptstyle n}\rbrace}
  Y_{\Phi} \overset{?}{=} Y_{\Theta} \qquad \text{for $\Theta\in\mathcal{Y}_{n-1}$}
\end{equation}
and how it fails to apply to Young projection
operators. We will examine two particular examples: We will assume that the
summation property~\eqref{eq:summation-ancestor-notYoung} holds for the
standard (not necessarily Hermitian) Young projection operators over
$\Pow{3}$ and $\Pow{4}$, which will force us to conclude that the corresponding
Young projectors are Hermitian in both cases -- clearly a false statement. 
This motivates us to check whether the summation
property~\eqref{eq:summation-ancestor-notYoung} does hold for
\emph{Hermitian} Young projection operators, thus completing part~\ref{itm:Goal1Part1} of
Goal~\ref{thm:OpsSpanGoal}.

Let us begin with our first example: In section~\ref{sec:IntroHermitianOps}
equation~\eqref{eq:Intro1}, we claimed that
\begin{equation}
  \label{eq:NonHermitian1}
  Y_{%
\begin{ytableau}
    \scriptstyle 1 & \scriptstyle 2 & \scriptstyle 3 %
\end{ytableau}} 
+ 
Y_{%
\begin{ytableau} \scriptstyle 1 & \scriptstyle 2 \\
    \scriptstyle 3 %
\end{ytableau}} 
\neq 
Y_{%
\begin{ytableau} 
 \scriptstyle 1 & \scriptstyle 2 
\end{ytableau}}
\ ,
\end{equation}
where $Y_{\Theta}$ is the Young projection operator
corresponding to the tableau $\Theta$. 
We have now acquired the necessary tools to show why the two sides
fail to match. Assuming equality in~\eqref{eq:NonHermitian1} and
adopting birdtrack notation, this relation takes the form 
\begin{equation}
  \FPic{3Sym123SN} + \sfrac{4}{3}\cdot\FPic{3Sym12ASym13} \stackrel{?}{=} \FPic{3Sym12SN}
\end{equation}
and would imply that
\begin{equation}
  \label{eq:NonHermitian2}
   \sfrac{4}{3}\cdot\FPic{3Sym12ASym13} \stackrel{?}{=} \FPic{3Sym12SN} - \FPic{3Sym123SN}
   \ .
\end{equation}
From section~\ref{sec:HermitianLinearMapsSection}, the right hand side
of \eqref{eq:NonHermitian2} is Hermitian, leading us to conclude that
the left hand side is Hermitian as well. This is clearly not true (as is
evident from the expansions in~\eqref{eq:Y13-2}, specifically the last
terms shown),
\begin{equation}
  Y_{\begin{ytableau} \scriptstyle 1 & \scriptstyle 2
    \\ \scriptstyle 3 \end{ytableau}} \neq Y_{\begin{ytableau} \scriptstyle 1 & \scriptstyle 2
    \\ \scriptstyle 3 \end{ytableau}}^{\dagger}
\ .
\end{equation}
Thus we have arrived at a contradiction. In a similar
way, it can be falsely concluded that $Y_{\begin{ytableau}
    \scriptstyle 1 & \scriptstyle 3 \\ \scriptstyle 2 \end{ytableau}}$
is Hermitian.

Let us now illustrate a slightly more advanced example involving an additional step which was
not present in the example for $n=3$. We will explicitly discuss the
case $n=4$. Let us assume
that equation~\eqref{eq:summation-ancestor-notYoung} holds for $n=3$,
\begin{equation}
  Y_{\begin{ytableau}
    \scriptstyle 1 & \scriptstyle 2 & \scriptstyle 3 \end{ytableau}} + 
Y_{\begin{ytableau} \scriptstyle 1 & \scriptstyle 2 \\
    \scriptstyle 3 \end{ytableau}} \stackrel{?}{=} Y_{\begin{ytableau} \scriptstyle 1 & \scriptstyle
    2 \end{ytableau}}
\qquad \text{and} \qquad
Y_{\begin{ytableau}
    \scriptstyle 1 \\ \scriptstyle 2 \\ \scriptstyle 3 \end{ytableau}} + 
Y_{\begin{ytableau} \scriptstyle 1 & \scriptstyle 3 \\
    \scriptstyle 2 \end{ytableau}} \stackrel{?}{=} Y_{\begin{ytableau} \scriptstyle 1 \\ \scriptstyle
    2 \end{ytableau}}
\end{equation}
and also for $n=4$,
\begin{subequations}
\begin{align}
  Y_{\begin{ytableau} \scriptstyle 1 & \scriptstyle 2 & \scriptstyle 3 \end{ytableau}} \; \stackrel{?}{=} \; &
  Y_{\begin{ytableau} \scriptstyle 1 & \scriptstyle 2 & \scriptstyle 3
      & \scriptstyle 4 \end{ytableau}} +
  Y_{\begin{ytableau} \scriptstyle 1 & \scriptstyle 2 & \scriptstyle 3 \\ \scriptstyle 4 \end{ytableau}} & \label{eq:AddY41}
  \\ 
  Y_{\begin{ytableau} \scriptstyle 1 & \scriptstyle 2 \\ \scriptstyle 3 \end{ytableau}} \; \stackrel{?}{=} \; &
  Y_{\begin{ytableau} \scriptstyle 1 & \scriptstyle 2 & \scriptstyle 4
      \\ \scriptstyle 3 \end{ytableau}} +
  Y_{\begin{ytableau} \scriptstyle 1 & \scriptstyle 2 \\ \scriptstyle 3 & \scriptstyle 4 \end{ytableau}} + 
  Y_{\begin{ytableau} \scriptstyle 1 & \scriptstyle 2 \\ \scriptstyle
      3 \\ \scriptstyle 4 \end{ytableau}} & \label{eq:AddY42}
  \\ 
  Y_{\begin{ytableau} \scriptstyle 1 & \scriptstyle 3 \\ \scriptstyle 2 \end{ytableau}} \; \stackrel{?}{=} \; &
  Y_{\begin{ytableau} \scriptstyle 1 & \scriptstyle 3 & \scriptstyle 4
      \\ \scriptstyle 2 \end{ytableau}} +
  Y_{\begin{ytableau} \scriptstyle 1 & \scriptstyle 3 \\ \scriptstyle 2 & \scriptstyle 4 \end{ytableau}} +
  Y_{\begin{ytableau} \scriptstyle 1 & \scriptstyle 3 \\ \scriptstyle
      2 \\ \scriptstyle 4 \end{ytableau}} & \label{eq:AddY43}
  \\ 
  Y_{\begin{ytableau} \scriptstyle 1 \\ \scriptstyle 2 \\ \scriptstyle 3 \end{ytableau}} \; \stackrel{?}{=} \; &
  Y_{\begin{ytableau} \scriptstyle 1 & \scriptstyle 4 \\ \scriptstyle 2 \\ \scriptstyle 3 \end{ytableau}} + 
  Y_{\begin{ytableau} \scriptstyle 1 \\ \scriptstyle 2 \\ \scriptstyle
      3 \\ \scriptstyle 4 \end{ytableau}} & \label{eq:AddY44}
\ .
\end{align}
\end{subequations}
Equations~\eqref{eq:AddY41} and~\eqref{eq:AddY44}, if indeed valid,
tell us that the operators $Y_{\begin{ytableau} \scriptstyle 1 &
    \scriptstyle 2 & \scriptstyle 3 \\ \scriptstyle 4 \end{ytableau}}$
and $Y_{\begin{ytableau} \scriptstyle 1 & \scriptstyle 4 \\
    \scriptstyle 2 \\ \scriptstyle 3 \end{ytableau}}$ are
Hermitian.\footnote{This can be concluded in the same way that we
  previously found that $Y_{\begin{ytableau} \scriptstyle 1 &
      \scriptstyle 2 \\ \scriptstyle 3 \end{ytableau}}$ is
  Hermitian.} To show that the remaining operators are Hermitian, we
notice that a similarity transformation with an element $\rho$ of
$S_4$ of the form
\begin{equation}
  Y_i \mapsto \rho^{\dagger} Y_i \rho
\end{equation}
does not change the Hermiticity of the operator $Y_i$. Thus, for
example the operator $(34) \cdot Y_{\begin{ytableau} \scriptstyle 1 & \scriptstyle 2 &
    \scriptstyle 3 \\ \scriptstyle 4 \end{ytableau}} \cdot (34)$, where
$(34)\in S_4$ is a transposition, is still Hermitian. It is now easy
to check (via direct calculation) that\footnote{In
  fact,~\cite{Tung:1985na} \emph{defines} the Young projection operator of a
  tableau $\Theta$ that can be obtained from $\Phi$ by reordering the
  entries of $\Phi$ according to a permutation $\rho$ as
  $Y_{\Theta}:=\rho^{\dagger} Y_{\Phi}\rho$.}
\begin{equation}
  (34) \cdot Y_{\begin{ytableau} \scriptstyle 1 & \scriptstyle 2 &
    \scriptstyle 3 \\ \scriptstyle 4 \end{ytableau}} \cdot (34) =
Y_{\begin{ytableau} \scriptstyle 1 & \scriptstyle 2 & \scriptstyle 4
    \\ \scriptstyle 3 \end{ytableau}} \qquad \text{and} \qquad (234) \cdot Y_{\begin{ytableau} \scriptstyle 1 & \scriptstyle 2 &
    \scriptstyle 3 \\ \scriptstyle 4 \end{ytableau}} \cdot (243) =
Y_{\begin{ytableau} \scriptstyle 1 & \scriptstyle 3 & \scriptstyle 4
    \\ \scriptstyle 2 \end{ytableau}}
\ ,
\end{equation}
where $(243)^{\dagger}=(234)$, and similarly
\begin{equation}
  (34) \cdot Y_{\begin{ytableau} \scriptstyle 1 & \scriptstyle 4 \\
      \scriptstyle 2 \\ \scriptstyle 3 \end{ytableau}} \cdot (34)
 = Y_{\begin{ytableau} \scriptstyle 1 & \scriptstyle 3 \\ \scriptstyle 2
    \\ \scriptstyle 4 \end{ytableau}}  \qquad \text{and} \qquad 
  (234) \cdot Y_{\begin{ytableau} \scriptstyle 1 & \scriptstyle 4 \\
      \scriptstyle 2 \\ \scriptstyle 3 \end{ytableau}} \cdot (243)
  = Y_{\begin{ytableau} \scriptstyle 1 & \scriptstyle 2 \\
    \scriptstyle 3 \\ \scriptstyle 4 \end{ytableau}}
\ .
\end{equation}
We therefore conclude that also the Young projection operators
$Y_{\begin{ytableau} \scriptstyle 1 & \scriptstyle 2 & \scriptstyle 4
    \\ \scriptstyle 3 \end{ytableau}}$, $Y_{\begin{ytableau}
    \scriptstyle 1 & \scriptstyle 3 & \scriptstyle 4 \\ \scriptstyle
    2 \end{ytableau}}$, $Y_{\begin{ytableau} \scriptstyle 1 &
    \scriptstyle 3 \\ \scriptstyle 2
    \\ \scriptstyle 4 \end{ytableau}}$ and $Y_{\begin{ytableau}
    \scriptstyle 1 & \scriptstyle 2 \\    \scriptstyle 3 \\
    \scriptstyle 4 \end{ytableau}}$ are Hermitian. The remaining two operators
$Y_{\begin{ytableau} \scriptstyle 1 & \scriptstyle 2 \\ \scriptstyle 3
    & \scriptstyle 4 \end{ytableau}}$ and $Y_{\begin{ytableau}
    \scriptstyle 1 & \scriptstyle 3 \\ \scriptstyle 2 & \scriptstyle
    4 \end{ytableau}}$ can thus be written as linear combinations of
Hermitian operators, 
using equations~\eqref{eq:AddY42} and~\eqref{eq:AddY43},
\begin{subequations}
\begin{align}
  Y_{\begin{ytableau} \scriptstyle 1 & \scriptstyle 2 \\ \scriptstyle 3
    & \scriptstyle 4 \end{ytableau}} = & 
  Y_{\begin{ytableau} \scriptstyle 1 & \scriptstyle 2 \\ \scriptstyle 3 \end{ytableau}} -
  \left( Y_{\begin{ytableau} \scriptstyle 1 & \scriptstyle 2 & \scriptstyle 4
      \\ \scriptstyle 3 \end{ytableau}}  + 
  Y_{\begin{ytableau} \scriptstyle 1 & \scriptstyle 2 \\ \scriptstyle
      3 \\ \scriptstyle 4 \end{ytableau}} \right) \\
Y_{\begin{ytableau} \scriptstyle 1 & \scriptstyle 3 \\ \scriptstyle 2
    & \scriptstyle 4 \end{ytableau}} = & 
  Y_{\begin{ytableau} \scriptstyle 1 & \scriptstyle 3 \\ \scriptstyle 2 \end{ytableau}} -
  \left( Y_{\begin{ytableau} \scriptstyle 1 & \scriptstyle 3 & \scriptstyle 4
      \\ \scriptstyle 2 \end{ytableau}}  +
  Y_{\begin{ytableau} \scriptstyle 1 & \scriptstyle 3 \\ \scriptstyle
      2 \\ \scriptstyle 4 \end{ytableau}} \right),
\end{align}
\end{subequations}
leading us to conclude that they are Hermitian as well. Thus, we have found that \emph{all} Young
projection operators corresponding to Young tableaux in
$\mathcal{Y}_4$ are Hermitian -- a contradiction. It should be noted
that, since the Littlewood-Young projection operators over $\Pow{m}$ reduce to the Young
projectors for $m\leq4$, we conclude that also the
LY-operators cannot satisfy the summation
property~\eqref{eq:summation-ancestor-notYoung} (at least for $m\leq4$).

In fact, one may continue this game for one more level (to the Young
projectors over $\Pow{5}$) before the tricks given in the above
examples seize to suffice and one has to come up with new tools. 

Nonetheless, the key message to take away from this section is that
the obstacle to summability is the lack of Hermiticity of the Young
operators. This provides a strong hint
that~\eqref{eq:summation-ancestor-notYoung} might hold for a Hermitian
version of the Young projection operators (as was already claimed in
section~\ref{sec:IntroHermitianOps}).  In
section~\ref{sec:HermitianSubspaceSpan}, we show explicitly that this
is true, completing part~\ref{itm:Goal1Part2} of
Goal~\ref{thm:OpsSpanGoal}.

% Lastly, we would like to remark that, since Young projection
% operators over $\Pow{m}$ do not satisfy the completeness
% relation~\eqref{eq:YoungCompleteness} beyond $m=4$, it is little
% surprising that they do not satisfy the summation
% property~\eqref{eq:summation-ancestor-notYoung}. It should however
% be noticed that this summation property fails to holds for Young
% projection operators exactly when the first non-Hermitian operators
% occur (at $\Pow{3}$).

In order to be able to do so, we first need to describe how to obtain
Hermitian Young projection operators. This will be the subject of the
following section.

\subsection{Hermitian Young projection operators: KS and beyond}

\subsubsection{KS construction principle}

A construction principle for Hermitian Young projection operators has
recently been found by Keppeler and
Sjödahl~\cite{Keppeler:2013yla}. We will now paraphrase their
construction method, see Theorem~\ref{thm:KSProjectors}, as it forms a
basis for proving that the summation property
eq.~\eqref{eq:summation-ancestor-notYoung} (resp. eq.~\eqref{eq:Outline1}) and its generalizations
indeed hold for Hermitian Young projectors, section
\ref{sec:HermitianSubspaceSpan}. We will further use
Theorem~\ref{thm:KSProjectors} as a starting point for a new
construction principle, which yields much more compact expressions for
Hermitian Young projection operators,
section~\ref{sec:CompactHermitianOps}. We will give Keppeler and
Sjödahl's algorithm without proof; a formal proof can be found
in~\cite{Keppeler:2013yla}.  
\ytableausetup {mathmode, boxsize=normal}

\begin{theorem}[KS Hermitian Young projectors~\cite{Keppeler:2013yla}]
\label{thm:KSProjectors}
  Let $\Theta \in \mathcal{Y}_n$ be a Young tableau. If $n \leq 2$,
  then the Hermitian Young projection operator $P_{\Theta}$
  corresponding to the tableau $\Theta$ is given by
  \begin{equation}
    \label{eq:KSLeq2}
P_{\Theta} := Y_{\Theta}.
  \end{equation}
This provides a termination criterion for an iterative
process that obtains $P_\Theta$ from $P_{\Theta_{(1)}}$
via\footnote{In~\cite{Keppeler:2013yla}, eq.~\eqref{eq:KSGreater2} is given as $P_{\Theta}=
  P_{\Theta_{(1)}}\otimes Y_{\Theta}\otimes P_{\Theta_{(1)}}$;
however, since the $P_i$ and $Y_j$ are understood to be linear maps on
the space $\Pow{n}$, this equation is merely a product of linear
maps. The authors therefore deem the tensor-product-notation
introduced by KS unnecessarily complicated, and
denote this product of linear maps as shown above.}
\begin{equation}
  \label{eq:KSGreater2}
P_{\Theta} :=  P_{\Theta_{(1)}} \;
Y_{\Theta} \; P_{\Theta_{(1)}}
\ ,
\end{equation}
once $n > 2$. In~\eqref{eq:KSGreater2} $P_{\Theta_{(1)}}$ is
understood to be canonically embedded in the algebra
$\API{\SUN,V^{\otimes n}}$. Thus, $P_{\Theta}$ is recursively obtained
from the full chain of its Hermitian ancestor operators
$P_{\Theta_{(m)}}$.\\

\noindent The above operators generalize
properties~\eqref{eq:YoungProperties} of Young projection operators to
all $n$:
\begin{subequations}
  \label{eq:KSProperties} 
\begin{alignat}{2}
 & \text{Idempotency:} \hspace{1.2cm}
    && P_{\Theta} \cdot P_{\Theta} = P_{\Theta} 
      \label{eq:KSIdempotency} \\
 &\text{Orthogonality:} \hspace{1.2cm}
    && P_{\Theta} \cdot P_{\Phi} = \delta_{\Theta \Phi} P_{\Theta}
      \label{eq:KSOrthogonality} \\
 &\text{Completeness:} \hspace{1.2cm}
    && \sum_{\Theta \in\mathcal{Y}_n} P_{\Theta} = \mathbb{1}_n
      \label{eq:KSCompleteness} 
\end{alignat}
\end{subequations}

\end{theorem}

\noindent As an example, consider the Young tableau
\begin{equation}
  \Theta =
  \begin{ytableau}
    1 & 2 & 4 \\
    3 & 5
  \end{ytableau}
\ ,
\end{equation}
with ancestor tableaux\footnote{We do not have to consider the ancestor $\Theta_{(4)}$, since
$\Theta_{(3)}\in\mathcal{Y}_2$ and thus terminates the recursion~\eqref{eq:KSGreater2}.}
\begin{equation}
  \Theta_{(1)} =
  \begin{ytableau}
    1 & 2 & 4 \\
    3
  \end{ytableau}\;, \quad \Theta_{(2)} = 
  \begin{ytableau}
    1 & 2 \\
    3 
  \end{ytableau} \quad \text{and} \quad \Theta_{(3)} =
  \begin{ytableau}
    1 & 2 
  \end{ytableau}
\ .
\end{equation}
When constructing the Hermitian Young projection operator $P_{\Theta}$
according to the KS-Theorem~\ref{thm:KSProjectors}, we first have to find
$P_{\Theta_{(3)}}$, $P_{\Theta_{(2)}}$ and $P_{\Theta_{(1)}}$. According to the Theorem,
$P_{\Theta_{(3)}}=Y_{\Theta_{(3)}}$, since
$\Theta_{(3)}\in\mathcal{Y}_2$. Then, following the iterative
procedure of the KS-Theorem, $P_{\Theta_{(2)}}$ and $P_{\Theta_{(1)}}$
are  given by
\begin{align}
  P_{\Theta_{(2)}} = & \; P_{\Theta_{(3)}} Y_{\Theta_{(2)}}
  P_{\Theta_{(3)}} = Y_{\Theta_{(3)}} Y_{\Theta_{(2)}}
  Y_{\Theta_{(3)}} \\
P_{\Theta_{(1)}} = & \; P_{\Theta_{(2)}} Y_{\Theta_{(1)}}
  P_{\Theta_{(2)}} = \underbrace{Y_{\Theta_{(3)}} Y_{\Theta_{(2)}}
  Y_{\Theta_{(3)}}}_{=P_{\Theta_{(2)}}} Y_{\Theta_{(1)}} \underbrace{Y_{\Theta_{(3)}} Y_{\Theta_{(2)}}
  Y_{\Theta_{(3)}}}_{=P_{\Theta_{(2)}}}.  
\end{align}
Then, the desired operator $P_{\Theta}$ is 
\begin{equation}
  P_{\Theta} = P_{\Theta_{(1)}} Y_{\Theta}
  P_{\Theta_{(1)}} = \underbrace{Y_{\Theta_{(3)}} Y_{\Theta_{(2)}}
  Y_{\Theta_{(3)}} Y_{\Theta_{(1)}} Y_{\Theta_{(3)}} Y_{\Theta_{(2)}}
  Y_{\Theta_{(3)}}}_{=P_{\Theta_{(1)}}} Y_{\Theta} \underbrace{Y_{\Theta_{(3)}} Y_{\Theta_{(2)}} Y_{\Theta_{(3)}} Y_{\Theta_{(1)}} Y_{\Theta_{(3)}} Y_{\Theta_{(2)}} Y_{\Theta_{(3)}}}_{=P_{\Theta_{(1)}}}.
\end{equation}
As a birdtrack, $P_{\Theta}$ can be written as
\begin{equation}
  \label{eq:KSOpsEx1}
    P_{\Theta} = \sfrac{128}{9} \cdot \scalebox{0.7}{$\underbrace{\underbrace{\FPic{5Sym12N}}_{\mbox{\normalsize
      $\bar{Y}_{\Theta_{(3)}}$}} \underbrace{\FPic{5Sym12N}\FPic{5s23N}\FPic{5ASym12N}\FPic{5s23N}}_{\mbox{\normalsize
      $\bar{Y}_{\Theta_{(2)}}$}}
\underbrace{\FPic{5Sym12N}}_{\mbox{\normalsize
      $\bar{Y}_{\Theta_{(3)}}$}}
\underbrace{\FPic{5s34N}\FPic{5Sym123N}\FPic{5s243N}\FPic{5ASym12N}\FPic{5s23N}}_{\mbox{\normalsize
      $\bar{Y}_{\Theta_{(1)}}$}}
\underbrace{\FPic{5Sym12N}}_{\mbox{\normalsize
      $\bar{Y}_{\Theta_{(3)}}$}} \underbrace{\FPic{5Sym12N}\FPic{5s23N}\FPic{5ASym12N}\FPic{5s23N}}_{\mbox{\normalsize
      $\bar{Y}_{\Theta_{(2)}}$}} \underbrace{\FPic{5Sym12N}}_{\mbox{\normalsize
      $\bar{Y}_{\Theta_{(3)}}$}}}_{\mbox{\normalsize
      $\bar{P}_{\Theta_{(1)}}$}}
\underbrace{\FPic{5s34N}\FPic{5Sym123Sym45N}\FPic{5s2453N}\FPic{5ASym12ASym34N}\FPic{5s23s45N}}_{\mbox{\normalsize
    $\bar{Y}_{\Theta}$}}
\underbrace{\underbrace{\FPic{5Sym12N}}_{\mbox{\normalsize
      $\bar{Y}_{\Theta_{(3)}}$}} \underbrace{\FPic{5Sym12N}\FPic{5s23N}\FPic{5ASym12N}\FPic{5s23N}}_{\mbox{\normalsize
      $\bar{Y}_{\Theta_{(2)}}$}}
\underbrace{\FPic{5Sym12N}}_{\mbox{\normalsize
      $\bar{Y}_{\Theta_{(3)}}$}}
\underbrace{\FPic{5s34N}\FPic{5Sym123N}\FPic{5s243N}\FPic{5ASym12N}\FPic{5s23N}}_{\mbox{\normalsize
      $\bar{Y}_{\Theta_{(1)}}$}}
\underbrace{\FPic{5Sym12N}}_{\mbox{\normalsize
      $\bar{Y}_{\Theta_{(3)}}$}} \underbrace{\FPic{5Sym12N}\FPic{5s23N}\FPic{5ASym12N}\FPic{5s23N}}_{\mbox{\normalsize
      $\bar{Y}_{\Theta_{(2)}}$}} \underbrace{\FPic{5Sym12N}}_{\mbox{\normalsize
      $\bar{Y}_{\Theta_{(3)}}$}}}_{\mbox{\normalsize
      $\bar{P}_{\Theta_{(1)}}$}}
$},
\end{equation}
where
\begin{equation}
  \frac{128}{9} = \left(\alpha_{\Theta_{(3)}}\right)^8
  \left(\alpha_{\Theta_{(2)}}\right)^4
  \left(\alpha_{\Theta_{(1)}}\right)^2 \alpha_{\Theta}
\end{equation}
is the appropriate normalization constant arising from the
KS-algorithm.

Let us emphasize that KS have proven that this or any other operator
constructed with their algorithm is Hermitian. The operator
\eqref{eq:KSOpsEx1} is however not symmetric under a flip about its
vertical axis, and thus Hermiticity is not visually obvious. An
additional advantage of the construction algorithm described in
section \ref{sec:CompactHermitianOps} is that it will necessarily
yield mirror-symmetric operators, making their Hermiticity immediately
visible.  \ytableausetup {mathmode, boxsize=0.7em}

\subsubsection{Beyond the KS construction}\label{sec:ShortKSOperators}

The results regarding Hermitian Young projection operators presented
up until now are all taken from~\cite{Keppeler:2013yla}. We will now
move beyond the established results and show that
\begin{enumerate}
\item the KS-operators can be simplified to yield more compact
  expressions, \emph{c.f.}
  Corollary~\ref{thm:ShortKSProjectors}\footnote{While the
    simplification in Corollary~\ref{thm:ShortKSProjectors} is already
    significant, the alternative construction given in
    section~\ref{sec:CompactHermitianOps} will be even more efficient.}
\item the KS-operators obey the summation
  property~\eqref{eq:summation-ancestor-notYoung}
  (resp.~\eqref{eq:Outline1})
  \begin{equation}
    \sum_{\Phi \in \lbrace\Theta\otimes\ybox{\scriptstyle n}\rbrace}
    P_{\Phi} = P_{\Theta}
\ ;
  \end{equation}
this will be shown in section~\ref{sec:HermitianSubspaceSpan}.
\end{enumerate}

In~\cite{Alcock-Zeilinger:2016bss}, we found
several simplification rules for birdtrack operators, some of which
are summarized in
section~\ref{sec:CancellationRules}. In particular, Theorem
\ref{thm:CancelWedgedParentOp} can be used to shorten the above operator~\eqref{eq:KSOpsEx1} to
\begin{align}
  P_{\Theta} = & \ \ Y_{\Theta_{(3)}} Y_{\Theta_{(2)}} Y_{\Theta_{(1)}}
  Y_{\Theta} Y_{\Theta_{(1)}} Y_{\Theta_{(2)}}
  Y_{\Theta_{(3)}} 
  \\
 = & \ 
 8 \cdot \scalebox{0.7}{$\underbrace{\FPic{5Sym12N}}_{\mbox{\normalsize
      $\bar{Y}_{\Theta_{(3)}}$}} \underbrace{\FPic{5Sym12N}\FPic{5s23N}\FPic{5ASym12N}\FPic{5s23N}}_{\mbox{\normalsize
      $\bar{Y}_{\Theta_{(2)}}$}}
  \underbrace{\FPic{5s34N}\FPic{5Sym123N}\FPic{5s243N}\FPic{5ASym12N}\FPic{5s23N}}_{\mbox{\normalsize
      $\bar{Y}_{\Theta_{(1)}}$}}
\underbrace{\FPic{5s34N}\FPic{5Sym123Sym45N}\FPic{5s2453N}\FPic{5ASym12ASym34N}\FPic{5s23s45N}}_{\mbox{\normalsize
    $\bar{Y}_{\Theta}$}}
\underbrace{\FPic{5s34N}\FPic{5Sym123N}\FPic{5s243N}\FPic{5ASym12N}\FPic{5s23N}}_{\mbox{\normalsize $\bar{Y}_{\Theta_{(1)}}$}} \underbrace{\FPic{5Sym12N}\FPic{5s23N}\FPic{5ASym12N}\FPic{5s23N}}_{\mbox{\normalsize
      $\bar{Y}_{\Theta_{(2)}}$}} \underbrace{\FPic{5Sym12N}}_{\mbox{\normalsize
      $\bar{Y}_{\Theta_{(3)}}$}}$}, &
\end{align}
where
$\left(\alpha_{\Theta_{(3)}}\alpha_{\Theta_{(2)}}\alpha_{\Theta_{(1)}}\right)^2\alpha_{\Theta}=8$. The
above expression for $P_{\Theta}$ is clearly considerably shorter than
the expression given in~\eqref{eq:KSOpsEx1}. In fact, Theorem
\ref{thm:CancelWedgedParentOp} allows us to systematically shorten the
KS-projection operators, exposing a new, much simpler general form:

\begin{corollary}[staircase form of Hermitian Young projectors]
\label{thm:ShortKSProjectors}
  Let $\Theta\in\mathcal{Y}_n$ be a Young tableau. Then, the 
  corresponding Hermitian Young projection operator $P_{\Theta}$ is
  given by
  \begin{equation}
    P_{\Theta} = \; Y_{\Theta_{(n-2)}}
    Y_{\Theta_{(n-3)}} Y_{\Theta_{(n-4)}} \ldots
    Y_{\Theta_{(2)}} Y_{\Theta_{(1)}} Y_{\Theta} \;
    Y_{\Theta_{(1)}} Y_{\Theta_{(2)}} \ldots
    Y_{\Theta_{(n-4)}} Y_{\Theta_{(n-3)}} Y_{\Theta_{(n-2)}}.
  \end{equation}
\end{corollary}
This result simply follows from a repeated application of
Theorem~\ref{thm:CancelWedgedParentOp}, where we notice that
$\Theta_{(n-2)}\in\mathcal{Y}_2$ necessarily.

Even though this simplification is already quite substantial, it is by
no means the simplest form achievable. We will present a new
construction principle in section~\ref{sec:CompactHermitianOps},
creating even more compact and thus easier usable Hermitian Young
projection operators. The proof of this construction will however make
use of the KS-Theorem~\ref{thm:KSProjectors}, see
app.~\ref{sec:Proofs}.

\subsection{Spanning subspaces with Hermitian
  operators}\label{sec:HermitianSubspaceSpan}

We are finally in a position to show that 
\begin{equation}
  \label{eq:SpanSubspaces1}
  \sum_{\Phi \in \lbrace \Theta \otimes \ybox{\scriptstyle n} \rbrace}
  P_{\Phi} = P_{\Theta} 
\end{equation}
holds for every $\Theta\in\mathcal{Y}_{n-1}$ if the $P_{\Xi}$ are the
Hermitian operators introduced previously. In
section~\ref{sec:IntroHermitianOps}, we gave the particular example
\begin{equation}
  P_{\begin{ytableau}
    \scriptstyle 1 & \scriptstyle 2 & \scriptstyle 3 \end{ytableau}} + 
P_{\begin{ytableau} \scriptstyle 1 & \scriptstyle 2 \\
    \scriptstyle 3 \end{ytableau}} = P_{\begin{ytableau} \scriptstyle 1 & \scriptstyle
    2 \end{ytableau}}
\end{equation}
which holds for the \emph{Hermitian} Young operators $P_{\Xi}$ but
fails to hold for their Young operator counterparts.

To prove~\eqref{eq:SpanSubspaces1} in general, we first need to show that a
projection operator $P_{\Theta}$ projects onto a subspace of the image
of an operator $P_{\Theta_{(m)}}$, where $\Theta_{(m)}$ is an ancestor
tableau of $\Theta$. In particular, this will mean that the image of
an operator $P_{\Theta}$ is a subset of the image of its parent
operator $P_{\Theta_{(1)}}$.

\begin{lemma}[Subspaces
  corresponding to Hermitian Young projection operators are nested]
  \label{thm:SubspaceProjectors} \-
  Let $\Theta\in\mathcal{Y}_n$ be a Young tableau and let
  $\Theta_{(m)}$ be its ancestor tableau, with $m<n$. Furthermore, let
  $P_{\Theta}$ and $P_{\Theta_{(m)}}$ be the Hermitian Young
  projection operators corresponding to these tableaux. Then, the
  image of $P_{\Theta}$ lies entirely in the image of
  $P_{\Theta_{(m)}}$, 
\begin{equation}
\label{eq:Inclusion-Herm-Ancestor}
    P_{\Theta} P_{\Theta_{(m)}} = P_{\Theta} = P_{\Theta_{(m)}}
    P_{\Theta}
\ .
  \end{equation}
\end{lemma}

An immediate consequence of this Lemma is that a Hermitian Young
projection operator $P_{\Theta}$ commutes with its ancestor operator
$P_{\Theta_{(m)}}$. In appendix~\ref{sec:Y-Ancestor-not-commute}, we
first exemplify that Young projectors do not necessarily obey the
analgous image inclusion properties $Y_{\Theta} Y_{\Theta_{(m)}} =
Y_{\Theta}$ and/or $Y_{\Theta_{(m)}} Y_{\Theta} = Y_{\Theta}$. We
prove that where the image inclusion fails to hold the associated
commutator $[Y_{\Theta_{(m)}}, Y_{\Theta}]$ does not vanish.

Image inclusion will play an integral part in
the proof of eq.~\eqref{eq:SpanSubspaces1} and thus highlights where
the proof would break down for the (not necessarily Hermitian) Young
projectors.

Before we give the proof of Lemma~\ref{thm:SubspaceProjectors}, We
wish to draw attention to how this proof makes use of some of the
simplification rules given in section~\ref{sec:CancellationRules}, as
this will be mirrored in the proofs of the main Theorems given in
appendix~\ref{sec:Proofs}.

\emph{Proof of Lemma~\ref{thm:SubspaceProjectors}:} To prove the
inclusion of the subspaces, it suffices to show that the product of
the operators satisfies eq.~\eqref{eq:Inclusion-Herm-Ancestor}
(\emph{c.f.} eq.~\eqref{eq:OperatorInclusion1}). What this relation
implies is that if we first act the product
$P_{\Theta_{(m)}}P_{\Theta}$ (or equivalently
$P_{\Theta}P_{\Theta_{(m)}}$) on an object $x$, we obtain the same
outcome as if we only act $P_{\Theta}$ on $x$. Hence, $P_{\Theta}$
must correspond to a smaller subspace than $P_{\Theta_{(m)}}$, and
this subspace must completely be contained in the subspace
corresponding to $P_{\Theta_{(m)}}$. From the shortened KS
construction, Corollary~\ref{thm:ShortKSProjectors}, the Hermitian
Young projection operators $P_{\Theta}$ and $P_{\Theta_{(m)}}$ are
given by
  \begin{align}
    P_{\Theta} & = Y_{\Theta_{(n-2)}} Y_{\Theta_{(n-3)}} \ldots
    Y_{\Theta_{(m+1)}} {\colorbox{red!20}{$Y_{\Theta_{(m)}}$}} \ldots
    Y_{\Theta_{(1)}} {\colorbox{blue!25}{$Y_{\Theta}$}} Y_{\Theta_{(1)}} \ldots {\colorbox{red!20}{$Y_{\Theta_{(m)}}$}} Y_{\Theta_{(m+1)}} \ldots Y_{\Theta_{(n-3)}}
    Y_{\Theta_{(n-2)}} \\
    P_{\Theta_{(m)}} & = Y_{\Theta_{(n-2)}} Y_{\Theta_{(n-3)}}
    \ldots Y_{\Theta_{(m+1)}} {\colorbox{red!20}{$Y_{\Theta_{(m)}}$}}
      Y_{\Theta_{(m+1)}} \ldots Y_{\Theta_{(n-2)}}
    Y_{\Theta_{(n-2)}}
\ .
  \end{align}
When forming the product $P_{\Theta}P_{\Theta_{(m)}}$, we
see a lot of cancellation of wedged ancestor operators due to Theorem~\ref{thm:CancelWedgedParentOp},
\begin{align}
 P_{\Theta} \cdot P_{\Theta_{(m)}} & =  
Y_{\Theta_{(n-2)}} 
\ldots
{\colorbox{red!20}{$Y_{\Theta_{(m)}}$}} 
\ldots
Y_{\Theta_{(1)}} 
{\colorbox{blue!25}{$Y_{\Theta}$}} 
Y_{\Theta_{(1)}} 
\ldots 
\underbrace{
{\colorbox{red!20}{$Y_{\Theta_{(m)}}$}} 
\ldots 
Y_{\Theta_{(n-2)}} 
\cdot 
Y_{\Theta_{(n-2)}} 
\ldots 
{\colorbox{red!20}{$Y_{\Theta_{(m)}}$}}
}_{=\colorbox{red!20}{$Y_{\Theta_{(m)}}$}}
 \ldots
    Y_{\Theta_{(n-2)}}
\nonumber \\
& = 
Y_{\Theta_{(n-2)}}
\ldots
{\colorbox{red!20}{$Y_{\Theta_{(m)}}$}} 
\ldots
Y_{\Theta_{(1)}} 
{\colorbox{blue!25}{$Y_{\Theta}$}} 
Y_{\Theta_{(1)}} 
\ldots 
{\colorbox{red!20}{$Y_{\Theta_{(m)}}$}} 
\ldots 
Y_{\Theta_{(n-2)}}
\ .
\end{align}
The above can easily be identified to be the operator
$P_{\Theta}$, yielding the first equality
$P_{\Theta}P_{\Theta_{(m)}}=P_{\Theta}$. The second
equality can similarly be shown, leading to the desired result. \qed

\ytableausetup
{mathmode, boxsize=normal}
\setul{}{0.8pt}

Let us now prove the summation property~\eqref{eq:SpanSubspaces1} for
Hermitian Young projection operators: Recall the completeness relation of Hermitian Young
projection operators, eq.~\eqref{eq:KSCompleteness},
\begin{equation}
  \label{eq:SpanSubspaces2a}
  \sum_{\Theta\in\mathcal{Y}_{n-1}} P_{\Theta} = \mathrm{id}_{n-1},
\end{equation}
where $\mathrm{id}_k$ is the identity operator on the space
$\Pow{k}$. Equation~\eqref{eq:SpanSubspaces2a} can be canonically
embedded into the space $\Pow{n}$ as was discussed in section
\ref{sec:OperatorsNotation}. In order to make the embedding of the
operator $P_{\Theta}$ explicit, we will -- for this section \emph{only}
-- make the identity operator on the last factor explicitly visible in
the birdtrack spirit and denote the embedded operator by the symbol
\ul{$P_{\Theta}$}.\footnote{In birdtrack notation, the canonically
  embedded operator \ul{$P_{\Theta}$} will be $P_{\Theta}$ with an
  extra index line on the bottom, making the notation
  \ul{$P_{\Theta}$} intuitive.} The embedded equation
\eqref{eq:SpanSubspaces2a} thus is
\begin{equation}
  \label{eq:SpanSubspaces3}
  \sum_{\Theta\in\mathcal{Y}_{n-1}} \text{\ul{$P_{\Theta}$}} = \mathrm{id}_{n}.
\end{equation}
Even though~\eqref{eq:SpanSubspaces3} is a decomposition of unity, a
finer decomposition of $\mathrm{id}_n$ (also using only orthogonal objects)
is obtained with Hermitian Young projection operators corresponding to Young
tableaux in $\mathcal{Y}_n$,
\begin{equation}
  \label{eq:SpanSubspaces4}
    \sum_{\Phi\in\mathcal{Y}_{n}} P_{\Phi} = \mathrm{id}_{n}.
\end{equation}
Since clearly $\mathcal{Y}_n$ is the union of all the sets
$\lbrace\Theta\otimes\ybox{n}\rbrace$, for all
$\Theta\in\mathcal{Y}_{n-1}$, the sum~\eqref{eq:SpanSubspaces4} can be
split into
\ytableausetup
{mathmode, boxsize=0.7em}
\begin{equation}
  \label{eq:SpanSubspaces5}
    \sum_{\Phi\in\mathcal{Y}_{n}} P_{\Phi} =
    \sum_{\Theta\in\mathcal{Y}_{n-1}}
    \left( \sum_{\Psi\in\lbrace\Theta\otimes\ybox{\scriptstyle n}\rbrace} P_{\Psi} \right) = \mathrm{id}_{n}.
\end{equation}
Since both~\eqref{eq:SpanSubspaces3} and~\eqref{eq:SpanSubspaces5} are
a decomposition of $\mathrm{id}_n$, they must be equal to each other,
yielding
\begin{equation}
  \label{eq:SpanSubspaces6}
  \sum_{\Theta\in\mathcal{Y}_{n-1}} \text{\ul{$P_{\Theta}$}} =
  \sum_{\Theta\in\mathcal{Y}_{n-1}} \left( \sum_{\Psi\in\lbrace\Theta\otimes\ybox{\scriptstyle
          n}\rbrace} P_{\Psi} \right).
\end{equation}
Let us now multiply the above equation with a particular operator
\ul{$P_{\Theta'}$} on $\Pow{n}$, where $\Theta'$ is a particular
tableau in $\mathcal{Y}_{n-1}$. Due to the orthogonality property
(eq.~\eqref{eq:KSOrthogonality}, Theorem~\ref{thm:KSProjectors}) and
the inclusion property (eq.~\eqref{eq:Inclusion-Herm-Ancestor},
Lemma~\ref{thm:SubspaceProjectors}) of Hermitian Young
projectors,\footnote{This is where the proof would break down for the
  standard Young projection operators even for $n\leq4$, where they explicitly do not
  satisfy the image inclusion
  property~\eqref{eq:Inclusion-Herm-Ancestor}, \emph{c.f.}
  appendix~\ref{sec:Y-Ancestor-not-commute}.} it follows that
\begin{align}
  \sum_{\Theta\in\mathcal{Y}_{n-1}}
  \delta_{\Theta\Theta'}
  \text{\ul{$P_{\Theta}$}} 
& = 
  \sum_{\Theta\in\mathcal{Y}_{n-1}} 
  \left( 
  \delta_{\Theta\Theta'}
  \sum_{\Psi\in\lbrace\Theta\otimes\ybox{\scriptstyle n}\rbrace}
    P_{\Psi} 
  \right) 
\label{eq:SpanSubspaces-last-step} \\
  \text{\ul{$P_{\Theta'}$}} 
& = 
  \sum_{\Psi\in\lbrace\Theta'\otimes\ybox{\scriptstyle n}\rbrace}
   P_{\Psi}  \label{eq:SpanSubspaces6b}
\ ,
\end{align}
yielding the desired equation~\eqref{eq:SpanSubspaces1}. This
concludes part~\ref{itm:Goal1Part2} of Goal~\ref{thm:OpsSpanGoal}.
\ytableausetup{mathmode, boxsize=1.2em}%

Since the Hermitian operators sum up to their Hermitian parent
operators (eq.~\eqref{eq:SpanSubspaces6b}), and these in turn sum to
their Hermitian parent operators, the summation property necessarily
holds over multiple generations. This statement also follows straight
from Lemma~\ref{thm:SubspaceProjectors}, which states that the image
of a Hermitian Young projection operator $P_{\Theta}$ is contained in
the image of its Hermitian ancestor operator $P_{\Theta_{(m)}}$, where
$m$ can be \emph{any} positive integer. Therefore, if
$\mathcal{Y}_{\Theta,n}:=\lbrace\Theta\otimes\ybox{m}\otimes\cdots\otimes\ybox{n}\rbrace$
is the subset of $\mathcal{Y}_n$ containing all tableaux that have
$\Theta\in\mathcal{Y}_{m-1}$ as their ancestor, then
\begin{equation}
  \label{eq:SpanSubspaces2}
 \sum_{\Phi \in \mathcal{Y}_{\Theta,n}} P_{\Phi} = P_{\Theta}
\ ,
\end{equation}
confirming eq.~\eqref{eq:Outline2}. For example, if
$\Theta = \begin{ytableau} 1 & 2 & 3 \end{ytableau}$, then
$P_{\Theta}$ can be written as a sum of the following Hermitian Young
projection operators corresponding to tableaux in $\mathcal{Y}_{5}$,
\ytableausetup {mathmode, boxsize=0.7em}
\begin{equation}
\label{eq:SpanSkipGenereationEx}
  \underbrace{P_{\begin{ytableau} \scriptstyle 1 & \scriptstyle 2 & \scriptstyle
      3 & \scriptstyle 4 & \scriptstyle 5 \end{ytableau}} + 
  P_{\begin{ytableau} \scriptstyle 1 & \scriptstyle 2 & \scriptstyle
      3 & \scriptstyle 4 \\ \scriptstyle 5 \end{ytableau}} +
  P_{\begin{ytableau} \scriptstyle 1 & \scriptstyle 2 & \scriptstyle
      3 & \scriptstyle 5 \\ \scriptstyle 4 \end{ytableau}} +
  P_{\begin{ytableau} \scriptstyle 1 & \scriptstyle 2 & \scriptstyle
      3 \\ \scriptstyle 4 & \scriptstyle 5 \end{ytableau}} +
  P_{\begin{ytableau} \scriptstyle 1 & \scriptstyle 2 & \scriptstyle
      3 \\ \scriptstyle 4 \\ \scriptstyle
      5 \end{ytableau}}}_{\sum_{\Phi \in \mathcal{Y}_{\Theta,5}} P_{\Phi}}
  = \underbrace{P_{\begin{ytableau} \scriptstyle 1 & \scriptstyle 2 & \scriptstyle
      3 \end{ytableau}}}_{P_{\Theta}}.
\end{equation}
This concludes the last part (part~\ref{itm:Goal1Part3}) of the first Goal of this paper.
\ytableausetup
{mathmode, boxsize=normal}

\begin{figure}[p]
  \centering
  \includegraphics[width=\textwidth]{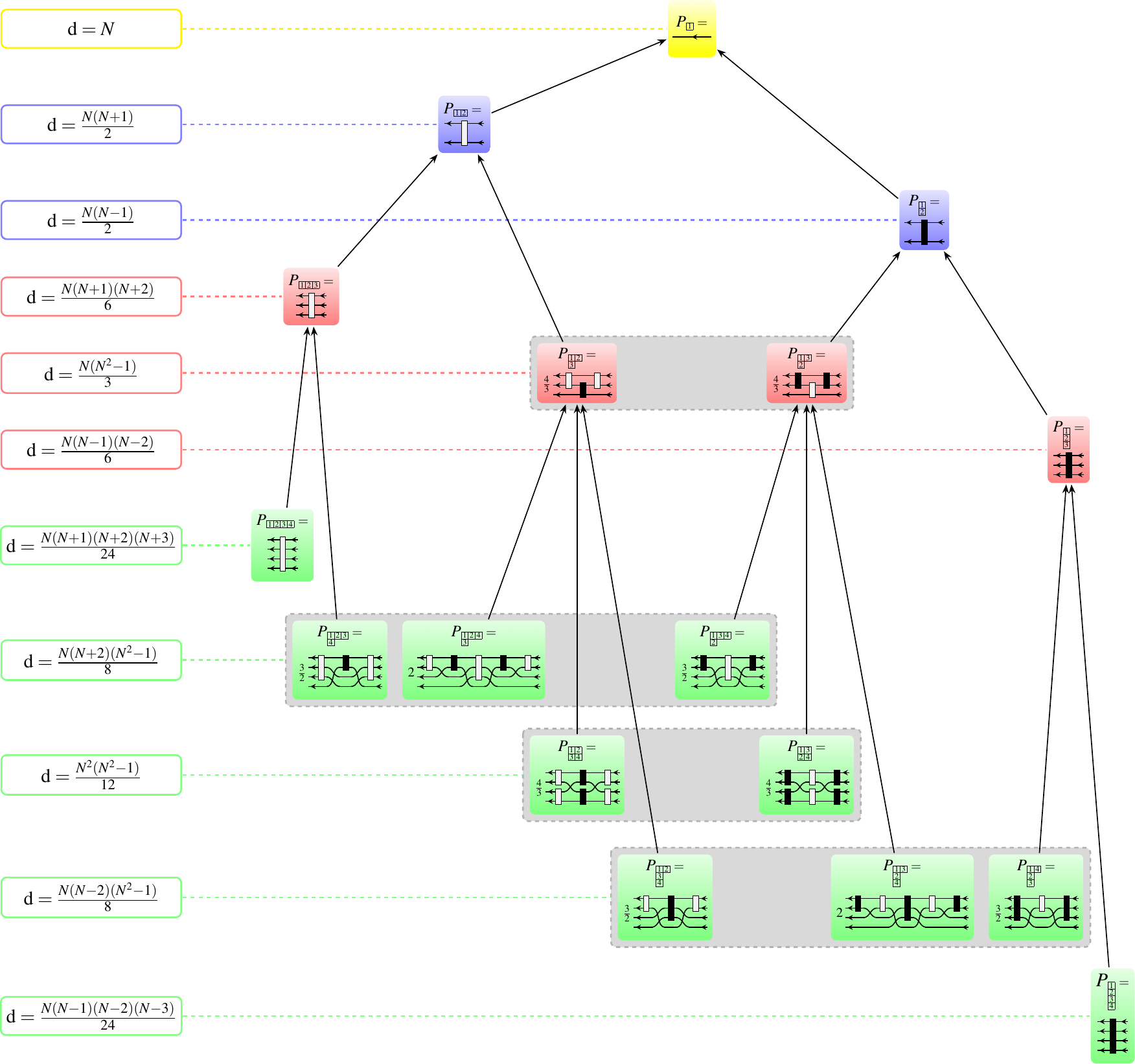}
  \caption{Hierarchy of Young tableaux and the associated nested
    Hermitian Young projector decompositions (in the sense of
    embeddings into $\API{\SUN,\Pow{4}}$): Projection operators that
    are contained in a gray box correspond to equivalent irreducible
    representations of $\SUN$, as their corresponding Young tableaux
    have the same shape~\cite{Cvitanovic:2008zz,Tung:1985na}. The
    dimension of the irreducible representation corresponding to a
    (set of) operators(s) is given on the
    left~\cite[fig. 9.1]{Cvitanovic:2008zz}. The arrows indicate which
    operators sum to which ancestor -- this summation property of
    Hermitian Young projection operators was not observed
    by~\cite{Cvitanovic:2008zz}.}
  \label{fig:hierarchy-4}
\end{figure}

Figure~\ref{fig:hierarchy-4} shows all Hermitian Young projectors
corresponding to Young tableaux in $\mathcal{Y}_n$ up to and including
$n=4$. The arrows in this figure indicate which operators sum to which
ancestor operators. The summation property of Hermitian Young
projectors was \emph{not} mentioned
in~\cite[fig. 9.1]{Cvitanovic:2008zz} where a virtually identical
figure can be found.

\section{An algorithm to construct compact expressions of Hermitian
  Young projection operators}\label{sec:CompactHermitianOps}

We will now come to Goal~\ref{thm:MOLDOpsGoal} of this paper and
provide a construction principle that allows us to directly arrive at
\emph{compact expressions} for Hermitian Young projection operators
(see Theorem \ref{thm:MOLDConstruction} below). This construction
yields much shorter expressions than the previously encountered KS
algorithm (Theorem~\ref{thm:KSProjectors}), or even the shortened
version of Theorem~\ref{thm:ShortKSProjectors}, as is exemplified in
Fig.~\ref{fig:MOLDAdvantage}.

\subsection{Lexically ordered Young tableaux}

It turns out that the ordering of the numbers within the Young tableau
plays a vital role in our algorithm. Thus, we will first establish
what we mean by the lexical order of a Young tableau.  To do so, we
will introduce column- and row-words:\footnote{It should be noted that
  Definition~\ref{def:TableauxWords} of the row-word is
  \emph{different} to the definition given in the standard literature
  such as~\cite{Sagan:2000,Fulton:1997}: there, the row word is read
  from bottom to top rather than from top to bottom. However, for the
  purposes of this paper, Definition~\ref{def:TableauxWords} is more
  useful than the standard definition.}

\begin{definition}[column- and row-words \& lexical ordering]
  \label{def:TableauxWords}
  Let $\Theta \in \mathcal{Y}_n$ be a Young tableau. We define the
  \emph{column-word} of $\Theta$, $\mathfrak{C}_{\Theta}$, to be the
  column vector whose entries are the entries of $\Theta$ as read
  column-wise from left to right. Similarly, the \emph{row-word} of
  $\Theta$, $\mathfrak{R}_{\Theta}$, is defined to be the row vector
  whose entries are those of $\Theta$ read row-wise from top to
  bottom.

  We will call a tableau $\Theta$ \emph{lexically ordered}, if either
  $\mathfrak{C}_{\Theta}$ or $\mathfrak{R}_{\Theta}$ or both are in
  lexical order.  In particular, we say that $\Theta$ is
  column-ordered (resp. row-ordered), if $\mathfrak{C}_{\Theta}$
  (resp. $\mathfrak{R}_{\Theta}$) is in lexical order.
\end{definition}

\noindent For example, the tableau 
\begin{equation}
\label{eq:ExWords1}
  \Phi :=
  \begin{ytableau}
    1 & 5 & 7 & 9 \\
    2 & 6 & 8 \\
    3 \\
    4
  \end{ytableau}
\end{equation}
has a column-word
\begin{equation}
 \mathfrak{C}_{\Phi} = (1,2,3,4,5,6,7,8,9)^t,
\end{equation}
and a row-word
\begin{equation}
  \mathfrak{R}_{\Phi} = (1,5,7,9,2,6,8,3,4).
\end{equation}
From this, we see that $\Phi$ in~\eqref{eq:ExWords1} is lexically
ordered. In particular, it is column-ordered (but not row-ordered).

% Thus, we will first establish
% what we mean by the lexical order of a Young tableau:\footnote{This is
% not to be confused with the order relation between tableaux given in
% Definition~\ref{def:OrderRelationTableaux} in section}

% \begin{definition}[lexical ordering]
%   \label{def:TableauOrder}
%   Let $\Theta \in \mathcal{Y}_n$ be a Young tableau and let
%   $\mathfrak{C}_{\Theta}$ and $\mathfrak{R}_{\Theta}$ denote the
%   column- and row-word of $\Theta$ according to
%   Def.~\ref{def:TableauxWords} respectively. We will call a tableau
%   $\Theta$ \emph{lexically ordered}, if either $\mathfrak{C}_{\Theta}$
%   or $\mathfrak{R}_{\Theta}$ or both are in lexical order.  In
%   particular, we say that $\Theta$ is column-ordered
%   (resp. row-ordered), if $\mathfrak{C}_{\Theta}$
%   (resp. $\mathfrak{R}_{\Theta}$) is in lexical order.
% \end{definition}

% For example, the tableau
% \begin{equation}
%   \begin{ytableau}
%     1 & 3 & 5 \\
%     2 & 4
%   \end{ytableau}
% \end{equation}
% is (column-) ordered, but the tableau
% \begin{equation}
%   \begin{ytableau}
%     1 & 2 & 4 \\
%     3 & 5
%   \end{ytableau}
% \end{equation}
% is not ordered.

In Theorem~\ref{thm:LexicalConstruction} we will describe a
construction principle for the Hermitian Young projection operators
corresponding to lexically ordered tableaux. This will form a starting
point for the general construction principle of the Hermitian Young
projectors given in section~\ref{sec:MOLD}, as is evident from their
proofs in appendix~\ref{sec:Proofs}.  It is clear
that Keppeler and Sjödahl had noticed that the projectors associated with
ordered tableaux are special: In the appendix
of~\cite{Keppeler:2013yla} they discuss two examples of Hermitian
Young projection operators (which happen to correspond to lexically
ordered tableaux) constructed according to the KS-Theorem, and argue
that these operators can be simplified quite drastically. The
procedure leads eventually to the same expressions that emerge
directly from Theorem~\ref{thm:LexicalConstruction}. However, Keppeler
and Sjödahl do not establish the connection to the lexical order of
the Young tableau and do not even hint at a general construction
principle.

\begin{theorem}[lexical Hermitian Young projectors]
\label{thm:LexicalConstruction}
  Let $\Theta \in \mathcal{Y}_n$ be \emph{row-ordered}. Then, the
  corresponding Hermitian Young projection operator $P_{\Theta}$ is
  given by
\begin{subequations}
\label{eq:LexicalProjector}
\begin{equation}
  \label{eq:RowOrderedProjector}
P_{\Theta} = \alpha_{\Theta} \cdot \bar{Y}_{\Theta} \bar{Y}_{\Theta}^{\dagger}.
  \end{equation}
On the other hand, if $\Theta \in
\mathcal{Y}_n$ is a \emph{column-ordered} tableau, then the corresponding
Hermitian Young projection operator $P_{\Theta}$ is given by
  \begin{equation}
    \label{eq:ColumnOrderedProjector}
P_{\Theta} = \alpha_{\Theta} \cdot \bar{Y}_{\Theta}^{\dagger} \bar{Y}_{\Theta}.
  \end{equation}
\end{subequations}
\end{theorem}

\noindent The proof of this Theorem is deferred to Appendix
\ref{sec:ProofsLexical}. It is directly evident from
eqns.~\eqref{eq:RowOrderedProjector} and
~\eqref{eq:ColumnOrderedProjector} that $P_\Theta$ is Hermitian in
both cases. Since Hermitian conjugation in birdtrack notation amounts
to reflection about a vertical axis,  the formulae also guarantee that
Hermiticity is directly visible as a reflection symmetry of the
associated birdtrack diagrams. 

As an example, consider the Young tableau 
\begin{equation}
  \Theta =
  \begin{ytableau}
    1 & 2 \\
    3
  \end{ytableau}
\end{equation}
which has a lexically ordered row-word $\mathfrak{R}_{\Theta} =
(1,2,3)$. The associated Hermitian Young projection operator $P_{\Theta}$
according to the Lexical-Theorem~\ref{thm:LexicalConstruction} is given by
\begin{equation}
  P_{\Theta} = \underbrace{\frac{4}{3}}_{\alpha_{\Theta}} \cdot
  \underbrace{\FPic{3Sym12ASym13}}_{\bar{Y}_{\Theta}}
  \underbrace{\FPic{3ASym13Sym12}}_{\bar{Y}_{\Theta}^{\dagger}} =
  \frac{4}{3}\cdot
  \underbrace{\FPic{3Sym12ASym23Sym12}}_{=:\bar{P}_{\Theta}}
\ .
\end{equation}
The Hermiticity of this operator is prominently visible in its mirror symmetry.

\subsection{Young tableaux with partial lexical order}\label{sec:MOLD}

We will now give a construction principle for compact expressions of
Hermitian Young projection operators corresponding to general, not
necessarily lexically ordered, tableaux. The goal is to use what
\emph{partial} order there is to a diagram to obtain an optimized
iterative procedure. As a first step we need to be able to quantify how
``un-ordered'' a Young tableau is; we define a \emph{Measure Of
  Lexical Disorder}:

\begin{definition}[measure of lexical disorder (MOLD)]\label{MOLDDef}
  Let $\Theta \in \mathcal{Y}_n$ be a Young tableau. We define its
  \emph{Measure Of Lexical Disorder} (MOLD) to be the smallest natural number
  $\mathcal{M}(\Theta) \in \mathbb{N}$ such that
  \begin{equation}
    \Theta_{\left(\mathcal{M}(\Theta)\right)} = \pi^{\mathcal{M}(\Theta)} \left(\Theta\right)
  \end{equation}
is a lexically ordered tableau. (Recall from Definition~\ref{ParentMap} that
$\pi^{\mathcal{M}(\Theta)}$ refers to $\mathcal{M}(\Theta)$ consecutive
applications of the parent map $\pi$ to the tableau $\Theta$.)
\end{definition}

\noindent We note that the MOLD of a Young tableau is a well-defined quantity, since one
will always eventually arrive at a lexically ordered tableau, as, for
example, all tableaux in $\mathcal{Y}_3$ are lexically ordered. This
then implies that the MOLD of a tableau $\Theta \in
\mathcal{Y}_n$ has an upper bound,
\begin{equation}
\label{eq:MOLDUpperBound}
\mathcal{M}(\Theta) \leq n-3,
\end{equation}
making it a well-defined quantity. As an example, consider the tableau
\begin{equation}
\Phi :=
  \begin{ytableau}
    1 & 2 & 4 \\
    3 & 5
  \end{ytableau}
\ .
\end{equation}
The MOLD of the above tableau is $\mathcal{M}(\Phi)=2$, since two
applications of the parent map generate a lexically ordered tableau,
but just one application of $\pi$ on $\Phi$ would not be sufficient,
\begin{equation}
  \underbrace{\begin{ytableau}
    1 & 2 & *(cyan!20) 4 \\
    3 & *(cyan!40) 5
  \end{ytableau}}_{\Phi} \quad \xlongrightarrow[]{\pi} \quad 
  \underbrace{\begin{ytableau}
    1 & 2 & *(cyan!20) 4 \\
    3 
  \end{ytableau}}_{\Phi_{(1)}} \quad \xlongrightarrow[]{\pi} \quad 
  \underbrace{\begin{ytableau}
    1 & 2 \\
    3 
  \end{ytableau}}_{\Phi_{(2)}}
\ .
\end{equation}

\noindent We will now give the main Theorem of this section, the
construction principle of Hermitian Young projection operators
corresponding to Young tableaux $\Theta$, using the MOLD of the
latter. To do so, we distinguish four cases; the reason
why they have to be dealt with separately is given in the analysis
following the Theorem, section~\ref{sec:MOLDAnalysis}.

\begin{theorem}[MOLD operators]
\label{thm:MOLDConstruction}
  Consider a Young tableau $\Theta  \in
  \mathcal{Y}_n$ with MOLD $\mathcal{M}(\Theta)=m$. Furthermore, suppose that $\Theta_{(m)}$
  has a lexically ordered \emph{row-word}. Then, the Hermitian Young
  projection operator corresponding to $\Theta$, $P_{\Theta}$, is,
  \emph{for $m$ even},
\begin{subequations}
\label{eq:MOLDConstruction}
  \begin{equation}\label{eq:MOLDHermitianOperatorConstruction1}
    P_{\Theta} \; = \beta_{\Theta} \cdot\mathbf{S}_{\Theta_{(m)}}
     \; \mathbf{A}_{\Theta_{(m-1)}}  \; \mathbf{S}_{\Theta_{(m-2)}}  \; \ldots
     \; \mathbf{S}_{\Theta_{(2)}}  \; \mathbf{A}_{\Theta_{(1)}}
     \; \colorbox{red!20}{$\bar{Y}_{\Theta} \bar{Y}_{\Theta}^\dagger$}  \; \mathbf{A}_{\Theta_{(1)}}  \; \mathbf{S}_{\Theta_{(2)}}  \; \ldots
     \; \mathbf{S}_{\Theta_{(m-2)}}  \; \mathbf{A}_{\Theta_{(m-1)}}
     \; \mathbf{S}_{\Theta_{(m)}},
\end{equation}
and, \emph{for $m$ odd},
\begin{equation}\label{eq:MOLDHermitianOperatorConstruction2}
P_{\Theta} \; = \beta_{\Theta} \cdot\mathbf{S}_{\Theta_{(m)}}
     \; \mathbf{A}_{\Theta_{(m-1)}}  \; \mathbf{S}_{\Theta_{(m-2)}}  \; \ldots
     \; \mathbf{A}_{\Theta_{(2)}}  \; \mathbf{S}_{\Theta_{(1)}}  \; 
    \colorbox{red!20}{$\bar{Y}_{\Theta}^\dagger \bar{Y}_{\Theta}$}
     \; \mathbf{S}_{\Theta_{(1)}}  \; \mathbf{A}_{\Theta_{(2)}}  \;
     \ldots \; 
     \; \mathbf{S}_{\Theta_{(m-2)}} \; \mathbf{A}_{\Theta_{(m-1)}}
     \; \mathbf{S}_{\Theta_{(m)}}.
  \end{equation}
Similarly, if $\Theta_{(m)}$ has a lexically ordered \emph{column-word},
$P_{\Theta}$ is, \emph{for $m$ even},
 \begin{equation}\label{eq:MOLDHermitianOperatorConstruction3}
    P_{\Theta} \; = \beta_{\Theta} \cdot\mathbf{A}_{\Theta_{(m)}}
     \; \mathbf{S}_{\Theta_{(m-1)}}  \; \mathbf{A}_{\Theta_{(m-2)}}  \; \ldots
     \; \mathbf{A}_{\Theta_{(2)}}  \; \mathbf{S}_{\Theta_{(1)}}
     \; \colorbox{red!20}{$\bar{Y}_{\Theta}^{\dagger} \bar{Y}_{\Theta}$}  \; \mathbf{S}_{\Theta_{(1)}}  \; \mathbf{A}_{\Theta_{(2)}}  \; \ldots
     \; \mathbf{A}_{\Theta_{(m-2)}}  \; \mathbf{S}_{\Theta_{(m-1)}}
     \; \mathbf{A}_{\Theta_{(m)}},
\end{equation}
and, \emph{for $m$ odd},
\begin{equation}\label{eq:MOLDHermitianOperatorConstruction4}
P_{\Theta} \; = \beta_{\Theta} \cdot\mathbf{A}_{\Theta_{(m)}} \; 
    \mathbf{S}_{\Theta_{(m-1)}}  \; \mathbf{A}_{\Theta_{(m-2)}}  \; \ldots
     \; \mathbf{S}_{\Theta_{(2)}}  \; \mathbf{A}_{\Theta_{(1)}} \; 
    \colorbox{red!20}{$\bar{Y}_{\Theta} \bar{Y}_{\Theta}^\dagger$}
     \; \mathbf{A}_{\Theta_{(1)}} \; \mathbf{S}_{\Theta_{(2)}}  \; \ldots \; 
     \; \mathbf{A}_{\Theta_{(m-2)}}  \; \mathbf{S}_{\Theta_{(m-1)}}
     \; \mathbf{A}_{\Theta_{(m)}}.
  \end{equation}
\end{subequations}
In the above, all symmetrizers and antisymmetrizers are understood to
be canonically embedded into $V^{\otimes n}$; $\beta_{\Theta}$ is a
\emph{non-zero constant} chosen such that $P_{\Theta}$ is idempotent.
\end{theorem}

\noindent The formal proof of this Theorem can be found in Appendix
\ref{sec:ProofsMOLD}. A comparative example of a Hermitian Young
projection operator constructed using MOLD and KS is given is
section~\ref{sec:MOLDAdvantage}, Fig~\ref{fig:MOLDAdvantage}. 

It should be noted that we have not provided an explicit expression
for the constant $\beta_{\Theta}$ in
Theorem~\ref{thm:MOLDConstruction}. This normalization-constant
however can easily be found for specific operators by direct
calculation since the MOLD-operators are very much suited for
automated calculations on a computer, as is described in
section~\ref{sec:MOLDAdvantage}. We would like to draw the reader's
attention to the fact that the symmetrizers and antisymmetrizers in
all four expressions of Theorem~\ref{thm:MOLDConstruction}
\emph{strictly alternate}, including those inside the Young
projectors.

As an example, consider the Young tableau
\begin{equation}
  \Theta :=
  \begin{ytableau}
    1 & 2 & 4 \\
    3 & 5
  \end{ytableau}.
\end{equation}
This tableau has MOLD $2$ (i.e. even MOLD), and $\Theta_{(2)}$ has a
lexically ordered row-word. Thus, we have to construct the Hermitian
Young projection operator $P_{\Theta}$ corresponding to $\Theta$
according to equation
\eqref{eq:MOLDHermitianOperatorConstruction1}. $P_{\Theta}$ is therefore
given by
\begin{align}
  P_{\Theta} & = \beta_{\Theta} \cdot \mathbf{S}_{\Theta_{(2)}} \;
  \mathbf{A}_{\Theta_{(1)}} \; \colorbox{red!20}{$\mathbf{S}_{\Theta} \;
  \mathbf{A}_{\Theta} \; \mathbf{S}_{\Theta}$} \;
  \mathbf{A}_{\Theta_{(1)}} \; \mathbf{S}_{\Theta_{(2)}} \nonumber \\
& = \beta_{\Theta} \cdot
\scalebox{0.75}{$\underbrace{\FPic{5Sym12SN}}_{\mbox{\normalsize
      $\mathbf{S}_{\Theta_{(2)}}$}}\underbrace{\FPic{5s23N}\FPic{5ASym12N}\FPic{5s23N}}_{\mbox{\normalsize
      $\mathbf{A}_{\Theta_{(1)}}$}}\colorbox{red!20}{$\underbrace{\FPic{5s34N}\FPic{5Sym123Sym45N}\FPic{5s34N}}_{\mbox{\normalsize
    $\mathbf{S}_{\Theta}$}}\underbrace{\FPic{5s23s45N}\FPic{5ASym12ASym34N}\FPic{5s23s45N}}_{\mbox{\normalsize
  $\mathbf{A}_{\Theta}$}}\underbrace{\FPic{5s34N}\FPic{5Sym123Sym45N}\FPic{5s34N}}_{\mbox{\normalsize
$\mathbf{S}_{\Theta}$}}$}\underbrace{\FPic{5s23N}\FPic{5ASym12N}\FPic{5s23N}}_{\mbox{\normalsize
$\mathbf{A}_{\Theta_{(1)}}$}}\underbrace{\FPic{5Sym12SN}}_{\mbox{\normalsize
$\mathbf{S}_{\Theta_{(2)}}$}}$}
\nonumber \\
& = \beta_{\Theta} \cdot  \scalebox{0.75}{\FPic{5Sym12N}\FPic{5s23N}\FPic{5ASym12N}\FPic{5s234N}\FPic{5Sym123Sym45N}\FPic{5s2453N}\FPic{5ASym12ASym34N}\FPic{5s2354N}\FPic{5Sym123Sym45N}\FPic{5s243N}\FPic{5ASym12N}\FPic{5s23N}\FPic{5Sym12N}}.
\end{align}
A direct calculation in \emph{Mathematica} reveals that
$\beta_{\Theta}\stackrel{!}{=}4$ for $P_{\Theta}$ to be idempotent.

\subsubsection{A Short Analysis of the MOLD-Theorem~\ref{thm:MOLDConstruction}}\label{sec:MOLDAnalysis}

We now pause for a moment to look at the four cases presented in
Theorem~\ref{thm:MOLDConstruction}  in
more detail and emphasize their differences. We hope to convey an intuitive
feel as to why the corresponding operators are constructed the way
they are.

First, let us look at the first two operators
\eqref{eq:MOLDHermitianOperatorConstruction1} and
\eqref{eq:MOLDHermitianOperatorConstruction2}. 
Both these operators $P_{\Theta}$ have a symmetrizer on the outside, namely $\mathbf{S}_{\Theta_{(m)}}$,
opposed to the operators~\eqref{eq:MOLDHermitianOperatorConstruction3} and
\eqref{eq:MOLDHermitianOperatorConstruction4} which have an
antisymmetrizer on the outside. This stems from the iterative
construction of Hermitian Young projection operators given by the
KS-Theorem~\ref{thm:KSProjectors}: By the Lexical-Theorem~\ref{thm:LexicalConstruction} we know that
$P_{\Theta_{(m)}}$ is given by
\begin{equation}
  \label{eq:MOLDHermAnalysis1}
  P_{\Theta_{(m)}} = \alpha_{\Theta} \cdot \bar{Y}_{\Theta_{(m)}} \bar{Y}_{\Theta_{(m)}}^{\dagger}
  = \alpha_{\Theta} \cdot  \mathbf{S}_{\Theta_{(m)}} \mathbf{A}_{\Theta_{(m)}} \mathbf{S}_{\Theta_{(m)}},
\end{equation}
since $P_{\Theta_{(m)}}$ is assumed to correspond to a row-ordered
Young tableau. When we thus construct $P_{\Theta}$ recursively
according to KS, \cite{Keppeler:2013yla}, we find that
\begin{equation}
\label{eq:MOLDHermAnalysis2}
  P_{\Theta} = P_{\Theta_{(m)}} \ldots Y_{\Theta} \ldots
  P_{\Theta_{(m)}} \ \propto \; \mathbf{S}_{\Theta_{(m)}} \mathbf{A}_{\Theta_{(m)}}
  \mathbf{S}_{\Theta_{(m)}} \ldots Y_{\Theta} \ldots
  \mathbf{S}_{\Theta_{(m)}} \mathbf{A}_{\Theta_{(m)}}
  \mathbf{S}_{\Theta_{(m)}}. 
\end{equation}
Thus, we expect there
to be symmetrizers on the outside of the operators $P_{\Theta}$ in
expressions~\eqref{eq:MOLDHermitianOperatorConstruction1} and
\eqref{eq:MOLDHermitianOperatorConstruction2}. Following a similar
logic, we expect there to be antisymmetrizers on the outside of
operators~\eqref{eq:MOLDHermitianOperatorConstruction3} and
\eqref{eq:MOLDHermitianOperatorConstruction4}.

Lastly, we discuss the importance of the distinction between even and
odd $m$ in the MOLD-Theorem~\ref{thm:MOLDConstruction}. In the
construction of \emph{all} $P_{\Theta}$ in the Theorem, we find that they
consist of products of alternating symmetrizers and antisymmetrizers
to more and more recent generations of $\Theta$ as we move further to
the center of $P_{\Theta}$. If the operator $P_{\Theta}$ thus starts
with $\mathbf{S}_{(m)}$ on the outside, as it does in
equations~\eqref{eq:MOLDHermitianOperatorConstruction1} and
\eqref{eq:MOLDHermitianOperatorConstruction2}, and the product has
alternating sets of symmetrizers and antisymmetrizers each going up
one generation, then the parity of $m$ will decide whether the set
corresponding to the tableau $\Theta_{(1)}$ in the product
$P_{\Theta}$ is a set of symmetrizers or antisymmetrizers. Thus, the
central three sets of symmetrizers and antisymmetrizers in the
product $P_{\Theta}$ will then either be
\begin{equation}
  \label{eq:MOLDHermAnalysis3}
  \mathbf{A}_{\Theta}  \; \mathbf{S}_{\Theta} \; \mathbf{A}_{\Theta}
  \; = \; \bar{Y}_{\Theta}^{\dagger}\bar{Y}_{\Theta} \qquad
  \text{or} \qquad \mathbf{S}_{\Theta}  \; \mathbf{A}_{\Theta} \;
  \mathbf{S}_{\Theta} \; = \;
  \bar{Y}_{\Theta}\bar{Y}_{\Theta}^{\dagger},
\end{equation}
dependent on the nature of the sets corresponding to $\Theta_{(1)}$,
but keeping the alternating structure of symmetrizers and antisymmetrizers.

The fact that the central sets of $P_{\Theta}$ in all four equations
of the above Theorem~\ref{thm:MOLDConstruction} are either product of
\eqref{eq:MOLDHermAnalysis3} opposed to simply $Y_{\Theta}$ or
$Y_{\Theta}^{\dagger}$ can be attributed to the fact the $P_{\Theta}$
is Hermitian and we would like its Hermiticity to be visually
explicit.

\subsection{The advantage of using MOLD}\label{sec:MOLDAdvantage}

The practical advantages of our construction opposed to the KS-Theorem
\ref{thm:KSProjectors} are striking. To illustrate this we return to
the same example used in~\cite{Alcock-Zeilinger:2016bss}
to demonstrate the effectiveness of the simplification rules derived
there. The Young tableau
\begin{equation}
  \label{eq:MOLDAdvantageEx1}
\Phi :=
\begin{ytableau}
  1 & 2 & 4 & 7 \\
  3 & 6 \\
  5 & 8 \\
  9
\end{ytableau}
\end{equation}
leads to an expression of the corresponding KS-projector with 127 symmetrizer- and antisymmetrizer-sets
which reduce to an object with only 13 such sets after applying the
cancellation and propagation rules
of~\cite{Alcock-Zeilinger:2016bss}. 

Both construction principles (KS and MOLD) are iterative in the sense
that they both require knowledge about the ancestor tableaux of a
tableau $\Theta \in \mathcal{Y}_n$. For the construction of KS as it
was originally described in \cite{Keppeler:2013yla}, one needs all
ancestor operators of $\Theta$ up until $\Theta_{(n-2)}$, while the
MOLD construction merely uses the ancestor tableaux up until
$\Theta_{\mathcal{M}(\Theta)}$, which is at most $\Theta_{(n-3)}$
according to~\eqref{eq:MOLDUpperBound}. This one tableau difference
does not seem excessive at first glance, but one should keep in mind
that the difference is \emph{at least} one tableau, often
more. However the bulk of the computing power used to generate
$\bar{P}_{\Phi}^{KS}$ comes from the fact that, in addition to the
ancestor tableaux of $\Theta$, one further requires information about
the explicit form of the ancestor Hermitian Young projectors
$P_{\Theta}$ all the way up to $P_{\Theta_{(2)}}$, which have to be
calculated separately. The MOLD-construction merely uses the Young
sets of symmetrizers or antisymmetrizers ($\mathbf{S}$ and
$\mathbf{A}$ respectively) of the ancestor tableaux of $\Theta$ up to
$\Theta_{\mathcal{M}(\Theta)}$, which can be immediately read off the
tableaux and thus needs minimal computing power.

Using the MOLD-Theorem~\ref{thm:MOLDConstruction}, one arrives at the
shorter version \emph{directly}, after a considerably shorter
recursive path and without the need for additional
simplifications. One bypasses a long repetitive list of steps
altogether!
\begin{figure}[h]\newlength\foo
  \begin{center}
\settototalheight\foo{\resizebox{\textwidth}{!}{  \diagram[height=.15cm]{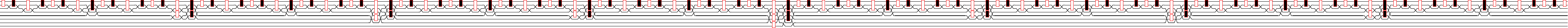}
}}
\resizebox{\textwidth}{!}{  \diagram[height=.15cm]{MOLDAdvantageEx2}
}
    \\
    \diagram[height=\foo]{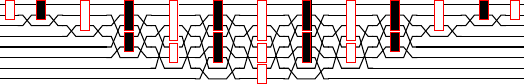}
  \end{center}
  \caption{For a size comparison, this figure shows
    $\bar{P}_{\Phi}^{\text{KS}}$ (top) and
    $\bar{P}_{\Phi}^{\text{MOLD}}$ (bottom) for the tableau $\Phi$ as
    defined in~\eqref{eq:MOLDAdvantageEx1} using two different
    constructions: The top operator was constructed using the
    KS-algorithm, while the bottom operator was constructed using
    MOLD. Both operators and the associated graphics were generated in
    \emph{Mathematica}.}
\label{fig:MOLDAdvantage}
\end{figure} 
The $127/13$ length ratio between $\bar{P}_{\Phi}^{\text{KS}}$ and
$\bar{P}_{\Phi}^{\text{MOLD}}$\footnote{It is important to note that
  $\bar{P}_{\Phi}^{KS}$ and $\bar{P}_{\Phi}^{MOLD}$ both differ from
  $P_{\Phi}$ by a constant, but this constant will depend on the
  construction principle used to find $\bar{P}_{\Phi}$. In that sense,
\begin{equation}
\beta_{\Phi}^{KS} \cdot \bar{P}_{\Phi}^{KS} =
    P_{\Phi} = \beta_{\Phi}^{MOLD} \cdot \bar{P}_{\Phi}^{MOLD}
\end{equation}
but $\beta_{\Phi}^{\text{KS}}\neq\beta_{\Phi}^{\text{MOLD}}$ in
general.} is strikingly apparent in Fig.~\ref{fig:MOLDAdvantage}. This
makes the MOLD algorithm a lot more practical to work with
analytically. The algorithm really comes into its own when used in
symbolic algebra programs: the MOLD construction allows us to
efficiently create projection operators for considerably larger Young
tableaux than the iterative KS equivalent. In particular, for the
example in Figure~\ref{fig:MOLDAdvantage}, the fact that the MOLD
algorithm simply avoids a long series of steps makes it over $18600$
times faster than its KS counterpart: It generates its result in
approximately $0.0038$ seconds, while KS takes approximately $71$
seconds (not even taking into account the cost of the simplification
steps to arrive at the final result) on a modern laptop.

Unlike $P_{\Theta}^{\text{KS}}$, $P_{\Theta}^{\text{MOLD}}$ is
\emph{obviously and visibly} Hermitian by
construction.\footnote{$P_{\Theta}^{KS}$ do not exhibit their
  Hermiticity directly since the center-piece of the $KS$ operator is
  the Young projection operator $\bar{Y}_{\Theta}$, which is
  inherently non-Hermitian. We need to rely on the proof
  given in~\cite{Keppeler:2013yla} to be assured of their Hermiticity.}
Given that birdtracks are meant to be a tool that makes dealing with
these operators \emph{visually} clear, this is an obvious advantage.

We have given a construction principle for compact expressions of
Hermitian Young projection operators, the MOLD-operators, in section
\ref{sec:MOLD}, and we have now seen that the MOLD-operators are
indeed more useful for practical calculations. We have thus achieved
Goal~\ref{thm:MOLDOpsGoal} of this paper.

\section{Conclusion \& outlook}

The representation theory of $\SUN$ is an old theory with many
successful applications in physics. Yet some of the tools remain
awkward and only applicable in specific situations, like the general
theory of angular momentum or the construction of Young projection
operators that lack Hermiticity. Newer tools like the birdtrack
formalism remain only partially connected with these time honored
results. We have a very specific interest in applications to QCD in
the JIMWLK context, in jet physics, in energy loss and generalized
parton distributions, so we have aimed at creating a set of tools that
we know will aid in these applications and, in the process have
pointed out where the existing tools fall short of our needs.
\begin{enumerate}
\item We have found that projection operators built on Young tableaux
  are uniquely suited to calculations that keep $N$ as a parameter.

  To simply list the irreducible multiplets in $\Pow{m}$, Young's
  procedure forms descendant tableaux of those representing the
  irreducibles contained in $\Pow{(m-1)}$, portraying an iterative
  procedure of ``adding a particle'' in each step.

  To parallel this in terms of projection operators and the associated
  subspaces (i.e. to implement eq.~\eqref{eq:SpanSubspaces2} which
  represents the general case of the summation relations collected in
  Fig.~\ref{fig:hierarchy-4}) -- as one needs to do to actually perform
  calculations in physics applications -- we have established that one
  needs Hermitian versions of these projection operators as
  those constructed earlier by Keppeler and
  Sjödahl~\cite{Keppeler:2013yla}. This sets the backdrop for the
  remaining developments.

\item Having motivated the necessity for Hermitian Young projection
  operators, we are faced with the fact that the KS algorithm quickly
  becomes unwieldy -- the iterative procedures are computationally
  expensive and produce long expressions
  (Fig.~\ref{fig:MOLDAdvantage}). We have earlier presented
  simplification rules~\cite{Alcock-Zeilinger:2016bss} to
  distill these down to more compact expressions, but that does not
  alleviate the computational cost.

  To address this issue, we have provided a new algorithm based on the
  Measure Of Lexical Disorder (MOLD) of a tableau in
  sec.~\ref{sec:MOLD} that drastically reduces the calculational
  footprint of the procedure compared to the KS method. It should be
  noted that the normalized operators constructed using the MOLD algorithm are
  equal to the KS-operators (as is evident from the proof
  given in appendix~\ref{sec:Proofs}) and thus inherit all properties
  of the KS-operators: idempotency, mutual orthogonality and
  completeness for all values of $m$.

  The MOLD algorithm almost completely incorporates the
  simplifications of~\cite{Alcock-Zeilinger:2016bss} at vastly reduced
  calculational cost (only isolated cases of MOLD-operators still
  allow for further simplification with the tools presented
  in~\cite{Alcock-Zeilinger:2016bss}). All the algorithms are
  eminently suited for implementation in symbolic algebra programs:
  all our explorations and examples have been generated in
  \emph{Mathematica}.

  In particular, the operators shown in Fig.~\ref{fig:MOLDAdvantage}
  were generated in \emph{Mathematica}: the operator on the top was
  constructed using the KS-Theorem, while the operator on the bottom
  resulted from the MOLD-construction. The automated calculation was
  significantly improved with the MOLD-algorithm, as the MOLD-operator
  in Fig.~\ref{fig:MOLDAdvantage} was obtained approximately $18600$
  times faster than the KS-equivalent on a modern laptop -- an
  improvement of $4$ orders of magnitude.
  
\end{enumerate}

Our own list of applications for the tools and insights presented in
this paper are QCD centric: Global singlet state projections of
Wilson-line operators that appear in a myriad of applications due to
factorization of hard and soft contributions help analyzing the
physics content in all of them; this will be explored further
in~\cite{AlcockZeilinger2016Singlets}. We hope that our presentation
is suitable to unify perspectives provided by the various approaches
to representation theory of $\SUN$ and that the results prove useful
beyond these immediate applications.

\paragraph{Acknowledgements:} H.W. is supported by South Africa's
National Research Foundation under CPRR grant nr 90509. J.A-Z. was
supported (in sequence) by the postgraduate funding office of the
University of Cape Town (2014), the National Research Foundation
(2015) and the Science Faculty PhD Fellowship of the University of
Cape Town (2016).

\appendix

\section{Littlewood-Young projection operators}\label{sec:littl-young-proj}

For completeness, this appendix summarizes the construction principle
of generalized Young projection operators due to
Littlewood~\cite[sec.~5.3]{Littlewood:1950}, which corrects for the
failure of Young projection operators over $\Pow{n}$ to be orthogonal
or complete beyond $n=4$. We shall call these the Littlewood-Young
(LY) projectors and denote them by $L$.

Before we give the construction principle of the LY-projectors, let us have
a closer look at products of Young projection operators and criteria which make
these products orthogonal. Consider two Young tableaux $\Theta$ and
$\Phi$ in $\mathcal{Y}_n$, then the product of their corresponding
Young projection operators is given by
\begin{equation}
  \label{eq:product-2Young-Ops}
Y_{\Theta} Y_{\Phi}
=
\alpha_{\Theta}\alpha_{\Phi} \cdot
\mathbf{S}_{\Theta}
\mathbf{A}_{\Theta}
\mathbf{S}_{\Phi}
\mathbf{A}_{\Phi}
\ .
\end{equation}
Clearly, if the product $\mathbf{A}_{\Theta}\mathbf{S}_{\Phi}$ vanishes, then so does
the product of the Young projectors, but if
$\mathbf{A}_{\Theta}\mathbf{S}_{\Phi}\neq0$, we have to conclude that
$Y_{\Theta}Y_{\Phi}\neq0$,
\begin{equation}
  \label{eq:product-2Young-Ops-nonzero}\begin{tikzpicture}[baseline=(current bounding box.west),
  every node/.style={inner sep=1pt,outer sep=-1pt}% ,
    % skip loop/.style={to path={-- ++(0,#1) -| (\tikztotarget)}}
    ]
    \matrix(ID)[
    matrix of math nodes,
    ampersand replacement=\&,
    row sep =0mm,
    column sep =0mm
    ]
    {Y_{\Theta} Y_{\Phi} = \alpha_{\Theta}\alpha_{\Phi} \cdot
      \& \mathbf{S}_{\Theta}
      \& \mathbf{A}_{\Theta}
      \& \mathbf{S}_{\Phi}
      \& \mathbf{A}_{\Phi}
      \& \; \Longrightarrow \;
      \& Y_{\Theta} Y_{\Phi} \neq 0
      \ .
      \\
};
% Braces %
\draw[decorate,decoration={brace,amplitude=4pt},thick] (ID-1-3.south east) --
(ID-1-2.south west) node[pos=.5,anchor=north,yshift=-2mm]
{\scriptsize$\neq 0$};
\draw[decorate,decoration={brace,amplitude=4pt},thick] (ID-1-5.south east) --
(ID-1-4.south west) node[pos=.5,anchor=north,yshift=-2mm]
{\scriptsize$\neq 0$};
\draw[decorate,decoration={brace,amplitude=4pt},thick] (ID-1-3.north west) --
(ID-1-4.north east) node[pos=.5,anchor=north,yshift=4mm]
{\scriptsize$\neq 0$};
\end{tikzpicture} 
\end{equation}
\ytableausetup{mathmode,boxsize=1.2em}%
For $\mathbf{A}_{\Theta}\mathbf{S}_{\Phi}$ to vanish, we merely
require a particular antisymmetrizer $\bm{A}_i\in\mathbf{A}_{\Theta}$
two have more than one leg in common with a symmetrizer
$\bm{S}_j\in\mathbf{S}_{\Phi}$.

If the tableaux $\Theta$ and $\Phi$ have different shapes, then this
is trivially given since there must exist at least one pair of boxes
$(\ybox{k},\ybox{l})$ that appear in the same column in $\Theta$ and
in the same row in $\Phi$ and vice versa, such
that~\cite[sec.~5.3,~Thm.~III]{Littlewood:1950}
\begin{equation}
  \label{eq:SAeq0-tableaux-diff-shape}
\mathbf{A}_{\Theta}
\mathbf{S}_{\Phi}
= 0 =
\mathbf{A}_{\Phi}
\mathbf{S}_{\Theta}
\qquad
\text{where $\Theta$ and $\Phi$ have different shapes}
\ .
\end{equation}

If the two tableaux $\Theta$ and $\Phi$ have the same shape, one must
work harder to see where this criterion applies to force
$\mathbf{A}_{\Theta}\mathbf{S}_{\Phi} = 0$. To see when a pair of
boxes $(\ybox{k},\ybox{l})$ appears in the same column in $\Theta$ and
in the same row in $\Phi$ for tableaux of the same shape, we need to
introduce an order relation between tableaux of the same shape using
their row-words (\emph{c.f.}  Definition~\ref{def:TableauxWords}):
\ytableausetup{mathmode,boxsize=normal}%

% Consider a particular Young tableau $\Theta$. We define the partition $p(\Theta)$
% corresponding to $\Theta$ to be the vector
% $(\lambda_1,\lambda_2,\ldots\lambda_k)$ where $\lambda_i$ gives the
% number of boxes in the $i^{th}$ row of $\Theta$,\footnote{In this
%   paper, we truncate the vector $p(\Theta)$ after the last row of
%   $\Theta$. In the literature, $p(\Theta)$ is often defined to be a
%   vector of length $m$ (for $\Theta\in\mathcal{Y}_m$) where each entry
%   corresponding to a row after the last row of $\Theta$ is $0$.}
% for example,
% \begin{equation}
% \Theta = 
%   \begin{ytableau}
%     1 & 3 & 5 & 6 \\
%     2 & 7 \\
%     4
%   \end{ytableau}
% \quad
% \text{has partition}
% \quad
% p(\Theta) = (4,2,1)
% \ .
% \end{equation}
% We may use the partition of a tableau to define an ordering between
% two tableaux $\Theta$ and $\Phi$ in $\mathcal{Y}_m$: We say that
% $\Theta$ precedes $\Phi$
% We further define an ordering between tableaux of a different shape: Le

\begin{definition}[order relation amongst tableaux of the same
  shape]\label{def:OrderRelationTableaux} \qquad
  Let $\Theta$ and $\Phi$ be two Young tableaux and let $\theta_{ij}$
  be the entry in the $i^{th}$ row and $j^{th}$ column of $\Theta$,
  and similarly for $\phi_{ij}$. Their corresponding row-words are
  given by
  $\mathfrak{R}_{\Theta}=(\theta_{11},\theta_{12},\ldots,\theta_{21},\ldots)$
  and
  $\mathfrak{R}_{\Phi}=(\phi_{11},\phi_{12},\ldots,\phi_{21},\ldots)$
  respectively. We say that $\Theta$ precedes $\Phi$ and write
  $\Theta\prec\Phi$ if $\theta_{ij}<\phi_{ij}$ for the leftmost $ij$
  where $\theta_{\ij}\in\mathfrak{R}_{\Theta}$ and
  $\phi_{ij}\in\mathfrak{R}_{\Phi}$ differ.\footnote{In words, we
    order a set of tableaux according to the relative lexical order of
  their associated row words. This concept is not to be confused with
  the lexical order \emph{within} a tableau introduced in Defintion~\ref{def:TableauxWords}.}
\end{definition}
\ytableausetup{mathmode,boxsize=0.8em}%
For example, the Young tableaux of shape $\ydiagram{3,2}$ can be
ordered as
\ytableausetup{mathmode,boxsize=normal}%
\begin{equation}
  \label{eq:32-block-tableau-ordering}
\underbrace{
  \begin{ytableau}
    1 & 2 & 3 \\
    4 & 5
  \end{ytableau}
}_{\mathfrak{R}_{\Theta}=(1,2,3,4,5)}
\quad \prec \quad
\underbrace{
  \begin{ytableau}
    1 & 2 & 4 \\
    3 & 5
  \end{ytableau}
}_{\mathfrak{R}_{\Theta}=(1,2,4,3,5)}
\quad \prec \quad
\underbrace{
  \begin{ytableau}
    1 & 2 & 5 \\
    3 & 4
  \end{ytableau}
}_{\mathfrak{R}_{\Theta}=(1,2,5,3,4)}
\quad \prec \quad
\underbrace{
  \begin{ytableau}
    1 & 3 & 4 \\
    2 & 5
  \end{ytableau}
}_{\mathfrak{R}_{\Theta}=(1,3,4,2,5)}
\quad \prec \quad
\underbrace{
  \begin{ytableau}
    1 & 3 & 5 \\
    2 & 4
  \end{ytableau}
}_{\mathfrak{R}_{\Theta}=(1,3,5,2,4)}
\ .
\end{equation}

It turns out that this order relation defines exactly when a pair of
boxes $(\ybox{k},\ybox{l})$ appearing in the same column of a tableau
$\Theta$ appear in the same row of a tableau $\Phi$ forcing the
product $\mathbf{A}_{\Theta}\mathbf{S}_{\Phi}$
to vanish~\cite[sec.~5.3,~Thm.~V]{Littlewood:1950}. We repeat the proof
given by~\cite{Littlewood:1950}: Let $\Theta$ and $\Phi$
be two Young tableaux of the same shape and let
$\Theta\prec\Phi$. Then, the first entry
$\theta_{ij}\in\mathfrak{R}_{\Theta}$ distinct from
$\phi_{ij}\in\mathfrak{R}_{\Phi}$ satisfies
$\theta_{ij}<\phi_{ij}$. Thus, the entry
$\phi_{kl}\in\mathfrak{R}_{\Phi}$ such that $\theta_{ij}=\phi_{kl}$
must appear in a row later than $i$, but, by definition of Young
tableaux, must be in a column before $j$,
\begin{equation}
  \theta_{ij} = \phi_{kl}
\qquad
\text{with $l<j$ and $k>i$}
\ .
\end{equation}
Since $l<j$, the entries $\theta_{il}$ and $\phi_{il}$ must be equal
(we assumed that the entries $\theta_{ij}$ and $\phi_{ij}$ were the
\emph{first} distinct entries appearing in the respective row-words),
$\theta_{il}=\phi_{il}$. Thus, the pair of entries
$(\theta_{ij}=\phi_{kl},\theta_{il}=\phi_{il})$ appears in the same
row in $\Theta$ (the $i^{th}$ row) and in the same column in $\Phi$
(the $l^{th}$ column), yielding
$\mathbf{A}_{\Phi}\mathbf{S}_{\Theta}=0$. Thus, we can say that
\begin{equation}
\label{eq:Y-Ops-same-shape-ON}
  Y_{\Phi} Y_{\Theta} = 0
\qquad 
\text{if $\Theta\prec\Phi$}
\ .
\end{equation}
It should be noted that the reverse not necessarily holds, that is 
\begin{equation}
\label{eq:Y-Ops-same-shape-notON}
  \Theta \prec \Phi
  \quad \notimplies \quad
  Y_{\Theta} Y_{\Phi} = 0
\ .
\end{equation}
Littlewood uses this one-sided orthogonality of Young projection
operators to create mutually orthogonal ones: 

Let $\lbrace\Theta_1,\Theta_2,\ldots\Theta_k\rbrace$ be the set of all
Young tableaux in $\mathcal{Y}_n$ with a particular shape and let them
be ordered such that $\Theta_i\prec\Theta_j$ whenever $i<j$. From
eq.~\eqref{eq:Y-Ops-same-shape-ON}, we have that
\begin{equation}
  \label{eq:YOps-orthogonal-larger}
Y_{\Theta_j} Y_{\Theta_i} = 0
\qquad
\text{whenever $i<j$} 
\ .
\end{equation}
The Littlewood-Young projection operators $L_{\Theta_i}$ corresponding to the
tableaux $\Theta_i\in\lbrace\Theta_1,\Theta_2,\ldots\Theta_k\rbrace\subset\mathcal{Y}_n$ are defined as
\begin{align}
  L_{\Theta_1} & = Y_{\Theta_1} \nonumber \\
  L_{\Theta_2} & = (1 - L_{\Theta_1}) Y_{\Theta_2} \nonumber \\
  \vdots & \nonumber \\
  L_{\Theta_k} & = (1 - L_{\Theta_1} - L_{\Theta_2} - \ldots -
                 L_{\Theta_{k-1}}) Y_{\Theta_k}
\label{eq:LYOps-construction}
\ .
\end{align}
\ytableausetup{mathmode, boxsize=0.7em}%

It should be noticed that the LY-operators corresponding to tableaux
in $\mathcal{Y}_n$ for $n\leq4$ reduce to the Young projectors, as in
these instances the Young projectors are mutually orthogonal. Even for
$n>5$, many LY-projectors reduce to the regular Young projectors, or
at least simplify drastically since most of the Young projectors
remain orthogonal even for large $n$. As an example, at $n=5$ the only
two Littlewood-Young projection operators that differ from their Young
counterpart are given by
\begin{equation}
L_{
\begin{ytableau} 
  \scriptstyle 1 & \scriptstyle 3 & \scriptstyle 5 \\ 
  \scriptstyle 2 & \scriptstyle 4 
\end{ytableau}} 
\; = \;
\left( 
1 - 
Y_{
\begin{ytableau} 
  \scriptstyle 1 & \scriptstyle 2 & \scriptstyle 3 \\ 
  \scriptstyle 4 & \scriptstyle 5 
\end{ytableau}} 
\right)
Y_{
\begin{ytableau} 
  \scriptstyle 1 & \scriptstyle 3 & \scriptstyle 5 \\ 
  \scriptstyle 2 & \scriptstyle 4 
\end{ytableau}} 
\qquad \text{and} \qquad
L_{
\begin{ytableau} 
  \scriptstyle 1 & \scriptstyle 4 \\ 
  \scriptstyle 2 & \scriptstyle 5 \\
  \scriptstyle 3
\end{ytableau}}
\; = \;
\left( 
1 - 
Y_{
\begin{ytableau} 
  \scriptstyle 1 & \scriptstyle 2 \\ 
  \scriptstyle 3 & \scriptstyle 4 \\
  \scriptstyle 5
\end{ytableau}}
\right)
Y_{
\begin{ytableau} 
  \scriptstyle 1 & \scriptstyle 4 \\ 
  \scriptstyle 2 & \scriptstyle 5 \\
  \scriptstyle 3
\end{ytableau}}
\ .
\end{equation}
\ytableausetup{mathmode, boxsize=normal}%

The LY-operators constructed according
to~\eqref{eq:LYOps-construction} are orthogonal for all values of $n$: If
two operators $L_{\Theta}$ and $L_{\Phi}$ correspond to tableaux of
different shapes, then $L_{\Theta}$ and $L_{\Phi}$ are orthogonal
since their respective constituent Young projection operators are
orthogonal. If $L_{\Theta_i}$ and $L_{\Theta_j}$ correspond to
tableaux of the same shape with $\Theta_i\prec\Theta_j$ ($i<j$), then
\begin{equation}
\label{eq:LjLi-ON}
  L_{\Theta_j} L_{\Theta_i} = 0
\qquad
\text{from eq.~\eqref{eq:YOps-orthogonal-larger}}
\end{equation}
To show the reverse ($L_{\Theta_i}L_{\Theta_j}=0$), it is highly
advantageous to note that  $L_{\Theta_j}$ as given
in~\eqref{eq:LYOps-construction} can be recast as
\begin{equation}
\label{eq:structural-form-LTheta}
  L_{\Theta_j} = 
(1 - Y_{\Theta_1})
(1 - Y_{\Theta_2})
\cdots
(1 - Y_{\Theta_{j-1}})
Y_{\Theta_j}
\ .
\end{equation}
This can be shown by induction: For $j=1,2$
eq.~\eqref{eq:structural-form-LTheta} trivially holds, we therefore go
through the case $j=3$ explicitly:
\begin{equation}
L_{\Theta_3} = 
\Big( 1 - 
\underbrace{
Y_{\Theta_1}
}_{=L_{\Theta_1}} -
\underbrace{
(1-Y_{\Theta_1}) Y_{\Theta_2}
}_{=L_{\Theta_2}}
\Big)
Y_{\Theta_3}
=
\big(
(1 - Y_{\Theta_1}) -
(1 - Y_{\Theta_1}) Y_{\Theta_2}
\big)
Y_{\Theta_3}
= (1 - Y_{\Theta_1}) (1 - Y_{\Theta_2}) Y_{\Theta_3}
\ .
\end{equation}
Suppose eq.~\eqref{eq:structural-form-LTheta} holds for all
$L_{\Theta_k}$ up some integer $k=j-1$. Then, by
eq.~\eqref{eq:LYOps-construction}, $L_{\Theta_j}$ is given by
\begin{align}
L_{\Theta_j} \; & =
\Big(
\underbrace{
1 - Y_{\Theta_1}
}_{= (1-Y_{\Theta_1})}
-
(1 - Y_{\Theta_1}) Y_{\Theta_2}
-
\ldots
-
(1 - Y_{\Theta_1})(1 - Y_{\Theta_2})\cdots(1 - Y_{\Theta_{j-2}})Y_{\Theta_{j-1}}
\Big)
Y_{\Theta_j} 
\nonumber \\
& =
(1 - Y_{\Theta_1})
\Big(
\underbrace{
1 - Y_{\Theta_2}
}_{= (1-Y_{\Theta_2})}
-
(1 - Y_{\Theta_2}) Y_{\Theta_3}
-
\ldots
-
(1 - Y_{\Theta_2})\cdots(1 - Y_{\Theta_{j-2}})Y_{\Theta_{j-1}}
\Big)
Y_{\Theta_j}
\nonumber \\
& =
(1 - Y_{\Theta_1}) (1 - Y_{\Theta_2})
\Big(
\underbrace{
1 - Y_{\Theta_3}
}_{= (1-Y_{\Theta_3})}
-
(1 - Y_{\Theta_3}) Y_{\Theta_4}
-
\ldots
-
(1 - Y_{\Theta_3})\cdots(1 - Y_{\Theta_{j-2}})Y_{\Theta_{j-1}}
\Big)
Y_{\Theta_j}
\nonumber \\
& \vdots
\nonumber \\
& =
(1 - Y_{\Theta_1})
(1 - Y_{\Theta_2})
\cdots
(1 - Y_{\Theta_{j-1}})
Y_{\Theta_j}
\ ,
\end{align}
confirming eq.~\eqref{eq:structural-form-LTheta}.

Let 
\begin{equation}
  M_{\Theta_{i-1}}
  :=
  (1-Y_{\Theta_1})(1-Y_{\Theta_2})\cdots(1-Y_{\Theta_{i-1}})
\end{equation}
such that $L_{\Theta_{i}}=M_{\Theta_{i-1}}Y_{\Theta_i}$. From
eq.~\eqref{eq:YOps-orthogonal-larger} is is clear that
\begin{equation}
  Y_{\Theta_k}(1-Y_{\Theta_l})
  =
  Y_{\Theta_k} - 0
  =
  Y_{\Theta_k}
  \qquad 
  \text{for every $l<k$}
  \ .
\end{equation}
Thus, if $i<j$ we have that
\begin{align}
  L_{\Theta_{i}}
  \cdot
  L_{\Theta_{j}}
  \; & =
  \underbrace{
  M_{\Theta_{i-1}}Y_{\Theta_i}
  }_{=L_{\Theta_{i}}}
  \cdot
  \underbrace{
  (1 - Y_{\Theta_1})
  \cdots
  (1 - Y_{\Theta_i})
  \cdots
  (1 - Y_{\Theta_{j-1}})
  Y_{\Theta_j}
  }_{=L_{\Theta_{j}}}
\nonumber \\
  & =
  M_{\Theta_{i-1}}
  \underbrace{
  Y_{\Theta_i}  
  (1 - Y_{\Theta_i})
  }_{=Y_{\Theta_i}-Y_{\Theta_i}^2=0}
  \cdots
  (1 - Y_{\Theta_{j-1}})
  Y_{\Theta_j}
\nonumber \\
  & = 0
  \ ,
\end{align}
where we used  the fact that Young projectors are idempotent
($Y_{\Theta_i}^2=Y_{\Theta_i}$). Thus, $L_{\Theta_{i}}$ and
$L_{\Theta_{j}}$ are orthogonal even if $i<j$.

We notice that the $L_{\Theta}$ remain idempotent just like
their Young counterparts,
\begin{equation}
  L_{\Theta_{i}} \cdot L_{\Theta_{i}} = 
L_{\Theta_i} 
\underbrace{
(1 - L_{\Theta_1} - L_{\Theta_2} - \ldots - L_{\Theta_{i-1}})
Y_{\Theta_1}
}_{=L_{\Theta_i}}
=
(L_{\Theta_i} - 0- 0- \ldots -0) Y_{\Theta_1}
=
L_{\Theta_i} Y_{\Theta_i}
=
L_{\Theta_i}
\ .
\end{equation}
Putting everything together, we can conclude that
\begin{equation}
\label{eq:LYOps-mutually-orthogonal}
  L_{\Theta} L_{\Phi} = \delta_{\Theta\Phi} L_{\Phi}
  \qquad
  \text{for all $\Theta,\Phi\in\mathcal{Y}_n$ for all values of $n$}
  \ ,
\end{equation}
confirming eqns.~\eqref{eq:LY-idem} and~\eqref{eq:LY-orth}.

Since the LY-operators are mutually orthogonal
(eq.~\eqref{eq:LYOps-mutually-orthogonal}), they can be simultaneously
diagonalised and have mutually orthogonal images. Since they are
idempotent their trace provides the dimension of their image:
\begin{equation}
  \Tr{L_{\Theta}} = \text{dim}(\Theta)
  \ .
\end{equation}

% Therefore, there exists a basis in which \emph{each} operator\tod{this paragraph is not very polished}
% $L_{\Theta}$ is given as a $N^n\times N^n$-matrix with $1$ on
% the diagonal of the subblock corresponding to the subspace onto
% which $L_{\Theta}$ projects and $0$ everywhere else. In this basis, it
% is clear that the dimension of the representation corresponding to
% $L_{\Theta}$, $\text{dim}(\Theta)$, is given by the trace:

Thus, if we can show that the dimensions of the subspaces
corresponding to the $L_{\Theta}$ sum up to the dimension of the whole
space $\text{dim}(\Pow{n})=N^n$, it must hold that the orthogonal
operators $L_{\Theta}$ sum up to the identity on $\Pow{n}$. To show
this, we notice that $L_{\Theta}$ has the same trace as its
counterpart $Y_{\Theta}$: Using the cyclic property of the trace, we
have
\begin{align}
  \Tr{L_{\Theta_i}} 
  & = 
  \Tr{
    (1 - L_{\Theta_1} - L_{\Theta_2} - \ldots - L_{\Theta_{i-1}})
    Y_{\Theta_i}
    } \nonumber \\
  & = 
  \Tr{
    Y_{\Theta_i}
    (1 - L_{\Theta_1} - L_{\Theta_2} - \ldots - L_{\Theta_{i-1}})
    } \nonumber \\
  & = 
  \Tr{
    Y_{\Theta_i} - 0 - 0 - \ldots - 0
    } \nonumber \\
  & = \Tr{Y_{\Theta_i}}
\end{align}
which implies that
\begin{equation}
  \label{eq:sum-dimension-LYOps}
  \sum\limits_{\Theta\in\mathcal{Y}_n} \Tr{L_{\Theta}}
  =
  \sum\limits_{\Theta\in\mathcal{Y}_n} \Tr{Y_{\Theta}}
  =
  N^n
  \qquad \text{for all $n$, where $N=\text{dim}(V)$}
  \ .
\end{equation}
Therefore, we conclude that the LY-operators sum up to the identity on
the space $\Pow{n}$
\begin{equation}
  \sum\limits_{\Theta\in\mathcal{Y}_n} L_{\Theta}
  = 
  \mathrm{id}_{n}
  \qquad
  \text{for all $n$}
  \ ,
\end{equation}
as was claimed in eq.~\eqref {eq:LY-decom-unity}.

\section{Young projectors that do not commute with their
  ancestors}\label{sec:Y-Ancestor-not-commute}

In this appendix, we prove that the standard Young projection operators  which do not project onto subspaces included in
the image of their ancestors (unlike
their Hermitian counterpart, \emph{c.f.}
Lemma~\ref{thm:YoungNonCommute}),
\begin{equation}
  \label{eq:image-inclusion-Youngs}
Y_{\Theta} Y_{\Theta_{(m)}} \neq Y_{\Theta}
\qquad \text{and/or} \qquad
Y_{\Theta_{(m)}}  Y_{\Theta} \neq Y_{\Theta}
\end{equation}
do not  commute with their
ancestor operators.
\ytableausetup{mathmode, boxsize=0.7em}%
While there exist Young projection operators for which one of the two
equations~\eqref{eq:image-inclusion-Youngs} yields an equality, for example
\begin{equation}
  Y_{\begin{ytableau} \scriptstyle 1 & \scriptstyle 2\end{ytableau}} \cdot 
Y_{\begin{ytableau} \scriptstyle 1 & \scriptstyle 2 \\
    \scriptstyle 3 \end{ytableau}}
\; = \;
Y_{\begin{ytableau} \scriptstyle 1 & \scriptstyle 2 \\
    \scriptstyle 3 \end{ytableau}}
\ ,
\end{equation}
for most Young projection operators neither of these conditions hold, e.g.
\begin{equation}
  Y_{\begin{ytableau} \scriptstyle 1 & \scriptstyle 2 \\
    \scriptstyle 3 & \scriptstyle 4 \end{ytableau}} \cdot
Y_{\begin{ytableau} \scriptstyle 1 & \scriptstyle 2 \\
    \scriptstyle 3 \end{ytableau}}
\; \neq \;
  Y_{\begin{ytableau} \scriptstyle 1 & \scriptstyle 2 \\
    \scriptstyle 3 & \scriptstyle 4 \end{ytableau}}
\qquad \text{and} \qquad
Y_{\begin{ytableau} \scriptstyle 1 & \scriptstyle 2 \\
    \scriptstyle 3 \end{ytableau}} \cdot
  Y_{\begin{ytableau} \scriptstyle 1 & \scriptstyle 2 \\
    \scriptstyle 3 & \scriptstyle 4 \end{ytableau}}
\; \neq \;
  Y_{\begin{ytableau} \scriptstyle 1 & \scriptstyle 2 \\
    \scriptstyle 3 & \scriptstyle 4 \end{ytableau}}
\ .
\end{equation}

\begin{lemma}[Young operators do not generally commute with their ancestor
  operators]\label{thm:YoungNonCommute} \-
  Let $\Theta$ be a Young tableau and $\Theta_{(m)}$ be its ancestor
  $m$ generations back, where $m$ is a strictly positive integer
  ($m>0$). If the images of $Y_{\Theta}$ and $Y_{\Theta_{(m)}}$ are
  not nested, i.e. $Y_\Theta Y_{\Theta_{(m)}} \neq Y_\Theta$ and/or
  $Y_{\Theta_{(m)}} Y_\Theta \neq Y_\Theta$, then $Y_{\Theta}$ and
  $Y_{\Theta_{(m)}}$ do not commute,
  \begin{equation}
    \label{eq:YoungNonCommute1}
    \left[ Y_{\Theta}, Y_{\Theta_{(m)}} \right] \neq 0
    \ .
  \end{equation}
  Conversely, a vanishing commutator $\left[ Y_{\Theta},
    Y_{\Theta_{(m)}} \right] = 0$ implies image inclusion $Y_\Theta
  Y_{\Theta_{(m)}} = Y_\Theta= Y_{\Theta_{(m)}} Y_\Theta$.
\end{lemma}

\ytableausetup{mathmode, boxsize=normal}%

\emph{Proof of Lemma~\ref{thm:YoungNonCommute}:} We present a proof by
contradiction: Suppose there exists a Young projection operator
$Y_{\Theta}$ which commutes with its ancestor operator
$Y_{\Theta_{(m)}}$ while $Y_{\Theta_{(m)}} Y_\Theta \neq Y_\Theta$,
  \begin{equation}
    \label{eq:YoungNonCommuteProof1}
    Y_{\Theta}Y_{\Theta_{(m)}} = Y_{\Theta_{(m)}}Y_{\Theta}
    \ .
  \end{equation}
  If we multiply equation~\eqref{eq:YoungNonCommuteProof1} with the
  operator $Y_{\Theta}$ on the right, and use
  Theorem~\ref{thm:CancelWedgedParentOp} to simplify the LHS of the
  resulting equation, we
  obtain
  \begin{equation}
     \label{eq:YoungNonCommuteProof2}
\underbrace{Y_{\Theta}Y_{\Theta_{(m)}}Y_{\Theta}}_{=Y_{\Theta}Y_{\Theta}=Y_{\Theta}}
     = \  Y_{\Theta_{(m)}}\underbrace{Y_{\Theta}Y_{\Theta}}_{=Y_{\Theta}} 
     \hspace{1cm}
    \Longrightarrow \hspace{1cm}  Y_{\Theta} =  Y_{\Theta_{(m)}}Y_{\Theta}
    \ ,
  \end{equation}
a contradiction. The case where $Y_\Theta Y_{\Theta_{(m)}} \neq
Y_\Theta$ follows from left multiplication with $Y_\Theta$.\qed
\ytableausetup{mathmode, boxsize=1em}%

We note that Lemma~\ref{thm:YoungNonCommute} is the reason why the
proof of the summation property for Hermitian Young projection
operators~\eqref{eq:SpanSubspaces1} breaks down for most Young projectors
at the last step (eq.~\eqref{eq:SpanSubspaces-last-step}), since
\begin{equation}
\left[ 
\sum_{\Theta\in\mathcal{Y}_{n-1}} 
\left( 
\sum_{\Psi\in\lbrace\Theta\otimes\ybox{\scriptstyle
        n}\rbrace} 
Y_{\Psi} 
\right) 
\right] 
\cdot 
Y_{\Theta'} 
\neq 
\sum_{\Theta\in\mathcal{Y}_{n-1}} 
\left( 
\delta_{\Theta\Theta'}
\sum_{\Psi\in\lbrace\Theta\otimes\ybox{\scriptstyle n}\rbrace}
Y_{\Psi} 
\right)
\ .
\end{equation}
\ytableausetup{mathmode, boxsize=normal}%

\section{Proofs of the construction principle for Hermitian Young
  projection operators}\label{sec:Proofs}

This appendix provides the proofs of the Theorems given in section~\ref{sec:CompactHermitianOps}.

\subsection{Proof of Theorem~\ref{thm:LexicalConstruction}
  \emph{``lexical Hermitian Young projectors''}}\label{sec:ProofsLexical}

The proof of Theorem~\ref{thm:LexicalConstruction} makes use of
propagation rules of birdtrack
operators~\cite{Alcock-Zeilinger:2016bss}. We thus
summarize the applicable rules
of~\cite{Alcock-Zeilinger:2016bss} in
section~\ref{sec:PropagationRules}, before giving the proof of the
Lexical-Theorem~\ref{thm:LexicalConstruction} in
section~\ref{sec:LexicalProofProof}.

\subsubsection{Propagation rules}\label{sec:PropagationRules}

We first require the definition of a new quantity,
an \emph{amputated} tableau:
\begin{definition}[amputated (Young) tableaux]\label{AmputatedTableaux}
  Let $\Theta$ be a (Young) tableau and let $\mathcal{C}$ denote a
  particular column in $\Theta$. We construct the row-amputated
  tableau of $\Theta$ according to $\mathcal{C}$,
  $\cancel{\Theta}_{r}\left[\mathcal{C}\right]$, by removing all rows
  of $\Theta$
  which do \emph{not} overlap with $\mathcal{C}$.\\
  Similarly, if $\mathcal{R}$ is a particular row in $\Theta$, we
  construct the column-amputated tableau of $\Theta$ according to
  $\mathcal{R}$, $\cancel{\Theta}_{c}\left[\mathcal{R}\right]$, by
  removing all columns that do not overlap with $\mathcal{R}$.
\end{definition}

For example, for the following tableau $\Theta$, the row amputated
tableau $\cancel{\Theta}_r$ according to column $(3,4,7)^t$ is
\begin{equation}
  \Theta =
  \begin{ytableau}
    1 & *(magenta!25) 3 & 5 & 9 \\
    2 & *(magenta!25) 4 & 8 & 10 \\
    6 & *(magenta!25) 7 & 13 \\
    11 \\
    12
  \end{ytableau} \quad \mapsto \quad \cancel{\Theta}_r \left[(3,4,7)^t\right] =
  \begin{ytableau}
    1 & *(magenta!25) 3 & 5 & 9 \\
    2 & *(magenta!25) 4 & 8 & 10\\
    6 & *(magenta!25) 7 & 13 
  \end{ytableau} \; ;
\end{equation}
the rows $(11)$ and $(12)$ were deleted since they did not overlap
with the shaded column $(3,4,7)^t$. Another example for column amputation
is shown in the step from eq.~\eqref{eq:col-to-amputate}
to~\eqref{eq:col-amputated}. The idea of amputated tableaux is
necessary to describe the following simplification rule for birdtrack
operators:

\begin{theorem}[propagation of (anti-) symmetrizers]
\label{thm:PropagateSyms}
  Let $\Theta$ be a Young tableau and $O$ be a birdtrack operator of the form
  \begin{equation}
    \label{eq:PropagateSyms1}
    O = \mathbf{S}_{\Theta} \; \mathbf{A}_{\Theta} \; 
    \mathbf{S}_{\Theta\setminus\mathcal R},
  \end{equation}
  in which the symmtrizer set $\mathbf{S}_{\Theta\setminus\mathcal R}$
  arises from $\mathbf{S}_{\Theta}$ by removing precisely one
  symmetrizer $\bm{S}_{\mathcal R}$. By definition $\bm{S}_{\mathcal
    R}$ corresponds to a row $\mathcal R$ in $\Theta$ such that
  $\mathbf{S}_{\Theta} = \mathbf{S}_{\Theta\setminus\mathcal R}
  \bm{S}_{\mathcal R} = \bm{S}_{\mathcal R}
  \mathbf{S}_{\Theta\setminus\mathcal R} $.

  If the column-amputated tableau of $\Theta$ according to the row
  $\mathcal{R}$, $\cancel{\Theta}_c\left[\mathcal{R}\right]$, is
  \textbf{rectangular}, then the symmetrizer $\bm{S}_{\mathcal R}$ may be
  propagated through the set $\mathbf{A}_{\Theta}$ from the left to
  the right, yielding
  \begin{equation}
    O=\mathbf{S}_{\Theta} \; \mathbf{A}_{\Theta} \;
    \mathbf{S}_{\Theta\setminus\mathcal R}= \mathbf{S}_{\Theta\setminus\mathcal R} \; \mathbf{A}_{\Theta} \;
    \mathbf{S}_{\Theta}
    \ ,
  \end{equation}
  which implies that $O$ is Hermitian. We may think of this procedure
  as moving the missing symmetrizer $\bm{S}_{\mathcal R}$ through the
  intervening antisymmetrizer set $\mathbf{A}_{\Theta}$. 
  
  Noting that $\mathbf{S}_{\Theta} =
  \mathbf{S}_{\Theta} \bm{S}_{\mathcal R} = \bm{S}_{\mathcal R}
  \mathbf{S}_{\Theta} $ we immediately augment this statement to
  \begin{equation}
    \label{eq:PropagateSyms2}
    \mathbf{S}_{\Theta} \; \mathbf{A}_{\Theta} \;
    \mathbf{S}_{\Theta\setminus\mathcal R}= \mathbf{S}_{\Theta\setminus\mathcal R} \; \mathbf{A}_{\Theta} \;
    \mathbf{S}_{\Theta}
    =
    \mathbf{S}_{\Theta} \; \mathbf{A}_{\Theta} \;\mathbf{S}_{\Theta}
    \ .
  \end{equation}

  If the roles of symmetrizers and antisymmetrizers are exchanged, we
  need to verify that the row-amputated tableau
  $\cancel{\Theta}_r\left[\mathcal{C}\right]$ with respect to a column
  $\mathcal C$ is rectangular to guarantee that
  \begin{equation}
    \label{eq:PropagateSyms3}
    \mathbf{A}_{\Theta} \; \mathbf{S}_{\Theta} \;
     \mathbf{A}_{\Theta\setminus\mathcal C}= \mathbf{A}_{\Theta\setminus\mathcal C} \; \mathbf{S}_{\Theta} \;
    \mathbf{A}_{\Theta} 
    =
    \mathbf{A}_{\Theta} \; \mathbf{S}_{\Theta} \;\mathbf{A}_{\Theta}
    \ .
  \end{equation}
  This amounts to moving the missing antisymmetrizer $\bm{A}_{\mathcal
    C}$ through the intervening symmetrizer set $\mathbf{S}_{\Theta}$.
\end{theorem}

\noindent As an example, consider the operator $Q$ given
by
\begin{equation}\label{eq:SimplificationEx1}
  Q := \;
  \scalebox{0.7}{$\underbrace{\FPic{7Sym123Sym45Sym67SN}}_{\mbox{\large$\mathbf{S}_{\Theta}$}}\underbrace{\FPic{7s24s367}\FPic{7ASym123ASym456N}\FPic{7s24s376}}_{\mbox{\large$\mathbf{A}_{\Theta}$}}\underbrace{\FPic{7Sym123Sym45SN}}_{\mbox{\large$\mathbf{S}_{\Theta\setminus\mathcal R}$}}$},
\end{equation}
To check if the amputated tableau is rectangular we first need to
reconstruct $\Theta$ with rows corresponding to $\mathbf{S}_{\Theta}$
and columns corresponding to $\mathbf{A}_{\Theta}$. Evidently,
\begin{equation}
  \label{eq:col-to-amputate}
  \Theta =
  \begin{ytableau}
    1 & 2 & 3 \\
    4 & 5 \\
    *(magenta!25) 6 & *(magenta!25) 7
  \end{ytableau}
\ .
\end{equation}
In $\Theta$, we have marked the row $(6,7)$ corresponding to the
symmetrizer $\bm{S}_{67}$, which we would like to propagate through to
the right. Thus, in accordance with Theorem \ref{thm:PropagateSyms},
we form the column-amputated tableau of $\Theta$ according to the row
$(6,7)$,
\begin{equation}\label{eq:col-amputated}
  \cancel{\Theta}_c \left[(6,7)\right] =
  \begin{ytableau}
    1 & 2 \\
    4 & 5 \\
    *(magenta!25) 6 & *(magenta!25) 7
  \end{ytableau}
\ ,
\end{equation}
and see that it is indeed a rectangular tableau. Thus, we may
propagate the symmetrizer $\bm{S}_{67}$ from the left to the right,
\begin{equation}
  Q = \;
  \scalebox{0.7}{$\FPic{7Sym123Sym45Sym67N}\FPic{7s24s367}\FPic{7ASym123ASym456N}\FPic{7s24s376}\FPic{7Sym123Sym45N}$}
    \; = \;
    \scalebox{0.7}{$\FPic{7Sym123Sym45N}\FPic{7s24s367}\FPic{7ASym123ASym456N}\FPic{7s24s376}\FPic{7Sym123Sym45Sym67N}$}
\; = \;
\scalebox{0.7}{$\FPic{7Sym123Sym45Sym67N}\FPic{7s24s367}\FPic{7ASym123ASym456N}\FPic{7s24s376}\FPic{7Sym123Sym45Sym67N}$}
\ .
\end{equation}

\subsubsection{Proof of Theorem~\ref{thm:LexicalConstruction}:}\label{sec:LexicalProofProof}

 We will present a proof by induction: First, we prove the
\emph{Base Step} for the projection operators of $\SUN$ over
$\Pow{3}$ (i.e with $3$ legs) since this is the smallest instant for
which the KS-algorithm produces a new operator (and also the first
instant for which \emph{non-Hermitian} Young projectors occur). Thereafter, we will
consider a general projection operator
corresponding to a Young tableau $\Theta\in\mathcal{Y}_{m+1}$ with
a lexically ordered column-word (the proof for operators corresponding
to row-ordered Young tableaux is very similar
and thus left as an exercise to the reader). We will assume that 
Theorem~\ref{thm:LexicalConstruction} is true for the Hermitian operator corresponding to its 
parent tableau $P_{\Theta_{(1)}}$, where $\Theta_{(1)} \in
\mathcal{Y}_m$; this is the \emph{Induction Hypothesis}. Then, we 
show that the projection operators obtained from the KS-Theorem reduce
to the expression given in the
Lexical-Theorem~\ref{thm:LexicalConstruction},
\begin{equation}
  \label{eq:HermitianLexicalOrderEndResult}
  P_{\Theta} = P_{\Theta_{(1)}}
  Y_{\Theta} P_{\Theta_{(1)}} = \alpha_{\Theta} \cdot \bar{Y}_{\Theta}^{\dagger} \bar{Y}_{\Theta},
\end{equation}
concluding the proof.

\bigskip

\textbf{The Base Step:} For the projection operators of $\SUN$ over
$\Pow{1}$ or $\Pow{2}$ (i.e. with $1$ or $2$ legs), the proof
of~\eqref{eq:HermitianLexicalOrderEndResult} is trivial since all
Young projection operators are automatically Hermitian; thus,
$\bar{Y}_{\Theta}^{\dagger}=\bar{Y}_{\Theta}$,
and~\eqref{eq:HermitianLexicalOrderEndResult} reduces to
\begin{equation}
\label{eq:YBar-quasi-idempotent}
 \alpha_{\Theta} \cdot \bar{Y}_{\Theta}^{\dagger}\bar{Y}_{\Theta} =
 \alpha_{\Theta}
 \underbrace{\bar{Y}_{\Theta}\bar{Y}_{\Theta}}_{\frac{1}{\alpha_{\Theta}}\Bar{Y}_{\Theta}}
 = \bar{Y}_{\Theta}
\ .
\end{equation}
Since all Young projection operators $Y_{\Theta}$ with $\Theta \in
\mathcal{Y}_{1,2}$ have normalization constant $1$ (as can easily be
checked by looking at all three of them explicitly), $Y_{\Theta} =
\bar{Y}_{\Theta}$ holds for these operators. Thus, the
Lexical-Theorem~\ref{thm:LexicalConstruction} returns the original,
already Hermitian operators as does the orignal KS-algorithm. The
first nontrivial differences occur for $n = 3$. We use this as the
base step. Here, we have the following Young projection operators
corresponding to their respective Young tableaux,
\begin{subequations}
  \label{eq:YoungOps-YoungTabl-S3}
  \begin{alignat}{7}
    \label{eq:YoungOpsS3} 
    & \FPic{3ASym123} && \qquad   && \sfrac{4}{3} \; \FPic{3Sym13ASym12} &&
    \qquad && 
    \sfrac{4}{3} \; \FPic{3Sym12ASym13} && \quad \text{and} \quad && 
     %\centermathcell{
       \FPic{3Sym123}
     %}  
  \\[3mm] 
  \label{eq:YoungOpsS3Tabl}
  & %\centermathcell{
  \begin{ytableau}
  1 \\
  2 \\
  3
  \end{ytableau} 
  % }
  && &&
   %\centermathcell{
  \begin{ytableau}
  1 & 3 \\
  2
  \end{ytableau} %} 
  && &&
   %\centermathcell{
  \begin{ytableau}
  1 & 2 \\
  3
  \end{ytableau} %} 
  && &&
  \begin{ytableau}
  1 & 2 & 3
  \end{ytableau}
  \ . 
  \end{alignat}
\end{subequations}
%   \begin{IEEEeqnarray}{rCCCCCCCl}
% \IEEEyesnumber\label{eq:YoungOps-YoungTabl-S3}\IEEEyessubnumber*
%     & \FPic{3ASym123} & \qquad & \sfrac{4}{3} \; \FPic{3Sym13ASym12} & \qquad &
%     \sfrac{4}{3} \; \FPic{3Sym12ASym13} & \quad \text{and} \quad & \FPic{3Sym123} & \label{eq:YoungOpsS3} \\
%   &&&& \nonumber \\
%   & \begin{ytableau}
%   1 \\
%   2 \\
%   3
%   \end{ytableau} & &
%   \begin{ytableau}
%   1 & 3 \\
%   2
%   \end{ytableau} & &
%   \begin{ytableau}
%   1 & 2 \\
%   3
%   \end{ytableau} & &
%   \begin{ytableau}
%   1 & 2 & 3
%   \end{ytableau}. \label{eq:YoungOpsS3Tabl}
%   \end{IEEEeqnarray}
  In~\eqref{eq:YoungOpsS3}, the first and last operator are already
  Hermitian and have normalization constant $1$. Therefore, the
  Lexical-Theorem will return these operators unchanged, \emph{c.f.} eq.~\eqref{eq:YBar-quasi-idempotent}.

  The second and third tableaux in~\eqref{eq:YoungOpsS3Tabl} are lexically
  column-ordered and row-ordered respectively. Thus, we must construct
  their Hermitian Young projection operators according to
  prescriptions~\eqref{eq:ColumnOrderedProjector}
  and~\eqref{eq:RowOrderedProjector} respectively.

Table~\ref{LexicalOrderHermitianInductionStepTable} shows that the
construction of the Hermitian projection operators corresponding to the tableaux
\eqref{eq:YoungOpsS3Tabl} obtained from the Lexical-Theorem
\ref{thm:LexicalConstruction} is equivalent that of the KS-Theorem
\ref{thm:KSProjectors},~\cite{Keppeler:2013yla}, thus concluding the base step
of this proof:
\begin{table}[H]
\centering
  \begin{tabular}[c]{c|c}
    \textbf{KS-Theorem~\ref{thm:KSProjectors},~\cite{Keppeler:2013yla}} & \textbf{Lexical-Theorem~\ref{thm:LexicalConstruction}} \\
    (Multiplying Hermitian parent on either side) & (Multiplying by
    Hermitian conjugate on appropriate side) \\
    \hline
    & \\
    $\FPic{3ASym12N}\FPic{3ASym123SN}\FPic{3ASym12N} \; = \;
    \FPic{3ASym123}$ & 
    $\FPic{3ASym123SN}\FPic{3ASym123SN} \; = \; \FPic{3ASym123}$ \\
    & \\
    $\lfrac{4}{3}\cdot
    \FPic{3ASym12N}\FPic{3Sym13ASym12}\FPic{3ASym12N} \; = \; \lfrac{4}{3}\cdot 
    \FPic{3ASym12Sym23ASym12}$ & 
    $\lfrac{4}{3}\cdot \FPic{3ASym12Sym13}\FPic{3IdN}\FPic{3Sym13ASym12}
    \; = \;
    \lfrac{4}{3}\cdot
    \FPic{3ASym12Sym23ASym12}$ \\
   & \\
   $\lfrac{4}{3}\cdot
   \FPic{3Sym12N}\FPic{3Sym12ASym13}\FPic{3Sym12N} \; = \; \lfrac{4}{3}\cdot
   \FPic{3Sym12ASym23Sym12}$ &
   $\lfrac{4}{3}\cdot \FPic{3Sym12ASym13}\FPic{3IdN}\FPic{3ASym13Sym12} \;
   = \; \lfrac{4}{3} \cdot \FPic{3Sym12ASym23Sym12}$ \\
   & \\
   $\FPic{3Sym12N}\FPic{3Sym123SN}\FPic{3Sym12N} \; = \;
    \FPic{3Sym123}$ &
    $\FPic{3Sym123SN}\FPic{3Sym123SN} \; = \; \FPic{3Sym123}$ \\
  \end{tabular}
  \caption{This table contrasts the construction of Hermitian Young
    projection operators according to the
    KS-Theorem~\ref{thm:KSProjectors} (left), with
    that according to the Lexical-Theorem~\ref{thm:LexicalConstruction} (right). 
    Despite visible algorithmic differences, the results are identical.}
\label{LexicalOrderHermitianInductionStepTable}
\end{table}

\textbf{The Induction Step:} Let $\Theta \in \mathcal{Y}_{m+1}$ be a
tableau with a lexically ordered column-word, and let $\Theta_{(1)}
\in \mathcal{Y}_m$ be its parent tableau. Clearly, the column-word of
$\Theta_{(1)}$ is also in lexical order. We will assume that the
Lexical-Theorem~\ref{thm:LexicalConstruction} holds for the Hermitian
Young projection operator $P_{\Theta_{(1)}}$, i.e. that
\begin{equation}
  \label{eq:HermitianLexicalOrderInductionHypothesis}
  P_{\Theta_{(1)}} = \alpha_{\Theta} \cdot \bar{Y}_{\Theta_{(1)}}^{\dagger} \bar{Y}_{\Theta_{(1)}},
\end{equation}
and we will refer to this condition as the \emph{Induction
  Hypothesis}. Thus, according to this induction hypothesis,
$P_{\Theta_{(1)}}$ can be written in terms of sets of symmetrizers and
antisymmetrizers corresponding to the tableau $\Theta_{(1)}$ as 
  \begin{equation}
    P_{\Theta_{(1)}} = \alpha_{\Theta_{(1)}} \cdot \mathbf{A}_{\Theta_{(1)}}
    \; \mathbf{S}_{\Theta_{(1)}} \; \mathbf{A}_{\Theta_{(1)}},
  \end{equation}
where we used the fact that $\mathbf{S}_{\Theta_{(1)}}\mathbf{S}_{\Theta_{(1)}}=\mathbf{S}_{\Theta_{(1)}}$.
 We now construct
$\bar{P}_{\Theta}$ from $\bar{P}_{\Theta_{(1)}}$ using the KS-Theorem
\ref{thm:KSProjectors}; we have
\begin{equation}
\label{eq:BaseStep-PTheta}
  \bar{P}_{\Theta} = \;
  \underbrace{\mathbf{A}_{\Theta_{(1)}} \;
    \mathbf{S}_{\Theta_{(1)}} \;
    \mathbf{A}_{\Theta_{(1)}}}_{\bar{P}_{\Theta_{(1)}}} \; 
  \underbrace{\mathbf{S}_{\Theta} \;
    \mathbf{A}_{\Theta}}_{\bar{Y}_{\Theta}} \; 
  \underbrace{\mathbf{A}_{\Theta_{(1)}} \; \mathbf{S}_{\Theta_{(1)}} \; \mathbf{A}_{\Theta_{(1)}}}_{\bar{P}_{\Theta_{(1)}}}.
\end{equation}
In the above, we have chosen to ignore the proportionality constant
for now, as carrying it with us would draw attention away from the important steps
of the proof. Once we have shown that $\bar{P}_{\Theta}=\bar{Y}_{\Theta}^{\dagger}\bar{Y}_{\Theta}$, we will show that the
proportionality constant $\alpha_{\Theta}$ given in
\eqref{eq:HermitianLexicalOrderInductionHypothesis} is indeed the one
we require for $P_{\Theta}$ to be idempotent.

Since $\Theta_{(1)}$ is the parent tableau of $\Theta$, the images of
all symmetrizers and antisymmetrizers in $Y_{\Theta}$ (and thus
$P_{\Theta}$) are contained in the images of the symmetrizers and
antisymmetrizers in $Y_{\Theta_{(1)}}$ respectively,\footnote{We use the notation introduced in section~\ref{sec:OperatorsNotation}.}
\begin{equation}
  \mathbf{S}_{\Theta} \subset \mathbf{S}_{\Theta_{(1)}} \qquad \text{and}
  \qquad \mathbf{A}_{\Theta} \subset
  \mathbf{A}_{\Theta_{(1)}}
\ ,
\end{equation}
and hence
\begin{equation}
  \label{eq:LexOrderProof3}
  \mathbf{S}_{\Theta_{(1)}} \; \mathbf{S}_{\Theta} =
  \mathbf{S}_{\Theta} = \mathbf{S}_{\Theta} \;
  \mathbf{S}_{\Theta_{(1)}} \qquad \text{and} \qquad
  \mathbf{A}_{\Theta_{(1)}} \; \mathbf{A}_{\Theta} =
  \mathbf{A}_{\Theta} = \mathbf{A}_{\Theta} \;
  \mathbf{A}_{\Theta_{(1)}}
\ ,
\end{equation}
\emph{c.f.} eq.~\eqref{eq:Inclusion-Ancestor-Ops}. Therefore, we are
able to factor $\mathbf{S}_{\Theta_{(1)}}$ out of
$\mathbf{S}_{\Theta}$ in~\eqref{eq:BaseStep-PTheta} to obtain 
\begin{equation}
\label{eq:LexOrderProof3b}
\begin{tikzpicture}[baseline=(current bounding box.west),
  every node/.style={inner sep=1pt,outer sep=-1pt}        ]
    \matrix(ID)[
    matrix of math nodes,
    ampersand replacement=\&,
    row sep =0mm,
    column sep =0mm
    ]
    {\bar{P}_{\Theta}  = 
      \& \mathbf{A}_{\Theta_{(1)}}
      \& \mathbf{S}_{\Theta_{(1)}} 
      \& \mathbf{A}_{\Theta_{(1)}} 
      \& \mathbf{S}_{\Theta_{(1)}}
      \& \mathbf{S}_{\Theta} 
      \& \mathbf{A}_{\Theta}  
      \& \mathbf{A}_{\Theta_{(1)}} 
      \& \mathbf{S}_{\Theta_{(1)}} 
      \& \mathbf{A}_{\Theta_{(1)}}
      \\
};
\draw[decorate,decoration={brace,amplitude=4pt},thick] (ID-1-3.south east) --
(ID-1-2.south west) node[pos=.5,anchor=north,yshift=-2mm]  {\scriptsize$= \bar{Y}_{\Theta_{(1)}}^{\dagger}$};
\draw[decorate,decoration={brace,amplitude=4pt},thick] (ID-1-5.south east) --
(ID-1-4.south west) node[pos=.5,anchor=north,yshift=-2mm]  {\scriptsize$= \bar{Y}_{\Theta_{(1)}}^{\dagger}$};
\draw[stealth-,thick,red] (ID-1-5.north) |- ++(0,+2mm)   -|  (ID-1-6.north) node[pos=.2,anchor=south,yshift=1mm]  {\scriptsize$\mathbf S_\Theta\to\mathbf S_{\Theta_{(1)}}\mathbf S_\Theta$};
\end{tikzpicture}
\ .
\end{equation}
Since
$Y_{\Theta_{(1)}}^{\dagger}=\alpha_{\Theta_{(1)}}\bar{Y}_{\Theta_{(1)}}^{\dagger}$
is a projection operator, it follows that
$\bar{Y}_{\Theta_{(1)}}^{\dagger}\bar{Y}_{\Theta_{(1)}}^{\dagger}\propto\bar{Y}_{\Theta_{(1)}}^{\dagger}$. Hence,
eq.~\eqref{eq:LexOrderProof3b} reduces to (making use of the
bar-notation to retain equality, \emph{c.f.}
eq.~\eqref{eq:Bar-Notation})
\begin{equation}
\begin{tikzpicture}[baseline=(current bounding box.west),
  every node/.style={inner sep=1pt,outer sep=-1pt}        ]
    \matrix(ID)[
    matrix of math nodes,
    ampersand replacement=\&,
    row sep =0mm,
    column sep =0mm
    ]
    {\bar{P}_{\Theta}  = 
      \& \mathbf{A}_{\Theta_{(1)}} 
      \& \cancel{\mathbf{S}_{\Theta_{(1)}}}
      \& \mathbf{S}_{\Theta} 
      \& \mathbf{A}_{\Theta}  
      \& \cancel{\mathbf{A}_{\Theta_{(1)}}}
      \& \mathbf{S}_{\Theta_{(1)}} 
      \& \mathbf{A}_{\Theta_{(1)}}
      \& = 
      \; \mathbf{A}_{\Theta_{(1)}} 
      \; \mathbf{S}_{\Theta} 
      \; \mathbf{A}_{\Theta} 
      \; \mathbf{S}_{\Theta_{(1)}} 
      \; \mathbf{A}_{\Theta_{(1)}}
      \ ,
      \\
};
\draw[decorate,decoration={brace,amplitude=4pt},thick] (ID-1-3.south east) --
(ID-1-2.south west) node[pos=.5,anchor=north,yshift=-2mm]  {\scriptsize$= \bar{Y}_{\Theta_{(1)}}^{\dagger}$};
\draw[-stealth,thick,red] (ID-1-3.north) |- ++(0,+2mm)   -|
(ID-1-4.north) node[pos=.2,anchor=south,yshift=1mm]
{\scriptsize$\mathbf{S}_{\Theta_{(1)}}\mathbf{S}_{\Theta}\to\mathbf{S}_{\Theta}$};
\draw[stealth-,thick,red] (ID-1-5.south) |- ++(0,-3mm)   -|  (ID-1-6.south) node[pos=.2,anchor=north,yshift=-1mm]  {\scriptsize$\mathbf{A}_{\Theta}\mathbf{A}_{\Theta_{(1)}}\to\mathbf{A}_{\Theta}$};
\end{tikzpicture} 
\end{equation}
where we used eq.~\eqref{eq:LexOrderProof3} to reabsorb
$\mathbf{S}_{\Theta_{(1)}}$ into $\mathbf{S}_{\Theta}$ and
$\mathbf{A}_{\Theta_{(1)}}$ into $\mathbf{A}_{\Theta}$. Thus
\begin{equation}
  \label{eq:LexOrderProof10}
  \bar{P}_{\Theta} = \; \mathbf{A}_{\Theta_{(1)}} \; \mathbf{S}_{\Theta} \; \mathbf{A}_{\Theta} \; \mathbf{S}_{\Theta_{(1)}} \;
    \mathbf{A}_{\Theta_{(1)}}
\ .
\end{equation}
\ytableausetup{mathmode, boxsize=2em}{}To complete the proof, we have to distinguish two cases: The case
where \ybox{\scriptstyle m+1} lies in the first row of $\Theta$, and
the case where it is positioned in any \emph{but} the first row.
\begin{enumerate}
\item\label{InductionHypothesis1} Suppose \ybox{\scriptstyle
    m+1} lies in the first row of $\Theta$. Since this is the box
  containing the highest value in the tableau $\Theta$, there is no
  box positioned below it (otherwise $\Theta$ would not be a Young
  tableau). Thus, the leg $(m+1)$ is not contained in any
  antisymmetrizer (of length $>1$), yielding the sets $\mathbf{A}_{\Theta_{(1)}}$ and
  $\mathbf{A}_{\Theta}$ identical,
  $\mathbf{A}_{\Theta_{(1)}} = \mathbf{A}_{\Theta}$,
\begin{equation}
  \bar{P}_{\Theta} = \; 
  \mathbf{A}_{\Theta_{(1)}} \;
  \mathbf{S}_{\Theta} \; 
  \mathbf{A}_{\Theta} \;
  \mathbf{S}_{\Theta_{(1)}} \;  
  \mathbf{A}_{\Theta_{(1)}}
\; = \; 
  \mathbf{A}_{\Theta} \;
\fcolorbox{red}{white}{$
  \mathbf{S}_{\Theta} \; 
  \mathbf{A}_{\Theta} \;
  \mathbf{S}_{\Theta_{(1)}} \;  
  \mathbf{A}_{\Theta}
$}
\ .
\end{equation}
We now apply the Cancellation-Theorem~\ref{thm:CancelMultipleSets} to
the part of $P_{\Theta}$ in the red box to obtain
\begin{equation}
  \label{eq:LexOrderProof15}
  \bar{P}_{\Theta} = \; \mathbf{A}_{\Theta} \;
  \mathbf{S}_{\Theta} \;  \mathbf{A}_{\Theta}
\end{equation}
as required.

\item\label{InductionHypothesis2} Suppose now that
  \ybox{\scriptstyle m+1} is situated in any but the first row
  of $\Theta$. In this case the leg $m+1$ does enter an
  antisymmetrizer (of length $>1$), thus $\mathbf{A}_{\Theta_{(1)}}
  \neq \mathbf{A}_{\Theta}$ -- a new strategy is needed. To understand
  the obstacles, let us once again look at the operator $P_{\Theta}$
  as described by equation~\eqref{eq:LexOrderProof10},
\begin{equation}
  \label{eq:LexOrderProof16}
  \bar{P}_{\Theta} = \; \colorbox{red!20}{$\mathbf{A}_{\Theta_{(1)}} \; \mathbf{S}_{\Theta} \; \mathbf{A}_{\Theta}$} \; \mathbf{S}_{\Theta_{(1)}} \;
    \mathbf{A}_{\Theta_{(1)}}.
\end{equation}
\emph{Describing the strategy:}
In~\eqref{eq:LexOrderProof16} we have suggestively marked a part of
$P_{\Theta}$ in colour: if we were allowed to \emph{exchange} the sets
$\mathbf{A}_{\Theta_{(1)}}$ and $\mathbf{A}_{\Theta}$, replacing
$\bar{P}_{\Theta}$ by
\begin{equation}
  \label{eq:LexOrderProofrev-order}
   \colorbox{red!20}{$\mathbf{A}_{\Theta} \; \mathbf{S}_{\Theta} \; \mathbf{A}_{\Theta_{(1)}}$} \; \mathbf{S}_{\Theta_{(1)}} \;
    \mathbf{A}_{\Theta_{(1)}}
    \ ,
\end{equation}
we would be able to factor the symmetrizer
$\mathbf{S}_{\Theta_{(1)}}$ out of $\mathbf{S}_{\Theta}$ by
relation~\eqref{eq:LexOrderProof3} and use the fact that
$Y_{\Theta_{(1)}}=\alpha_{\Theta_{(1)}}\bar{Y}_{\Theta_{(1)}}$ is a
projection operator to obtain
\begin{equation}
\label{eq:LexOrderProof-shorten-PTheta}
\begin{tikzpicture}[baseline=(current bounding box.west),
  every node/.style={inner sep=1pt,outer sep=-1pt}        ]
    \matrix(ID)[
    matrix of math nodes,
    ampersand replacement=\&,
    row sep =0mm,
    column sep =0mm
    ]
    {\eqref{eq:LexOrderProofrev-order}  = 
      \& \mathbf{A}_{\Theta}
      \& \mathbf{S}_{\Theta} 
      \& \mathbf{S}_{\Theta_{(1)}} 
      \& \mathbf{A}_{\Theta_{(1)}} 
      \& \mathbf{S}_{\Theta_{(1)}}
      \& \mathbf{A}_{\Theta_{(1)}} 
      \& \propto \; \mathbf{A}_{\Theta} 
      \; \mathbf{S}_{\Theta} 
      \; \mathbf{S}_{\Theta_{(1)}} 
      \; \mathbf{A}_{\Theta_{(1)}}
      \ .
      \\
};
\draw[-stealth,thick,red] (ID-1-3.north) |- ++(0,+2mm)   -|  (ID-1-4.north) node[pos=.2,anchor=south,yshift=1mm]  {\scriptsize$\mathbf{S}_\Theta\to\mathbf{S}_{\Theta}\mathbf{S}_{\Theta_{(1)}}$};
\draw[decorate,decoration={brace,amplitude=4pt},thick] (ID-1-5.south east) --
(ID-1-4.south west) node[pos=.5,anchor=north,yshift=-2mm]
{\scriptsize$= \bar{Y}_{\Theta_{(1)}}$};
\draw[decorate,decoration={brace,amplitude=4pt},thick] (ID-1-7.south east) --
(ID-1-6.south west) node[pos=.5,anchor=north,yshift=-2mm]  {\scriptsize$= \bar{Y}_{\Theta_{(1)}}$};
\end{tikzpicture}
\end{equation}
Re-absorbing $\mathbf{S}_{\Theta_{(1)}}$ into $\mathbf{S}_{\Theta}$ yields
\begin{equation}
\label{eq:LexOrderProofrev-order-simplify}
\begin{tikzpicture}[baseline=(current bounding box.west),
  every node/.style={inner sep=1pt,outer sep=-1pt}        ]
    \matrix(ID)[
    matrix of math nodes,
    ampersand replacement=\&,
    row sep =0mm,
    column sep =0mm
    ]
    {\eqref{eq:LexOrderProofrev-order}  = \;
      \& \mathbf{A}_{\Theta}
      \& \mathbf{S}_{\Theta} 
      \& \cancel{\mathbf{S}_{\Theta_{(1)}}}
      \& \mathbf{A}_{\Theta_{(1)}} 
      \& = \; \mathbf{A}_{\Theta} 
      \;  \mathbf{S}_{\Theta} 
      \; \mathbf{A}_{\Theta_{(1)}}
      \ .
      \\
};
\draw[stealth-,thick,red] (ID-1-3.north) |- ++(0,+2mm)   -|  (ID-1-4.north) node[pos=.2,anchor=south,yshift=1mm]  {\scriptsize$\mathbf{S}_{\Theta}\mathbf{S}_{\Theta_{(1)}}\to\mathbf{S}_{\Theta}$};
\end{tikzpicture}
\end{equation}
From there, a similar argument as is needed to justify the missing
step from~\eqref{eq:LexOrderProof16}
to~\eqref{eq:LexOrderProofrev-order} can be used to show that
\begin{equation}
\label{eq:LexOrderProof-formP}
  \mathbf{A}_{\Theta} \;
  \mathbf{S}_{\Theta} \; \mathbf{A}_{\Theta_{(1)}} \; = \; \mathbf{A}_{\Theta} \;
  \mathbf{S}_{\Theta} \; \mathbf{A}_{\Theta}
\ ,
\end{equation}
yielding the desired
form of $\bar{P}_{\Theta}$. The main obstacle in achieving
this result thus lies in the  justification of the exchange
of antisymmetrizers in the step from~\eqref{eq:LexOrderProof16}
to~\eqref{eq:LexOrderProofrev-order}.

\emph{The full argument:} We will accomplish this exchange of
$\mathbf{A}_{\Theta_{(1)}}$ and $\mathbf{A}_{\Theta}$ within the
marked region of~\eqref{eq:LexOrderProof16} in the following way:
Consider the Young tableaux $\Theta_{(1)}$ and $\Theta$ as depicted in
Figure~\ref{fig:LexOrderProof1}:
\begin{figure}[H]
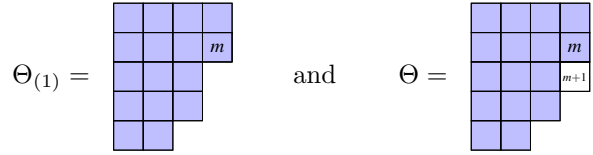

  \centering
  $\Theta_{(1)} =$ \; 
\scalebox{0.9}{\FPic{YTLexical-Theta-1}} \qquad {\normalsize and} \qquad
{\normalsize $\Theta =$} \;
\scalebox{0.9}{\FPic{YTLexical-Theta}}
\caption{This figure gives a schematic depiction of the Young tableaux
  $\Theta_{(1)}$ and $\Theta$. The boxes that are common in the two
  tableaux have been marked in colour. The box with entry $(m+1)$ has to
  lie in the bottom-most position of the last column of $\Theta$, as
  otherwise the column-word of $\Theta$, $\mathfrak{C}_{\Theta}$,
  would not be in lexical order, contradictory to our initial
  assumption. The requirement that $\mathfrak{C}_{\Theta}$ is
  lexically ordered therefore also uniquely determines the position of
  the box $m$, as is indicated in this figure.}
  \label{fig:LexOrderProof1}
\end{figure}
Since, by assumption, \ybox{\scriptstyle m+1} does \emph{not}
lie in the first row of $\Theta$, the leg $(m+1)$ is contained in an
antisymmetrizer (of length $>1$) in $\mathbf{A}_{\Theta}$, as was
already mentioned previously. Let us denote this antisymmetrizer by
$\bm{A}^{m+1}_{\Theta} \in \mathbf{A}_{\Theta}$. Furthermore, let
$\bm{A}^{m}_{\Theta_{(1)}}$ be the corresponding antisymmetrizer of
the tableau $\Theta_{(1)}$: in other words,
$\bm{A}^{m}_{\Theta_{(1)}}$ is the antisymmetrizer
$\bm{A}^{m+1}_{\Theta}$ with the leg $m+1$ removed. Hence
$\bm{A}^{m}_{\Theta_{(1)}}\supset\bm{A}^{m+1}_{\Theta}$, using the
notation introduced in section~\ref{sec:OperatorsNotation}.

Since $\Theta_{(1)}$ is the parent tableau of $\Theta$, all columns
but the last will be identical in the two tableaux, see
Figure~\ref{fig:LexOrderProof1}. Thus, the antisymmetrizers
corresponding to any but the last row will be contained in both sets
$\mathbf{A}_{\Theta_{(1)}}$ and $\mathbf{A}_{\Theta}$, which in
particular implies that
\begin{equation}
  \label{eq:LexOrderProof17}
  \mathbf{A}_{\Theta} =
  \mathbf{A}_{\Theta_{(1)}}\;\bm{A}^{m+1}_{\Theta}
\end{equation}
since $\bm{A}^{m}_{\Theta_{(1)}}\supset\bm{A}^{m+1}_{\Theta}$.
Thus, if we were able to commute the
antisymmetrizer $\bm{A}^{m+1}_{\Theta}$ through the set
$\mathbf{S}_{\Theta}$ from the right to the left (and then absorb
$\bm{A}^{m}_{\Theta_{(1)}}$ into $\bm{A}^{m+1}_{\Theta}$), we could cast $P_{\Theta}$ into the desired
form~\eqref{eq:LexOrderProof-formP}. In fact, this is exactly what we
will do: According to Theorem~\ref{thm:PropagateSyms}, the antisymmetrizer
$\bm{A}^{m+1}_{\Theta}$ can be propagated through the set
$\mathbf{S}_{\Theta}$ if the row-amputated Young tableau
$\cancel{\Theta}_{r}$ according to the last column of $\Theta$ is
rectangular. Thus, let us form this amputated tableau,
\begin{equation}
  \cancel{\Theta}_{r} = \;
\scalebox{0.9}{\FPic{YTLexical-Theta-Ampc}}
\ .
\end{equation}
This tableau is indeed rectangular,\footnote{It is important to note
  that this amputated tableau would not necessarily be rectangular if
  $\Theta$ were not lexically ordered, as then
  \ybox{\scriptscriptstyle m+1}\ytableausetup{mathmode,
    boxsize=normal} could be situated in a column other than the
  last one. Thus, for non-lexically ordered tableaux, the proof breaks
  down at this point.} allowing us to propagate the antisymmetrizer
$\bm{A}^{m+1}_{\Theta}$ from the right to the left, yielding
\begin{equation}
  \bar{P}_{\Theta} 
  = \; 
  \colorbox{red!20}{$\mathbf{A}_{\Theta} \; 
    \mathbf{S}_{\Theta} \; \mathbf{A}_{\Theta_{(1)}}$} \; 
  \mathbf{S}_{\Theta_{(1)}} \;
    \mathbf{A}_{\Theta_{(1)}}
 \ .
\end{equation}
Having demonstrated, that $\mathbf{A}_{\Theta_{(1)}}$ and
$\mathbf{A}_{\Theta}$ may be swapped, it is possible to simplify
$\bar{P}_{\Theta}$ as shown in~\eqref{eq:LexOrderProof-shorten-PTheta}--\eqref{eq:LexOrderProofrev-order-simplify},
\begin{equation}
\begin{tikzpicture}[baseline=(current bounding box.west),
  every node/.style={inner sep=1pt,outer sep=-1pt}        ]
    \matrix(ID)[
    matrix of math nodes,
    ampersand replacement=\&,
    row sep =0mm,
    column sep =0mm
    ]
    {\bar{P}_{\Theta}  = 
      \& \mathbf{A}_{\Theta}
      \& \mathbf{S}_{\Theta} 
      \& \mathbf{S}_{\Theta_{(1)}} 
      \& \mathbf{A}_{\Theta_{(1)}} 
      \& \mathbf{S}_{\Theta_{(1)}}
      \& \mathbf{A}_{\Theta_{(1)}} 
      \& = 
      \& \mathbf{A}_{\Theta}
      \& \mathbf{S}_{\Theta} 
      \& \cancel{\mathbf{S}_{\Theta_{(1)}}} 
      \& \mathbf{A}_{\Theta_{(1)}} 
      \& = 
      \& \mathbf{A}_{\Theta}
      \& \mathbf{S}_{\Theta} 
      \& \mathbf{A}_{\Theta_{(1)}} 
      \ .
      \\
};
\draw[-stealth,thick,red] (ID-1-3.north) |- ++(0,+2mm)   -|
(ID-1-4.north) node[pos=.2,anchor=south,yshift=1mm]
{\scriptsize$\mathbf{S}_\Theta\to\mathbf{S}_{\Theta}\mathbf{S}_{\Theta_{(1)}}$};
\draw[stealth-,thick,red] (ID-1-10.north) |- ++(0,+2mm)   -|
(ID-1-11.north) node[pos=.2,anchor=south,yshift=1mm]
{\scriptsize$\mathbf{S}_\Theta \mathbf{S}_{\Theta_{(1)}}\to\mathbf{S}_{\Theta}$};
\draw[decorate,decoration={brace,amplitude=4pt},thick] (ID-1-5.south east) --
(ID-1-4.south west) node[pos=.5,anchor=north,yshift=-2mm]
{\scriptsize$= \bar{Y}_{\Theta_{(1)}}$};
\draw[decorate,decoration={brace,amplitude=4pt},thick] (ID-1-7.south east) --
(ID-1-6.south west) node[pos=.5,anchor=north,yshift=-2mm]
{\scriptsize$= \bar{Y}_{\Theta_{(1)}}$};
\draw[decorate,decoration={brace,amplitude=4pt},thick] (ID-1-12.south east) --
(ID-1-11.south west) node[pos=.5,anchor=north,yshift=-2mm]  {\scriptsize$= \bar{Y}_{\Theta_{(1)}}$};
\end{tikzpicture}
\end{equation}
We once again use Theorem~\ref{thm:PropagateSyms} to obtain the
desired form of $\Bar{P}_{\Theta}$,
\begin{equation}
  \label{eq:LexOrderProof21}
\Bar{P}_{\Theta} =  
      \; \mathbf{A}_{\Theta} 
      \; \mathbf{S}_{\Theta}
      \; \mathbf{A}_{\Theta_{(1)}}
      \; \xlongequal{\text{Thm.~\ref{thm:PropagateSyms}}}  
      \; \mathbf{A}_{\Theta} 
      \; \mathbf{S}_{\Theta}
      \; \mathbf{A}_{\Theta}
      \ .
\end{equation}

\end{enumerate}
It remains to show that the normalization constant given in
\eqref{eq:HermitianLexicalOrderEndResult} is the right one: that is,
we will show that $P_{\Theta}=\alpha_{\Theta}\bar{P}_{\Theta}$, where
$\bar{P}_{\Theta}=\bar{Y}_{\Theta}^{\dagger}\bar{Y}_{\Theta}=\mathbf{A}_{\Theta}\mathbf{S}_{\Theta}\mathbf{A}_{\Theta}$
(as was found in ~\eqref{eq:LexOrderProof15} and
\eqref{eq:LexOrderProof21}), is indeed a projection operator. We will
establish this by simply squaring $P_{\Theta}$ and requiring that it
is idempotent:
\begin{equation}
\label{eq:LexOrderProof-norm-constant1}
  P_{\Theta} P_{\Theta} = \alpha_{\Theta}^2 \cdot \left( \mathbf{A}_{\Theta} \; \mathbf{S}_{\Theta}
   \; \mathbf{A}_{\Theta} \right) \left( \mathbf{A}_{\Theta} \; \mathbf{S}_{\Theta}
   \; \mathbf{A}_{\Theta} \right) = \alpha_{\Theta}^2 \cdot \mathbf{A}_{\Theta} \; \underbrace{\mathbf{S}_{\Theta} \; \mathbf{A}_{\Theta}}_{=\bar{Y}_{\Theta}} \; \underbrace{\mathbf{S}_{\Theta}
   \; \mathbf{A}_{\Theta}}_{=\bar{Y}_{\Theta}}
\ ,
\end{equation}
where we have used the fact that
$\mathbf{A}_{\Theta}\mathbf{A}_{\Theta}=\mathbf{A}_{\Theta}$. By the
idempotency of Young projection operators $Y_{\Theta}$, it follows that 
$\bar{Y}_{\Theta}\bar{Y}_{\Theta}=\nicefrac{1}{\alpha_{\Theta}}\bar{Y}_{\Theta}$,
simplifying~\eqref{eq:LexOrderProof-norm-constant1} as
\begin{equation}
\label{eq:LexOrderProof-norm-constant2}
P_{\Theta} P_{\Theta} = \frac{\alpha_{\Theta}^2}{\alpha_{\Theta}}
\cdot \mathbf{A}_{\Theta} \; \underbrace{\mathbf{S}_{\Theta} \;
  \mathbf{A}_{\Theta}}_{=\bar{Y}_{\Theta}} = \alpha_{\Theta} \cdot \mathbf{A}_{\Theta} \; \mathbf{S}_{\Theta}
   \; \mathbf{A}_{\Theta} = P_{\Theta}
\ ;
\end{equation}
this concludes the proof of this Theorem~\ref{thm:LexicalConstruction}. \qed

\subsection{Proof of Theorem~\ref{thm:MOLDConstruction}
  \emph{``partially lexical Hermitian Young projectors''} (or
  MOLD-Theorem)}\label{sec:ProofsMOLD}

We now present a proof of the MOLD-Theorem~\ref{thm:MOLDConstruction}
by induction, using the Lexical-Theorem~\ref{thm:LexicalConstruction}
as a base step:

Consider a Young tableau $\Theta$ with MOLD $\mathcal{M}(\Theta)$
          such that $\Theta_{(\mathcal{M}(\Theta))}$ has a lexically ordered row-word; the proof
  for $\Theta_{(\mathcal{M}(\Theta))}$ having lexically ordered column-word is very similar
  and thus left as an exercise to the reader. We will provide a
  \emph{Proof by Induction} on the MOLD of $\Theta$,
  $\mathcal{M}(\Theta)$. Furthermore, we will for now ignore the
  proportionality constant $\beta_{\Theta}$ and concentrate on the
  birdtrack $\bar{P}_{\Theta}$ only. From the steps in the following
  proof, it will become evident that $\beta_{\Theta}\neq 0$ and $\beta_{\Theta}<\infty$ (as is
  explicitly discussed at the appropriate places within the proof),
  ensuring that $P_{\Theta}:=\beta_{\Theta}\bar{P}_{\Theta}$ is a
  \emph{non-trivial} (i.e. non-zero) projection operator.

\bigskip

\textbf{The Base Step:} Suppose that $\mathcal{M}(\Theta)=0$. In that
case, $\Theta$ itself has a lexically ordered row-word. Then, by the
MOLD-Theorem, $\bar{P}_{\Theta}$ must be of the following form
\begin{equation}
  \bar{P}_{\Theta} = \; \underbrace{\mathbf{S}_{\Theta} \;
    \mathbf{A}_{\Theta} \; \mathbf{S}_{\Theta}}_{=\bar{Y}_{\Theta} \bar{Y}_{\Theta}^{\dagger}}; \label{eq:MOLDHermProof2}
\end{equation}
this agrees with the result we obtained from the Lexical-Theorem
\ref{thm:LexicalConstruction} for which we have already given a full proof
in the Appendix~\ref{sec:ProofsLexical}. Also, the normalization
constant $\beta_{\Theta}=\alpha_{\Theta}\neq 0$, as required by the
MOLD-Theorem. Thus, the base step of the induction is fulfilled.

\bigskip

\textbf{The Induction Step:} Let us now consider a Young tableau
$\Theta$, such that the MOLD-Theorem holds for its parent tableau
$\Theta_{(1)}$. Further, assume that $\mathcal{M}(\Theta_{(1)}) = m$,
for some positive integer $m$ with $\Theta_{(m)}$ being row-ordered;
thus, we have that $\mathcal{M}(\Theta) = m+1$. We can now have one of
two situations, either $m$ is even, or $m$ is odd. First, suppose that
$\bm{m}$ \textbf{is even}. Then, according to the MOLD-Theorem, the
birdtrack of the projection operator $P_{\Theta_{(1)}}$ is given by
(\emph{c.f.} eq.~\eqref{eq:MOLDHermitianOperatorConstruction1})
  \begin{equation}
    \bar{P}_{\Theta_{(1)}} = \; \mathcal{C} 
      \;
     \underbrace{\colorbox{red!20}{$\mathbf{S}_{\Theta_{(1)}}  \; \mathbf{A}_{\Theta_{(1)}}
       \; \mathbf{S}_{\Theta_{(1)}}$}}_{=\bar{Y}_{\Theta_{(1)}} \bar{Y}_{\Theta_{(1)}}^{\dagger}} \; 
   \mathcal{C}^\dagger
\ , 
  \end{equation}
where we defined $\mathcal{C}$ to be
\begin{equation}
  \mathcal{C} :=
  \mathbf{S}_{\Theta_{(m+1)}}\mathbf{A}_{\Theta_{(m)}}\mathbf{S}_{\Theta_{(m-1)}}\ldots
 \mathbf{S}_{\Theta_{(3)}}\mathbf{A}_{\Theta_{(2)}}.
\end{equation}
We will now construct the birdtrack $\bar{P}_{\Theta}$ according to
the KS-Theroerm~\ref{thm:KSProjectors}~\cite{Keppeler:2013yla}; this yields
\begin{align}
  \bar{P}_{\Theta} 
& \;  =  \; \bar{P}_{\Theta_{(1)}} \; \colorbox{blue!25}{$\bar{Y}_{\Theta}$} \;
  \bar{P}_{\Theta_{(1)}} 
  \nonumber \\  
&
\begin{tikzpicture}[baseline=(current bounding box.west),
  every node/.style={inner sep=2pt,outer sep=2pt}        ]
    \matrix(ID)[
    matrix of math nodes,
    ampersand replacement=\&,
    row sep =0mm,
    column sep =0mm
    ]
    { \; = \; \mathcal{C} \;
      \& \mathbf{S}_{\Theta_{(1)}}
      \& \mathbf{A}_{\Theta_{(1)}} 
      \& \mathbf{S}_{\Theta_{(1)}} 
      \& \cdots
      \& \mathbf{A}_{\Theta_{(m)}}
      \& \cancel{\mathbf{S}_{\Theta_{(m+1)}}}
      \& \mathbf{S}_{\Theta}
      \& \mathbf{A}_{\Theta} 
      \& \mathbf{S}_{\Theta_{(m+1)}}
      \& \cdots
      \& \mathbf{S}_{\Theta_{(1)}}
      \& \mathbf{A}_{\Theta_{(1)}} 
      \& \mathbf{S}_{\Theta_{(1)}} 
      \& \; \mathcal{C}^{\dagger} \ ,
      \\
};
\draw[draw=none,fill=red!20,thick] ($(ID-1-2.north west) + (4pt,0)$) rectangle
($(ID-1-4.south east) + (-4pt,0pt)$) node[midway] {$\mathbf{S}_{\Theta_{(1)}}\;\mathbf{A}_{\Theta_{(1)}}\;\mathbf{S}_{\Theta_{(1)}}$};
\draw[draw=none,fill=red!20,thick] ($(ID-1-12.north west) + (4pt,0)$) rectangle
($(ID-1-14.south east) + (-4pt,0pt)$) node[midway] {$\mathbf{S}_{\Theta_{(1)}}\;\mathbf{A}_{\Theta_{(1)}}\;\mathbf{S}_{\Theta_{(1)}}$};
\draw[draw=none,fill=blue!25,thick] ($(ID-1-8.north west) + (4pt,0)$) rectangle
($(ID-1-9.south east) + (-4pt,0pt)$) node[midway] {$\mathbf{S}_{\Theta}\;\mathbf{A}_{\Theta}$};
\draw[color=gray!60,thick] ($(ID-1-2.north west) + (0pt,2.5pt)$) rectangle
($(ID-1-14.south east) + (-1pt,-2pt)$);
\draw[-stealth,thick,red] (ID-1-7.south) |- ++(0,-3mm)   -|
(ID-1-8.south) node[pos=.2,anchor=north,yshift=0mm]   {\scriptsize$\mathbf{S}_{\Theta_{(m+1)}}\mathbf{S}_{\Theta}\to\mathbf{S}_{\Theta}$};
\end{tikzpicture}
\label{eq:MOLDHermProof7PBar}
\end{align}
where we absorbed $\mathbf{S}_{\Theta_{(m+1)}}$ into $\mathbf{S}_{\Theta}$.
We notice that the parts of $\Bar P_{\Theta}$ denoted by $\mathcal{C}$
are already in the form that we want them to be. We thus focus our
attention on the part of $\Bar P_{\Theta}$ inside the gray box.  If we
can show that this part can be written as
\begin{equation}
\bar{P}_{\Theta} \overset{?}= \; \mathcal{C}\;\fcolorbox{gray!60}{white}{$\mathbf{S}_{\Theta_{(1)}}\underbrace{\mathbf{A}_{\Theta}\mathbf{S}_{\Theta}\mathbf{A}_{\Theta}}_{Y_{\Theta}^{\dagger} Y_{\Theta}}\mathbf{S}_{\Theta_{(1)}}$}\;
\mathcal{C}^{\dagger},
\end{equation}
then we have completed the proof. We will accomplish this goal in two
steps:
\begin{enumerate}
\item We will use the Cancellation-Theorem~\ref{thm:CancelMultipleSets} (see
  section~\ref{sec:CancellationRules}) to cancel the wedged ancestor sets of
  (anti-) symmetrizers in the gray box and thus reduce $\bar{P}_{\Theta}$ to
  \begin{equation}
    \label{eq:MOLDHermProofGoal1}
\bar{P}_{\Theta} = \; \mathcal{C}\;\fcolorbox{gray!60}{white}{$\mathbf{S}_{\Theta_{(1)}}{\color{blue}\mathbf{A}_{\Theta_{(1)}}}\mathbf{S}_{\Theta}{\color{red}\mathbf{A}_{\Theta}}\mathbf{S}_{\Theta_{(1)}}$}\;\mathcal{C}^{\dagger}.
\end{equation}
\item We then make use of the Hermiticity of $P_{\Theta}$ to show that 
  \begin{equation}
    \label{eq:MOLDHermProofGoal2}
\bar{P}_{\Theta} = \; \mathcal{C}\;\fcolorbox{gray!60}{white}{$\mathbf{S}_{\Theta_{(1)}}{\color{red}\mathbf{A}_{\Theta}}\mathbf{S}_{\Theta}{\color{red}\mathbf{A}_{\Theta}}\mathbf{S}_{\Theta_{(1)}}$}\;\mathcal{C}^{\dagger}.
\end{equation}
\end{enumerate}

Let us start the two-step-process:
\begin{enumerate}
\item\label{itm:MOLDHermProof1} The first step is easily
  accomplished: We factor a set $\mathbf{S}_{\Theta_{(1)}}$ out of
  $\mathbf{S}_{\Theta}$ and a set $\mathbf{A}_{\Theta_{(1)}}$ out of
  $\mathbf{A}_{\Theta}$,
\begin{equation}
\label{eq:MOLDProof-Simplify-Gray-Box1}
\begin{tikzpicture}[baseline=(current bounding box.west),
  every node/.style={inner sep=2pt,outer sep=2pt}        ]
    \matrix(ID)[
    matrix of math nodes,
    ampersand replacement=\&,
    row sep =0mm,
    column sep =0mm
    ]
    {\bar{P}_{\Theta} = \; \mathcal{C} \;
      \& \mathbf{S}_{\Theta_{(1)}}
      \& \mathbf{A}_{\Theta_{(1)}} 
      \& \mathbf{S}_{\Theta_{(1)}} 
      \& \cdots
      \& \mathbf{A}_{\Theta_{(m+2)}}
      \& \mathbf{S}_{\Theta_{(1)}}
      \& \mathbf{S}_{\Theta}
      \& \mathbf{A}_{\Theta} 
      \& \mathbf{A}_{\Theta_{(1)}}
      \& \mathbf{S}_{\Theta_{(m+1)}}
      \& \cdots
      \& \mathbf{S}_{\Theta_{(1)}}
      \& \mathbf{A}_{\Theta_{(1)}} 
      \& \mathbf{S}_{\Theta_{(1)}} 
      \& \; \mathcal{C}^{\dagger}
      \ .
      \\
};
\draw[draw=none,fill=red!20,thick] ($(ID-1-2.north west) + (4pt,0)$) rectangle
($(ID-1-4.south east) + (-4pt,0pt)$) node[midway] {$\mathbf{S}_{\Theta_{(1)}}\;\mathbf{A}_{\Theta_{(1)}}\;\mathbf{S}_{\Theta_{(1)}}$};
\draw[draw=none,fill=red!20,thick] ($(ID-1-13.north west) + (4pt,0)$) rectangle
($(ID-1-15.south east) + (-4pt,0pt)$) node[midway] {$\mathbf{S}_{\Theta_{(1)}}\;\mathbf{A}_{\Theta_{(1)}}\;\mathbf{S}_{\Theta_{(1)}}$};
\draw[draw=none,fill=blue!25,thick] ($(ID-1-8.north west) + (4pt,0)$) rectangle
($(ID-1-9.south east) + (-4pt,0pt)$) node[midway] {$\mathbf{S}_{\Theta}\;\mathbf{A}_{\Theta}$};
\draw[color=gray!60,thick] ($(ID-1-2.north west) + (0pt,2.5pt)$) rectangle
($(ID-1-15.south east) + (-1pt,-2pt)$);
\draw[stealth-,thick,red] (ID-1-7.north) |- ++(0,+2mm)   -|
(ID-1-8.north) node[pos=.2,anchor=south,yshift=-1mm]
{\scriptsize$\mathbf{S}_{\Theta}\to\mathbf{S}_{\Theta_{(1)}}\mathbf{S}_{\Theta}$};
\draw[-stealth,thick,red] (ID-1-9.south) |- ++(0,-3mm)   -|  (ID-1-10.south) node[pos=.2,anchor=north,yshift=0mm]   {\scriptsize$\mathbf{A}_{\Theta}\to\mathbf{A}_{\Theta}\mathbf{A}_{\Theta_{(1)}}$};
\end{tikzpicture}
\end{equation}
We now encounter sets of symmetrizers and antisymmetrizers
corresponding to ancestor tableaux $\Theta_{(k)}$ with $1\leq k\leq m$
wedged between sets belonging to the tableau $\Theta_{(1)}$. Thus, we
may use Theorem~\ref{thm:CancelMultipleSets} to simplify the operator
$\bar{P}_{\Theta}$ (once again making use of the bar-notation,
\emph{c.f.}  eq.~\eqref{eq:Bar-Notation}, to retain equality in the
process)
\begin{align}
\label{eq:MOLDProof-Simplify-Gray-Box2}
 &
\begin{tikzpicture}[baseline=(current bounding box.west),
  every node/.style={inner sep=2pt,outer sep=2pt}        ]
    \matrix(ID)[
    matrix of math nodes,
    ampersand replacement=\&,
    row sep =0mm,
    column sep =0mm
    ]
    {\bar{P}_{\Theta} = \; \mathcal{C} \;
      \& \mathbf{S}_{\Theta_{(1)}}
      \& \mathbf{A}_{\Theta_{(1)}} 
      \& \mathbf{S}_{\Theta_{(1)}} 
      \& \cdots
      \& \mathbf{A}_{\Theta_{(m+2)}}
      \& \mathbf{S}_{\Theta_{(1)}}
      \& \mathbf{S}_{\Theta}
      \& \mathbf{A}_{\Theta} 
      \& \mathbf{A}_{\Theta_{(1)}}
      \& \mathbf{S}_{\Theta_{(m+1)}}
      \& \cdots
      \& \mathbf{S}_{\Theta_{(1)}}
      \& \mathbf{A}_{\Theta_{(1)}} 
      \& \mathbf{S}_{\Theta_{(1)}} 
      \& \; \mathcal{C}^{\dagger}
      \\
};
% Boxes
\draw[draw=none,fill=red!20,thick] ($(ID-1-2.north west) + (4pt,0)$) rectangle
($(ID-1-4.south east) + (-4pt,0pt)$) node[midway] {$\mathbf{S}_{\Theta_{(1)}}\;\mathbf{A}_{\Theta_{(1)}}\;\mathbf{S}_{\Theta_{(1)}}$};
\draw[draw=none,fill=red!20,thick] ($(ID-1-13.north west) + (4pt,0)$) rectangle
($(ID-1-15.south east) + (-4pt,0pt)$) node[midway] {$\mathbf{S}_{\Theta_{(1)}}\;\mathbf{A}_{\Theta_{(1)}}\;\mathbf{S}_{\Theta_{(1)}}$};
\draw[draw=none,fill=blue!25,thick] ($(ID-1-8.north west) + (4pt,0)$) rectangle
($(ID-1-9.south east) + (-4pt,0pt)$) node[midway] {$\mathbf{S}_{\Theta}\;\mathbf{A}_{\Theta}$};
\draw[color=gray!60,thick] ($(ID-1-2.north west) + (0pt,2.5pt)$) rectangle
($(ID-1-15.south east) + (-1pt,-2pt)$);
% Braces
\draw[decorate,decoration={brace,amplitude=12pt},thick] (ID-1-7.south east) --
(ID-1-3.south west) node[pos=.5,anchor=north,yshift=-3.5mm]
{\scriptsize$\propto\mathbf{A}_{\Theta_{(1)}}\mathbf{S}_{\Theta_{(1)}}$};
\draw[decorate,decoration={brace,amplitude=12pt},thick] (ID-1-15.south east) --
(ID-1-10.south west) node[pos=.5,anchor=north,yshift=-3.5mm]
{\scriptsize$\propto\mathbf{A}_{\Theta_{(1)}}\mathbf{S}_{\Theta_{(1)}}$};
\end{tikzpicture}
\nonumber \\
& \phantom{\bar{P}_{\Theta}\; }
\begin{tikzpicture}[baseline=(current bounding box.west),
  every node/.style={inner sep=2pt,outer sep=2pt}        ]
    \matrix(ID)[
    matrix of math nodes,
    ampersand replacement=\&,
    row sep =0mm,
    column sep =0mm
    ]
    { = \; \mathcal{C} \;
      \& \mathbf{S}_{\Theta_{(1)}}
      \& \mathbf{A}_{\Theta_{(1)}} 
      \& \mathbf{S}_{\Theta_{(1)}} 
      \& \mathbf{S}_{\Theta}
      \& \mathbf{A}_{\Theta} 
      \& \mathbf{A}_{\Theta_{(1)}}
      \& \mathbf{S}_{\Theta_{(1)}}
      \& \; \mathcal{C}^{\dagger}
      \\
};
% Boxes
\draw[draw=none,fill=blue!25,thick] ($(ID-1-5.north west) + (4pt,0)$) rectangle
($(ID-1-6.south east) + (-4pt,0pt)$) node[midway] {$\mathbf{S}_{\Theta}\;\mathbf{A}_{\Theta}$};
\draw[color=gray!60,thick] ($(ID-1-2.north west) + (0pt,2.5pt)$) rectangle
($(ID-1-8.south east) + (-1pt,-2pt)$);
\end{tikzpicture}
\ .
\end{align}
Re-absorbing $\mathbf{S}_{\Theta_{(1)}}$ into $\mathbf{S}_{\Theta}$ and
$\mathbf{A}_{\Theta_{(1)}}$ into $\mathbf{A}_{\Theta}$ yields the
desired result,
\begin{align}
 &
\begin{tikzpicture}[baseline=(current bounding box.west),
  every node/.style={inner sep=1pt,outer sep=-1pt}        ]
    \matrix(ID)[
    matrix of math nodes,
    ampersand replacement=\&,
    row sep =0mm,
    column sep =0mm
    ]
    { \bar{P}_{\Theta}= \; \mathcal{C} \;\: 
      \& \: \mathbf{S}_{\Theta_{(1)}}
      \& \: \mathbf{A}_{\Theta_{(1)}} 
      \& \: \cancel{\mathbf{S}_{\Theta_{(1)}}} \: 
      \& \mathbf{S}_{\Theta}
      \& \: \mathbf{A}_{\Theta} 
      \& \: \cancel{\mathbf{A}_{\Theta_{(1)}}}
      \& \: \mathbf{S}_{\Theta_{(1)}}
      \& \;\:  \mathcal{C}^{\dagger}
      \\
};
% Boxes
\draw[draw=none,fill=blue!25,thick] ($(ID-1-5.north west) + (-2.5pt,4pt)$) rectangle
($(ID-1-6.south east) + (2pt,-4pt)$) node[midway] {$\mathbf{S}_{\Theta}\;\mathbf{A}_{\Theta}$};
\draw[color=gray!60,thick] ($(ID-1-2.north west) + (-3pt,6pt)$) rectangle
($(ID-1-8.south east) + (2.5pt,-4pt)$);
% Arrows
\draw[-stealth,thick,red] (ID-1-4.north) |- ++(0,+3mm)   -|
(ID-1-5.north) node[pos=.2,anchor=south,yshift=1mm]
{\scriptsize$\mathbf{S}_{\Theta_{(1)}}\mathbf{S}_{\Theta}\to\mathbf{S}_{\Theta}$};
\draw[stealth-,thick,red] (ID-1-6.south) |- ++(0,-3mm)   -|
(ID-1-7.south) node[pos=.2,anchor=north,yshift=-1mm]
{\scriptsize$\mathbf{A}_{\Theta}\mathbf{A}_{\Theta_{(1)}}\to\mathbf{A}_{\Theta}$};
\end{tikzpicture} 
\nonumber \\
& \phantom{\bar{P}_{\Theta}\; }
\begin{tikzpicture}[baseline=(current bounding box.west),
  every node/.style={inner sep=1pt,outer sep=-1pt}        ]
    \matrix(ID)[
    matrix of math nodes,
    ampersand replacement=\&,
    row sep =0mm,
    column sep =0mm
    ]
    { = \; \mathcal{C} \;\: 
      \& \: \mathbf{S}_{\Theta_{(1)}}
      \& \: {\color{blue}\mathbf{A}_{\Theta_{(1)}}}
      \& \: \mathbf{S}_{\Theta}
      \& \: {\color{red}\mathbf{A}_{\Theta}}
      \& \: \mathbf{S}_{\Theta_{(1)}}
      \& \: \; \mathcal{C}^{\dagger}
      \\
};
\draw[color=gray!60,thick] ($(ID-1-2.north west) + (-3pt,6pt)$) rectangle
($(ID-1-6.south east) + (2.5pt,-6pt)$);
\end{tikzpicture} 
\label{eq:MOLDHermProof-Short-PBar}
\ ,
\end{align}
thus concluding this step of the proof.

\item\label{itm:MOLDHermProof2} For the second step of the proof, we first
  notice that the operator obtained in the previous step,
  operator~\eqref{eq:MOLDHermProof-Short-PBar}, is Hermitian; this is due to
  the fact that $\bar{P}_{\Theta}$ as given
  in~\eqref{eq:MOLDHermProof7PBar} was constructed according to the
  iterative method described in the
  KS-Theorem~\ref{thm:KSProjectors},~\cite{Keppeler:2013yla}. In
  particular, this implies that
  $\bar{P}_{\Theta} = \bar{P}_{\Theta}^{\dagger}$, yielding
\begin{equation}
  \label{eq:MOLDHermProof19}
\bar{P}_{\Theta} \; = \; \mathcal{C}\;\fcolorbox{gray!60}{white}{$\mathbf{S}_{\Theta_{(1)}}{\color{blue}\mathbf{A}_{\Theta_{(1)}}}\mathbf{S}_{\Theta}{\color{red}\mathbf{A}_{\Theta}}\mathbf{S}_{\Theta_{(1)}}$}\;\mathcal{C}^{\dagger}
\; = \;
\mathcal{C}\;\fcolorbox{gray!60}{white}{$\mathbf{S}_{\Theta_{(1)}}{\color{red}\mathbf{A}_{\Theta}}\mathbf{S}_{\Theta}{\color{blue}\mathbf{A}_{\Theta_{(1)}}}\mathbf{S}_{\Theta_{(1)}}$}\;\mathcal{C}^{\dagger}
\; = \; \bar{P}_{\Theta}^{\dagger}.
\end{equation}
When we gave a proof of the Lexical-Theorem~\ref{thm:LexicalConstruction}
in Appendix~\ref{sec:ProofsLexical}, we were able to
prove that $\mathbf{A}_{\Theta_{(1)}}$ can be extended to become
$\mathbf{A}_{\Theta}$ by using techniques described in Appendix~\ref{sec:PropagationRules}. Now however, we are
no longer able to use these techniques, as most amputated tableaux
$\cancel{\Theta_r}$ or $\cancel{\Theta_c}$ would \emph{not} be
rectangular (as can be easily verified by an example). We therefore need a different strategy to
arrive at the desired form for $\bar{P}_{\Theta}$.

In addition to $\bar{P}_{\Theta}$ as given in
\eqref{eq:MOLDHermProof-Short-PBar}, let us define the operator $\bar{O}$ by
\begin{equation}\label{eq:MOLDHermProof19c}
  \bar{O} := \;
\mathcal{C}\;\fcolorbox{gray!60}{white}{$\mathbf{S}_{\Theta_{(1)}}{\color{red}\mathbf{A}_{\Theta}}\mathbf{S}_{\Theta}{\color{red}\mathbf{A}_{\Theta}}\mathbf{S}_{\Theta_{(1)}}$}\;\mathcal{C}^{\dagger};
\end{equation}
clearly, this operator is Hermitian by construction due to its symmetry.
We seek to show that $\bar{P}_{\Theta}=\bar{O}$ 
in order to conclude the second step of this proof. This will be 
accomplished by showing that 
\begin{equation}
  \label{eq:MOLDHermProofIncl1}
  \bar{O} \subset \bar{P}_{\Theta} \qquad \text{and} \qquad \bar{P}_{\Theta}
  \subset \bar{O},
\end{equation}
where we use the notation introduced in
section~\ref{sec:OperatorsNotation}. These inclusions will then lead us
to conclude that the subspaces onto which $\bar{O}$ and $\bar{P}_{\Theta}$
project are equal, rendering the two operators equal,
$\bar{O}=\bar{P}_{\Theta}$.

Let us prove the two inclusions
\eqref{eq:MOLDHermProofIncl1}: As discussed in section
\ref{sec:OperatorsNotation}, the first inclusion holds if and only if
$\bar{O}\cdot\bar{P}_{\Theta}=\bar{P}_{\Theta}=\bar{P}_{\Theta}\cdot\bar{O}$,
\emph{c.f.} equation~\eqref{eq:OperatorInclusion1}. We thus
need to examine the product of $\bar{O}$ and $\bar{P}_{\Theta}$. We
consider
\begin{equation}
  \bar{O} \cdot \bar{P}_{\Theta} \; = \;
  \mathcal{C}\;\fcolorbox{gray!60}{white}{$\mathbf{S}_{\Theta_{(1)}}{\color{red}\mathbf{A}_{\Theta}}
    \mathbf{S}_{\Theta}{\color{red}\mathbf{A}_{\Theta}}\mathbf{S}_{\Theta_{(1)}}$}\;\mathcal{C}^{\dagger}\cdot\mathcal{C}\;\fcolorbox{gray!60}{white}{$\mathbf{S}_{\Theta_{(1)}}{\color{blue}\mathbf{A}_{\Theta_{(1)}}}
    \mathbf{S}_{\Theta}{\color{red}\mathbf{A}_{\Theta}}\mathbf{S}_{\Theta_{(1)}}$}\;\mathcal{C}^{\dagger}
\ .
\end{equation}
Similar to what was done in part~\ref{itm:MOLDHermProof1}, we use
Theorem~\ref{thm:CancelMultipleSets} to simplify the central part of
the product $\bar{O} \cdot \bar{P}_{\Theta}$
\begin{equation}
  \bar{O} \cdot \bar{P}_{\Theta} \; = \;
  \mathcal{C}\;
\mathbf{S}_{\Theta_{(1)}}
{\color{red}\mathbf{A}_{\Theta}}
\underbrace{
\mathbf{S}_{\Theta}
{\color{red}\mathbf{A}_{\Theta}}
\mathbf{S}_{\Theta_{(1)}}
\;\mathcal{C}^{\dagger}\cdot\mathcal{C}\;
\mathbf{S}_{\Theta_{(1)}}
{\color{blue}\mathbf{A}_{\Theta_{(1)}}}
 \mathbf{S}_{\Theta}
{\color{red}\mathbf{A}_{\Theta}}
}_{\propto\mathbf{S}_{\Theta}{\color{red}\mathbf{A}_{\Theta}}}
\mathbf{S}_{\Theta_{(1)}}
\;\mathcal{C}^{\dagger}
\ ,
\end{equation}
yielding (making use of the bar-notation)
\begin{equation}
  \bar{O} \cdot \bar{P}_{\Theta} 
\; = \;
  \mathcal{C}\;
\mathbf{S}_{\Theta_{(1)}}
{\color{red}\mathbf{A}_{\Theta}}
\underbrace{
\mathbf{S}_{\Theta}
{\color{red}\mathbf{A}_{\Theta}}
}_{=\bar{Y}_{\Theta}}
\mathbf{S}_{\Theta_{(1)}}
\;\mathcal{C}^{\dagger}
\; = \;
\bar{O}
\ .
\end{equation}
Hence, we found that
$\bar{O}\cdot\bar{P}_{\Theta}=\bar{O}$. Recalling that both operators
$\bar{O}$ and $\bar{P}_{\Theta}$ are Hermitian, it follows that
\begin{equation} \bar{O}=\bar{O}^{\dagger}=\left(\bar{O}\cdot\bar{P}_{\Theta}\right)^{\dagger}=\bar{P}_{\Theta}^{\dagger}\cdot\bar{O}^{\dagger}=\bar{P}_{\Theta}\cdot\bar{O}.
\end{equation}
Thus, we have shown that both equalities,
$\bar{O}\cdot\bar{P}_{\Theta}=\bar{O}$ and
$\bar{P}_{\Theta}\cdot\bar{O}=\bar{O}$, hold, implying the first
inclusion $\bar{O}\subset\bar{P}_{\Theta}$.

To prove the second inclusion in~\eqref{eq:MOLDHermProofIncl1}, we
need to consider the product $\bar{P}_{\Theta}\cdot\bar{O}$,
\begin{equation}
  \bar{P}_{\Theta} \cdot \bar{O} \; = \; \mathcal{C}\;\fcolorbox{gray!60}{white}{$\mathbf{S}_{\Theta_{(1)}}{\color{blue}\mathbf{A}_{\Theta_{(1)}}} \mathbf{S}_{\Theta}{\color{red}\mathbf{A}_{\Theta}}\mathbf{S}_{\Theta_{(1)}}$}\;\mathcal{C}^{\dagger}\cdot\mathcal{C}\;\fcolorbox{gray!60}{white}{$\mathbf{S}_{\Theta_{(1)}}{\color{red}\mathbf{A}_{\Theta}} \mathbf{S}_{\Theta}{\color{red}\mathbf{A}_{\Theta}}\mathbf{S}_{\Theta_{(1)}}$}\;\mathcal{C}^{\dagger}.
\end{equation}
Once again, we may use Theorem~\ref{thm:CancelMultipleSets} to
simplify this product as
\begin{IEEEeqnarray}{rCl}
  \bar{P}_{\Theta} \cdot \bar{O} \; & = & \;
  \mathcal{C}\;
\mathbf{S}_{\Theta_{(1)}}
{\color{blue}\mathbf{A}_{\Theta_{(1)}}}
\underbrace{
\mathbf{S}_{\Theta}
{\color{red}\mathbf{A}_{\Theta}}
\mathbf{S}_{\Theta_{(1)}}
\;\mathcal{C}^{\dagger}\cdot\mathcal{C}\;
\mathbf{S}_{\Theta_{(1)}}
{\color{red}\mathbf{A}_{\Theta}}
\mathbf{S}_{\Theta}
{\color{red}\mathbf{A}_{\Theta}}
}_{\propto\mathbf{S}_{\Theta}{\color{red}\mathbf{A}_{\Theta}}}
\mathbf{S}_{\Theta_{(1)}}
\;\mathcal{C}^{\dagger}
  \nonumber \\
& = &
\mathcal{C}\;
\mathbf{S}_{\Theta_{(1)}}
{\color{blue}\mathbf{A}_{\Theta_{(1)}}}
\underbrace{
\mathbf{S}_{\Theta}
{\color{red}\mathbf{A}_{\Theta}}
}_{=\bar{Y}_{\Theta}}
\mathbf{S}_{\Theta_{(1)}}
\;\mathcal{C}^{\dagger}.
 \label{eq:MOLDHermProofIncl12}
\end{IEEEeqnarray}
We recognize the right hand side of the above equation
\eqref{eq:MOLDHermProofIncl12} to be the operator $\bar{P}_{\Theta}$. We
thus found that $\bar{P}_{\Theta}\cdot\bar{O}=\bar{P}_{\Theta}$.
Once again, we make use of the Hermiticity of the operators
$\bar{O}$ and $\bar{P}_{\Theta}$ to see that
\begin{equation}
  \bar{P}_{\Theta} 
= 
\bar{P}_{\Theta}^{\dagger} 
= 
\left( 
\bar{P}_{\Theta} 
\cdot
\bar{O} \right)^{\dagger} 
= 
\bar{O}^{\dagger}
\cdot
\bar{P}_{\Theta}^{\dagger} 
=
\bar{O}
\cdot
\bar{P}_{\Theta}
\ ,
\end{equation}
yielding the desired inclusion $\bar{P}_{\Theta}\subset\bar{O}$.  We
have thus managed to prove both inclusions in
\eqref{eq:MOLDHermProofIncl1}, forcing us to conclude that the two
operators $\bar{O}$ and $\bar{P}_{\Theta}$ are equal,
$\bar{O}=\bar{P}_{\Theta}$, yielding
\begin{equation}
  \label{eq:MOLDHermProof20}
\bar{P}_{\Theta} 
= 
\bar{O} 
=
\mathcal{C}\;
\fcolorbox{gray!60}{white}{$
\mathbf{S}_{\Theta_{(1)}}
{\color{red}\mathbf{A}_{\Theta}}
\mathbf{S}_{\Theta}
{\color{red}\mathbf{A}_{\Theta}}
\mathbf{S}_{\Theta_{(1)}}
$}
\;\mathcal{C}^{\dagger}
\ ,
\end{equation}
as desired by eq.~\eqref{eq:MOLDHermProof-Short-PBar}.
\end{enumerate}

\noindent Suppose now that $\bm{m}$ \textbf{is odd}. The proof for odd
$m$ will also be conducted in two steps, just as for even $m$. We will only give
an outline of this proof, as the steps are very similar to those
for even $m$.

By the induction hypothesis, the projection operator $P_{\Theta_{(1)}}$ is of the form
  \begin{equation}
    \bar{P}_{\Theta_{(1)}} 
= 
\; \mathbf{S}_{\Theta_{(m+1)}}
\; \mathbf{A}_{\Theta_{(m)}}  
\; \ldots
\; \mathbf{S}_{\Theta_{(2)}}
\; \underbrace{
\colorbox{red!20}{$
\mathbf{A}_{\Theta_{(1)}}  
\; \mathbf{S}_{\Theta_{(1)}}
\; \mathbf{A}_{\Theta_{(1)}}
$}
}_{=\bar{Y}_{\Theta_{(1)}}^{\dagger} \bar{Y}_{\Theta_{(1)}}} 
\; \mathbf{S}_{\Theta_{(2)}}  
\; \ldots
\; \mathbf{A}_{\Theta_{(m)}} 
\; \mathbf{S}_{\Theta_{(m+1)}}
\ .
  \end{equation}
Constructing the birdtrack of the Hermitian Young projection operator
$P_{\Theta}$, $\bar{P}_{\Theta}$, according
to the KS-Theorem~\ref{thm:KSProjectors}~\cite{Keppeler:2013yla} gives
\begin{multline}
\label{eq:MOLDHermProof23}
  \bar{P}_{\Theta} 
=
\bar{P}_{\Theta_{(1)}} 
\colorbox{blue!25}{$\bar{Y}_{\Theta}$} 
\; \bar{P}_{\Theta_{(1)}}
 \\
=
\mathcal{C}_{\Theta}
\; \fcolorbox{gray!60}{white}{
\colorbox{red!20}{
$\mathbf{A}_{\Theta_{(1)}}\mathbf{S}_{\Theta_{(1)}}\mathbf{A}_{\Theta_{(1)}}
$}
$\mathbf{S}_{\Theta_{(2)}}\ldots\mathbf{S}_{\Theta_{(m+1)}}$
\colorbox{blue!25}{
$\mathbf{S}_{\Theta}\mathbf{A}_{\Theta}$
}
$\mathbf{S}_{\Theta_{(m+1)}}\ldots\mathbf{S}_{\Theta_{(2)}}$
\colorbox{red!20}{
$\mathbf{A}_{\Theta_{(1)}}\mathbf{S}_{\Theta_{(1)}}\mathbf{A}_{\Theta_{(1)}}$
}}
\; \mathcal{C}_{\Theta}^{\dagger}
\ , 
\end{multline}
where $\mathcal{C}_{\Theta}:=\mathbf{S}_{\Theta_{(m+1)}}\ldots\mathbf{S}_{\Theta_{(2)}}$.
We again use Theorem~\ref{thm:CancelMultipleSets} to simplify the operator~\eqref{eq:MOLDHermProof23},
\begin{equation}
   \bar{P}_{\Theta} 
= 
\; \mathbf{S}_{\Theta_{(m+1)}}
\ldots
\mathbf{S}_{\Theta_{(2)}}
\fcolorbox{gray!60}{white}{$
\mathbf{A}_{\Theta_{(1)}}
{\color{red}\mathbf{S}_{\Theta}}
\mathbf{A}_{\Theta}
{\color{blue}\mathbf{S}_{\Theta_{(1)}}}
\mathbf{A}_{\Theta_{(1)}}
$}
\;\mathbf{S}_{\Theta_{(2)}}
\ldots
\mathbf{S}_{\Theta_{(m+1)}}
\ .
\end{equation}
We then define an operator $\bar{O}$ by
\begin{equation}
  \bar{O}
 := 
\; \mathbf{S}_{\Theta_{(m+1)}}
\ldots
\mathbf{S}_{\Theta_{(2)}}
\fcolorbox{gray!60}{white}{$
\mathbf{A}_{\Theta_{(1)}}
{\color{red}\mathbf{S}_{\Theta}}
\mathbf{A}_{\Theta}
{\color{red}\mathbf{S}_{\Theta}}
\mathbf{A}_{\Theta_{(1)}}
$}
\; \mathbf{S}_{\Theta_{(2)}}
\ldots
\mathbf{S}_{\Theta_{(m+1)}}
\ .
\end{equation}
Using Theorem
\ref{thm:CancelMultipleSets} as well as the fact that both
$\bar{P}_{\Theta}$ and $\bar{O}$ are Hermitian by construction, we may show the inclusions
$\bar{P}_{\Theta}\subset\bar{O}$ and $\bar{O}\subset\bar{P}_{\Theta}$,
to conclude that $\bar{P}_{\Theta}=\bar{O}$, yielding the desired result
\begin{equation}
   \bar{P}_{\Theta} = \;  \mathbf{S}_{\Theta_{(m+1)}}\ldots\mathbf{S}_{\Theta_{(2)}}\fcolorbox{gray!60}{white}{$\mathbf{A}_{\Theta_{(1)}}\underbrace{\mathbf{S}_{\Theta}\mathbf{A}_{\Theta}\mathbf{S}_{\Theta}}_{=\bar{Y}_{\Theta}
       \bar{Y}_{\Theta}^{\dagger}}\mathbf{A}_{\Theta_{(1)}}$}\;\mathbf{S}_{\Theta_{(2)}}\ldots\mathbf{S}_{\Theta_{(m+1)}}
\ .
\end{equation}

The proof of equations~\eqref{eq:MOLDHermitianOperatorConstruction3}
and~\eqref{eq:MOLDHermitianOperatorConstruction4} in the MOLD-Theorem follows the same
steps as the proof of
equations~\eqref{eq:MOLDHermitianOperatorConstruction1} and~\eqref{eq:MOLDHermitianOperatorConstruction2} given above and is thus left as an exercise to the reader.

Lastly, we notice that the idempotency of $P_{\Theta}$ in each of the
cases~\eqref{eq:MOLDConstruction} can be verified by using the
Cancellation-Theorem~\ref{thm:CancelMultipleSets}: For example if
$P_{\Theta}$ is constructed according
to~\eqref{eq:MOLDHermitianOperatorConstruction1}, it follows that
\begin{IEEEeqnarray*}{rCl}
  P_{\Theta}\cdot P_{\Theta} & = & \beta_{\Theta}^2 \cdot \mathbf{S}_{\Theta_{(m)}}
     \; \ldots 
     \; \underbrace{\colorbox{red!20}{$\mathbf{S}_{\Theta} \; \mathbf{A}_{\Theta}
       \; \mathbf{S}_{\Theta}$}  \; \ldots
     \; \mathbf{S}_{\Theta_{(m)}} \cdot \mathbf{S}_{\Theta_{(m)}}
     \; \ldots 
     \; \colorbox{red!20}{$\mathbf{S}_{\Theta} \; \mathbf{A}_{\Theta}
       \; \mathbf{S}_{\Theta}$}}_{= \lambda \cdot \colorbox{red!20}{$\mathbf{S}_{\Theta} \; \mathbf{A}_{\Theta}
       \; \mathbf{S}_{\Theta}$}}
     \; \ldots 
     \; \mathbf{S}_{\Theta_{(m)}} \\
& = & \beta_{\Theta}^2 \lambda \cdot \mathbf{S}_{\Theta_{(m)}}
     \; \ldots 
     \; \colorbox{red!20}{$\mathbf{S}_{\Theta} \; \mathbf{A}_{\Theta}
       \; \mathbf{S}_{\Theta}$}  \; \ldots
     \; \mathbf{S}_{\Theta_{(m)}},
\end{IEEEeqnarray*}
where $\lambda$ is a \emph{non-zero} constant, since all the cancelled
sets can be absorbed into $\mathbf{S}_{\Theta}$ and
$\mathbf{A}_{\Theta}$ respectively (\emph{c.f.}
Theorem~\ref{thm:CancelMultipleSets}). Thus, defining
  \begin{equation}
    \beta_{\Theta} := \frac{1}{\lambda} < \infty
  \end{equation}
ensures that $P_{\Theta}$ is indeed a projection operator. \qed

%%%%%%%%%%%%%%%%%%%%%%%%%%%%%%%%%%%%%%%%%%%%%%%%%%%%%%%%%%%%%%%%%%%%%%
 \bibliographystyle{utphys}
 \bibliography{PaperLibrary,BookLibrary,GroupTheory}
%\input{HermitianProjPaper-v17.bbl}
%%%%%%%%%%%%%%%%%%%%%%%%%%%%%%%%%%%%%%%%%%%%%%%%%%%%%%%%%%%%%%%%%%%%%%

\end{document}